\documentclass{aa}
\pdfoutput=1
\usepackage{txfonts}
\usepackage{url}
\usepackage{hyperref}
\usepackage{natbib}
\usepackage{graphicx}

\newcommand{\x}{\vec{x}}
\renewcommand{\d}[0]{{\rm d}}
\newcommand{\rmn}[1]{{\mathrm{#1}}}

\def\bigstrut{\vrule width0pt height1.1em depth0.6em}
\def\smallstrut{\vrule width0pt height0.9em depth0.4em}

\begin{document}

\defcitealias{hildebrandt/etal:2017}{H17}
\defcitealias{fenechconti/etal:2017}{FC17}
\defcitealias{wright/etal:2018}{W18}
\defcitealias{kannawadi/etal:2018}{K19}

\title{KiDS+VIKING-450: Cosmic shear tomography with optical+infrared data}
\author
{H.~Hildebrandt
\inst{1,2}
\and
F.~K\"ohlinger\inst{3}
\and
J.~L.~van~den~Busch\inst{2,1}
\and
B.~Joachimi\inst{4}
\and
C.~Heymans\inst{5,1}
\and
A.~Kannawadi\inst{6}
\and
A.~H.~Wright\inst{2,1}
\and
M.~Asgari\inst{5}
\and
C.~Blake\inst{7}
\and
H.~Hoekstra\inst{6}
\and
S.~Joudaki\inst{8}
\and
K.~Kuijken\inst{6}
\and
L.~Miller\inst{8}
\and
C.~B.~Morrison\inst{9}
\and
T.~Tröster\inst{5}
\and
A.~Amon\inst{5}
\and
M.~Archidiacono\inst{10}
\and
S.~Brieden\inst{10,11,12}
\and
A.~Choi\inst{13}
\and
J.~T.~A.~de~Jong\inst{14}
\and
T.~Erben\inst{2}
\and
B.~Giblin\inst{5}
\and
A.~Mead\inst{15}
\and
J.~A.~Peacock\inst{5}
\and
M.~Radovich\inst{16}
\and
P.~Schneider\inst{2}
\and
C.~Sif\'on\inst{17,18}
\and
M.~Tewes\inst{2}
}

\authorrunning{The KiDS collaboration}

\institute{
Astronomisches Institut, Ruhr-Universit\"at Bochum,
Universit\"atsstr. 150, 44801, Bochum, Germany,\\ \email{hendrik@astro.ruhr-uni-bochum.de}
\and
Argelander-Institut f\"ur Astronomie, Universit\"at Bonn, Auf dem H\"ugel 71, 53121 Bonn, Germany
\and
Kavli Institute for the Physics and Mathematics of the Universe (WPI), The University of Tokyo Institutes for Advanced Study, The University of Tokyo, Kashiwa, Chiba 277-8583, Japan
\and
Department of Physics and Astronomy, University College London, Gower Street, London WC1E 6BT, UK
\and
Institute for Astronomy, University of Edinburgh, Royal Observatory, Blackford Hill, Edinburgh EH9 3HJ, UK
\and
Leiden Observatory, Leiden University, Niels Bohrweg 2, 2333 CA
Leiden, the Netherlands
\and
Centre for Astrophysics \& Supercomputing, Swinburne University of
Technology, PO Box 218, Hawthorn, VIC 3122, Australia
\and
Department of Physics, University of Oxford, Denys Wilkinson Building, Keble Road, Oxford OX1 3RH, U.K.
\and
Department of Astronomy, University of Washington, Box 351580,
Seattle, WA 98195, USA
\and
Institute for Theoretical Particle Physics and Cosmology (TTK), RWTH Aachen University, 52056 Aachen, Germany
\and
ICC, University of Barcelona, IEEC-UB, Mart\'i i Franqu\'es, 1,
08028 Barcelona, Spain
\and
Dept. de F\'isica Qu\`antica i Astrof\'isica, Universitat de Barcelona, Mart\'i i Franqu\'es 1, 08028 Barcelona, Spain
\and
Center for Cosmology and AstroParticle Physics, The Ohio State University, 191 West Woodruff Avenue, Columbus, OH 43210, USA
\and
Kapteyn Astronomical Institute, University of Groningen, 9700AD
Groningen, the Netherlands
\and
Department of Physics and Astronomy, University of British Columbia, 6224 Agricultural Road, Vancouver, BC V6T 1Z1, Canada
\and
INAF -- Osservatorio Astronomico di Padova, via dell'Osservatorio 5, 35122 Padova, Italy
\and
Department of Astrophysical Sciences, Peyton Hall, Princeton
University, Princeton, NJ 08544, USA
\and
Instituto de F\'isica, Pontificia Universidad Cat\'olica de Valpara\'iso, Casilla 4059, Valpara\'iso, Chile
}

\date{Received December 14, 2018; accepted November 15, 2019}
\abstract{ We present a tomographic cosmic shear analysis of the
  Kilo-Degree Survey (KiDS) combined with the VISTA Kilo-Degree
  Infrared Galaxy Survey (VIKING). This is the first time that a full
  optical to near-infrared data set has been used for a wide-field
  cosmological weak lensing experiment. This unprecedented data,
  spanning 450~deg$^2$, allows us to improve significantly the
  estimation of photometric redshifts, such that we are able to
  include robustly higher-redshift sources for the lensing
  measurement, and – most importantly – solidify our knowledge of the
  redshift distributions of the sources. Based on a flat $\Lambda$CDM
  model we find
  $S_8\equiv\sigma_8\sqrt{\Omega_{\rm m}/0.3}=0.737_{-0.036}^{+0.040}$
  in a blind analysis from cosmic shear alone. The tension between
  KiDS cosmic shear and the Planck-Legacy CMB measurements remains in
  this systematically more robust analysis, with $S_8$ differing by
  $2.3\sigma$. This result is insensitive to changes in the priors on
  nuisance parameters for intrinsic alignment, baryon feedback, and
  neutrino mass. KiDS shear measurements are calibrated with a new,
  more realistic set of image simulations and no significant B-modes
  are detected in the survey, indicating that systematic errors are
  under control. When calibrating our redshift distributions by
  assuming the 30-band COSMOS-2015 photometric redshifts are correct
  (following the Dark Energy Survey and the Hyper Suprime-Cam Survey),
  we find the tension with Planck is alleviated. The robust
  determination of source redshift distributions remains one of the
  most challenging aspects for future cosmic shear surveys. Data
  products from this analysis are available at
  \url{http://kids.strw.leidenuniv.nl}.}
\maketitle

\section{Introduction}
\label{sec:intro}

Observational cosmology is progressing at a fast pace. Increasingly precise measurements test the predictions of the standard $\Lambda$CDM cosmological model from multiple angles. The main cosmological parameters have been determined with great precision through various missions measuring fluctuations in the Cosmic Microwave Background (CMB) radiation, most recently by the ESA Planck satellite \citep{planck/cosmo:2018}. These measurements mainly probe the Universe's physics at a redshift of $z\sim 1100$. If the underlying assumptions of $\Lambda$CDM are correct, the values of the parameters estimated from CMB measurements should agree with results from lower-redshift probes. Several such cosmological measurements at later cosmic times have been established over the past couple of decades, ranging from the Hubble diagram of supernovae of type Ia \citep[e.g.][]{betoule/etal:2014} over galaxy redshift surveys \citep[e.g.][]{alam/etal:2017}, determinations of the galaxy cluster mass function \citep[e.g.][]{bocquet/etal:2018}, to measurements of gravitational lensing \citep{jee/etal:2016,hildebrandt/etal:2017,troxel/etal:2018b,hikage/etal:2018}.

In general, the agreement between these -- quite different -- probes is surprisingly good, increasing the confidence that $\Lambda$CDM indeed yields a correct description of reality. The sheer number of consistent results means that any single mildly discrepant result should be regarded with a healthy dose of scepticism. A falsification of the extremely successful $\Lambda$CDM paradigm would certainly require very convincing evidence. The greatest parameter discrepancy within $\Lambda$CDM, one whose statistical significance has been growing over the past few years, is the difference in the value of the Hubble constant determined from Planck and from distance ladder measurements \citep[see][quoting a significance of $3.8\sigma$]{riess/etal:2018}. Here we explore another test of the model, the growth of large-scale structure.

It is not expected that measurements of primary CMB anisotropies from near-future experiments will lead to much greater precision in measurements of key parameters like the matter density, $\Omega_{\rm m}$, the amplitude of the matter power spectrum, $\sigma_8$,\footnote{Linear-theory root mean square fluctuations of the matter density contrast in spheres with a radius of 8~$h^{-1}$Mpc at redshift $z=0$.} or the Hubble constant, $H_0$. Most information about these parameters has already been optimally extracted from the Planck data \citep{planck/cosmo:2018}. Also the CMB alone cannot constrain the dark energy equation-of-state $w=p/\rho$ very precisely as the effects of the accelerating expansion only become important at late cosmic times. While ongoing ground-based CMB experiments will yield very interesting insights into small-scale fluctuations and measure CMB polarisation with unprecedented precision, those new measurements will not add much statistical power to the measurements of $\Omega_{\rm m}$, $\sigma_8$, $H_0$, and $w$. Hence, in order to provide a further challenge to the standard model, other probes have to push the envelope.

Weak gravitational lensing by the large-scale structure of the Universe \citep[also known as cosmic shear; see][for reviews]{kilbinger:2015,mandelbaum:2018} is one of these probes that is currently making rapid progress with increasingly large, dedicated experiments coming online. This delicate measurement of millions -- or in the near-future billions -- of galaxy ellipticities and redshifts has to be understood in such a way that systematic errors remain subdominant to the quickly decreasing statistical uncertainties.

Recently we presented one of the most robust cosmic shear analyses to date \citep[][hereafter H17]{hildebrandt/etal:2017} based on data from the European Southern Observatory's Kilo-Degree Survey \citep[KiDS; ][]{kuijken/etal:2015,dejong/etal:2015,dejong/etal:2017}. Using $\sim450~{\rm deg}^2$ of four-band ($ugri$) data (hence the name `KiDS-450') we measured $S_8\equiv\sigma_8\sqrt{\Omega_{\rm m}/0.3}$ with a relative error of $\sim5\%$. This uncertainty was estimated from a comprehensive and redundant analysis of, and subsequent marginalisation over, all known systematic errors. A blinding scheme was used to suppress confirmation biases and yield an objective result. 

Surprisingly the measurements were found to be discrepant at the $2.3\sigma$ level with results from the Planck CMB experiment \citep{planck/cosmo:2015}. While this mild disagreement might very well be a chance fluctuation it could also hint at some systematic problem with either or both of the two experiments. Another more far-reaching possibility that could explain these results would be a deviation from $\Lambda$CDM \citep[for an example of an extended cosmological model that eases this tension see][]{joudaki/etal:2017b}. However, it is clear that a $2.3\sigma$ `detection' is not convincing enough to make such a radical claim. There are other low-redshift large-scale structure probes that also measure lower values of $S_8$ than Planck \citep[for an overview see][]{mccarthy/etal:2018}, but currently it is not clear yet if these $S_8$ discrepancies between early and late Universe probes are due to unknown systematics or -- perhaps in combination with the $H_0$ tension described above -- hint at a fundamental problem with the cosmological standard model.

The rapid progress in cosmic shear surveys makes it possible to improve on this situation in the near future with more precise measurements. The Dark Energy Survey \citep[DES,][]{flaugher/etal:2015} as well as the Hyper Suprime-Cam (HSC) Wide Survey \citep{aihara/etal:2017} have recently reached a statistical power that surpasses the measurement by \citetalias{hildebrandt/etal:2017}. Given that the systematic and statistical errors in \citetalias{hildebrandt/etal:2017} were very similar in size, as their data volumes increase, the challenge for all three surveys will be to control their systematic errors such that they do not compromise their unprecedented statistical power in the future.

The cosmic shear results from the first year of DES observations \citep[DESy1,][]{troxel/etal:2018a} as well as the first data release of HSC \citep[HSC-DR1,][]{hikage/etal:2018} are fully consistent with the KiDS results, but both show a somewhat higher value for $S_8$. Their results lie in-between the KiDS-450 measurement and the Planck-2015 value. Several aspects of the DESy1 as well as HSC-DR1 cosmic shear analyses differ from the analysis presented in \citetalias{hildebrandt/etal:2017}, where some of these differences are explored in \citet{troxel/etal:2018b}. We would argue that the most important difference, namely the way the different surveys estimate their redshift distributions, has not received as much attention, however, and we address this issue -- amongst other things -- in this work.

KiDS observations are still ongoing so that future cosmic shear measurements with this survey will beat down statistical noise. But it is the systematic side of the error budget where KiDS has the greatest potential. One important difference between KiDS on the one hand and DES and HSC on the other hand is that KiDS is observed with a dedicated weak lensing telescope with more benign point-spread-function (PSF) distortions. This is partly due to the fact that its camera is located in the Cassegrain instead of the prime focus as for e.g. DES and HSC. Another unique aspect of KiDS is that it fully overlaps with a well-matched (in terms of depth) infrared survey, the VISTA Kilo-Degree Infrared Galaxy Survey (VIKING). This additional near-infrared (NIR) imaging data helps in determining more accurate photometric redshifts (photo-$z$), one indispensable requirement for cosmic shear measurements. The infrared data improve the performance of these photo-$z$ in the high-redshift regime so that higher-$z$ sources can be selected and exploited for the lensing measurement. Hence, adding VIKING to KiDS means that cosmic shear results become not only more robust but also more precise, and probing structures at slightly higher redshifts.

In this paper we present an updated cosmic shear tomography measurement based on the integration of VIKING imaging into the KiDS-450 data set, dubbed KV450, which represents the first time that cosmic shear tomography has been measured from a combined optical+NIR data set over hundreds of square degrees. The KiDS optical and VIKING NIR data and their reduction are briefly described in Sect.~\ref{sec:data}. The tomographic binning and in particular the calibration of the redshifts are covered in Sect.~\ref{sec:photo}. Galaxy shape measurements are discussed in Sect.~\ref{sec:shapes}. The estimation of correlation functions and their covariance are described in Sect.~\ref{sec:cov} and the details of the theoretical model are introduced in Sect.~\ref{sec:theory}. Cosmological results are presented in Sect.~\ref{sec:results} and discussed in Sect.~\ref{sec:discussion}. The paper is summarised and an outlook to future work is given in in Sect.~\ref{sec:summary}. 

For the expert reader who is familiar with the analysis presented in \citetalias{hildebrandt/etal:2017}, Appendix~\ref{app:changes} presents a concise list of the changes included in this analysis. Some of the more technical aspects of this work are then presented in further appendices, where Appendix~\ref{app:triangle} presents the posterior distributions for the full set of cosmological parameters, Appendix~\ref{app:photoz_tests} details redundant techniques to determine redshift distributions and some consistency checks, Appendix~\ref{app:codecomp} shows a comparison of results from different cosmology codes on the data, and Appendix~\ref{app:history} reports the timeline of this project, in particular the handling of the blinding.

\section{Data}
\label{sec:data}

\subsection{Imaging Data}
In this work, we utilise the combined KiDS+VIKING-450 (KV450) data set described in \citet[][hereafter W18]{wright/etal:2018}. The optical data, object detection, optical photometry, and ellipticity measurements are unchanged compared to \citetalias{hildebrandt/etal:2017}. Forced matched-aperture photometry on the VIKING NIR data is extracted with the \textsc{GAaP} \citep[Gaussian Aperture and PSF;][]{kuijken:2008,kuijken/etal:2015} method from individual exposures. This 5-band NIR photometry is combined with the 4-band optical photometry to estimate new, more accurate photo-$z$. For KV450 we use a newer version of the \textsc{BPZ} (Bayesian Photometric Redshift) photo-$z$ code \citep[v1.99.3;][]{benitez:2000,coe/etal:2006} and an improved redshift prior \citep{raichoor/etal:2014}. Details on the data reduction, multi-band photometry, and photo-$z$ performance are covered in \citetalias{wright/etal:2018}.

The main properties of the combined KV450 data set are:
\begin{itemize}
\item The effective, unmasked area reduces from 360.3~deg$^2$ to 341.3~deg$^2$ due to incomplete coverage of VIKING. We only use the area that is fully covered in all nine bands.
\item As some of the VIKING data were taken under poor seeing conditions the \textsc{GAaP} photometry failed in some fields and bands for the smallest objects. This is due to the aperture being chosen based on the good-seeing KiDS $r$-band image, which can lead to apertures that are too small for fluxes to be extracted from the worst-seeing VIKING images. We do not use these objects in the analysis, but this decision results in a source density that is varying more strongly than for KiDS-450. Details about this can be found in \citetalias{wright/etal:2018}
\item The photo-$z$ improve considerably as detailed in \citetalias{wright/etal:2018}. In particular, the performance at high redshifts is dramatically improved, with photo-$z$ scatter and outlier rates being smaller by a factor of $\sim2$ at $z>1$, so that we can reliably select high-redshift galaxies for our cosmic shear measurement.
\end{itemize}

\subsection{Spectroscopic data}
\label{sec:spec}

The KV450 photo-$z$ calibration (see Sect.~\ref{sec:DIR}) relies heavily on spectroscopic surveys. We distinguish between deep, pencil-beam surveys that are used for the weighted, direct calibration (DIR, Sect.~\ref{sec:DIR}) and wide, shallow spectroscopic redshift (spec-$z$) surveys that are only used for photo-$z$ calibration with small-scale cross-correlations (CC, Appendix~\ref{sec:CC}) and a complementary large-scale clustering-redshift estimate from an optimal quadratic estimator (OQE, Appendix~\ref{app:OQE}), with some of the deep, pencil-beam surveys also contributing to the CC technique. 

The deep spec-$z$ surveys employed for the KV450 photo-$z$ calibration are:
\begin{itemize}

\item zCOSMOS \citep{lilly/etal:2009}: Here we use a non-public, deep zCOSMOS catalogue that was kindly provided to us by the zCOSMOS team for KiDS photo-$z$ calibration. We measure CC over an area of $\sim0.5$~deg$^2$ with this data set. For DIR we use a slightly larger catalogue. These additional spec-$z$ from zCOSMOS cannot be used for CC because of their more inhomogeneous spatial distribution at the edge of the zCOSMOS observing area due to incomplete targeting, which biases angular correlation function measurements. While the COSMOS field is observed by KiDS, it is not in the VIKING footprint because very deep VISTA data in the $YJHK_{\rm s}$ bands are available in this field through the UltraVISTA project \citep{mccracken/etal:2012}. We add $z$-band data from the CFHTLS-Deep project \citep{hudelot/etal:2012} to complete the filter set.\footnote{Note that the MegaCam@CFHT $z$-band filter is similar to the VIRCAM@VISTA $z$-band filter. We ignore the subtle differences here as they do not play any role at the signal-to-noise level of our lensing galaxies.} 

\item DEEP2 Redshift Survey \citep{newman/etal:2013}: While KV450 itself does not overlap with DEEP2 we obtained KiDS- and VIKING-like data in two of the DEEP2 fields (one KiDS/VIKING pointing of $\sim1$~deg$^2$ each) so that these very rich spectroscopic fields can be used for CC as well as DIR. DEEP2 is colour-selected in these two equatorial fields and provides mostly information in the crucial redshift range $0.5\la z \la 1.5$.

\item VVDS \citep[VIMOS VLT Deep Survey,][]{lefevre/etal:2013}: Similarly to DEEP2 we obtained KiDS- and VIKING-like data on the VVDS-Deep equatorial field at $\mathrm{RA}\approx2$h. This very deep field that was not available for \citetalias{hildebrandt/etal:2017} reduces sample variance and susceptibility to selection effects in the CC and DIR calibrations and adds some very faint, high-$z$ galaxies to the calibration sample.

\item GAMA-G15Deep \citep{kafle/etal:2018}: An area of $\sim1$~deg$^2$ was observed to greater depth in the GAMA survey \citep{driver/etal:2011}. Targets were selected down to an $r$-band magnitude of $r<22$ instead of $r<19.8$ as in the rest of the survey. This deep GAMA field called G15Deep is part of the KV450 footprint and is used for DIR. It contains mostly galaxies with $z\la0.7$

\item CDFS (Chandra Deep Field South): We use the combined spec-$z$ catalogue provided by ESO\footnote{\url{http://www.eso.org/sci/activities/garching/projects/goods/MasterSpectroscopy.html}}, which adds some very faint objects to the DIR calibration sample. Most of the spec-$z$ used in CDFS come from either VVDS \citep{lefevre/etal:2013} or ESO-GOODS \citep{popesso/etal:2009,balestra/etal:2010,vanzella/etal:2008}. KiDS-like imaging data were obtained from the VOICE project \citep{vaccari/etal:2016}, and VISTA-VIDEO \citep{jarvis/etal:2013} data were degraded to VIKING depth in this field.
\end{itemize}

The wide area spec-$z$ surveys that we employ are GAMA \citep[Galaxy and Mass Assembly,][]{driver/etal:2011}, SDSS \citep[Sloan Digital Sky Survey,][]{alam/etal:2015}, 2dFLenS \citep[2-degree Field Lensing Survey,][]{blake/etal:2016}, and WiggleZ Dark Energy Survey \citep{drinkwater/etal:2010}. These are described in more detail in Appendix~\ref{sec:CC}. Properties of the spec-$z$ samples used for calibration are summarised in Table~\ref{tab:specz}. We only use highly secure redshift measurements corresponding to an estimated confidence of at least 95\%\footnote{This corresponds to quality flags 3 and 4 for zCOSMOS, VVDS, and DEEP2.}. Note that most objects have more secure redshift estimates so that the total fraction of spec-$z$ failures will be $\ll5\%$, more around $\sim1\%$.

\begin{table}
\caption{\label{tab:specz}Spectroscopic redshift surveys used for the calibration of KV450 photo-$z$. }
\centering
\begin{tabular}{lrrrrl}
\hline
\hline
Survey & Area      & No. of   & $z$-max & $r_\mathrm{lim}$ & Used for \smallstrut\\
       & [deg$^2$] & spec-$z$ &         &                & \smallstrut\\
\hline
SDSS$^*$    &119.2 & 15564 & 0.7 &      & CC/OQE \smallstrut\\
GAMA$^*$    & 75.9 & 79756 & 0.4 & 19.8 & CC/OQE \smallstrut\\
2dFLenS$^*$ & 61.2 &  3914 & 0.8 &      & CC/OQE \smallstrut\\
WiggleZ$^*$ & 60.1 & 19968 & 1.1 &      & CC/OQE \smallstrut\\
zCOSMOS     &  0.7 &  9930 & 1.0 & 24   & CC/DIR\smallstrut\\
DEEP2       &  0.8 &  6919 & 1.5 & 24.5 & CC/DIR\smallstrut\\
VVDS$^*$    &  1.0 &  4688 & 1.3 & 25   & CC/DIR\smallstrut\\
G15Deep$^*$ &  1.0 &  1792 & 0.7 & 22   & DIR\smallstrut\\
CDFS        &  0.1 &  2044 & 1.4 & 25   & DIR\smallstrut\\
\hline
\end{tabular}
\tablefoot{The second column contains the overlap area used for calibration after quite conservative masking for good, homogeneous coverage (by the spec-$z$ survey as well as KiDS and VIKING). The numbers in the third column correspond to the objects with secure spectroscopic redshift measurements in the overlap area. The maximum redshift in the fourth column is an approximate estimate up to which redshift data from a particular survey contribute significantly to the calibration. The last column reports which redshift calibration techniques make use of the different samples. An asterisk in the first column indicates new calibration data that were not used in \citetalias{hildebrandt/etal:2017}.}
\end{table}

\section{Tomographic bins \& redshift calibration}
\label{sec:photo}

The KV450 data set presented here is unique because never before has a combined optical+NIR data set been used for cosmic shear tomography over hundreds of square degrees. It is hence the KV450 photometric redshifts that represent the most important improvement compared to previous work. In this section we detail how we select galaxies in tomographic bins (Sect.~\ref{sec:zbinning}) and estimate their redshift distributions (Sect.~\ref{sec:DIR}). For the latter task we use the well-established weighted direct calibration technique with deep spectroscopic redshifts catalogues, which was already used in \citetalias{hildebrandt/etal:2017}, with some crucial improvements. The systematic robustness of the resulting redshift distributions is tested by looking at subsamples of the spectroscopic calibration sample, an independent high-quality photo-$z$ calibration sample from the COSMOS field, a post-processing step to suppress residual large-scale structure (Appendix~\ref{sec:sDIR}), and precise clustering-redshift techniques (Appendices~\ref{sec:CC}~\&~\ref{app:OQE}) that are completely independent and conceptually very different from the fiducial method. Thus, there is a great level of redundancy in this redshift calibration that should increase the reliability of the cosmological conclusions based on these redshift distributions.

\subsection{Photo-$z$ binning}
\label{sec:zbinning}
We bin galaxies in five tomographic redshift bins according to their photo-$z$ estimate $z_{\rm B}$ (most probable Bayesian redshift from \textsc{BPZ}). As in \citetalias{hildebrandt/etal:2017}, we define four bins of width $\Delta z_{\rm B}=0.2$ over the range $0.1<z_{\rm B}\le 0.9$. A fifth bin including all galaxies with $0.9<z_{\rm B}\le1.2$ is added here thanks to the greatly improved high-redshift performance of the 9-band photo-$z$ and improved shear calibration (see Sect.~\ref{sec:m_bias}). Properties of the galaxies in the different bins are summarised in Table~\ref{tab:tomobins}.

\begin{table*}
\caption{\label{tab:tomobins}Properties of the galaxies in the five tomographic redshift bins used for the KV450 cosmic shear measurements.}
\centering
\begin{tabular}{ccrrrrr}
\hline
\hline
Bin & $z_{\rm B}$ range & No. of & $n_{\rm eff}$ H12 & $\sigma_\epsilon$ & $\left<z_{\rm DIR}\right>$ & $m$-bias \smallstrut\\
   && objects &  [arcmin$^{-2}$] & & &\smallstrut\\
\hline
1   & $0.1<z_{\rm B}\le 0.3$ & 1\,027\,504 & 0.80 & 0.276 & $0.394 \pm 0.039$ & $-0.017 \pm 0.02$ \smallstrut\\
2   & $0.3<z_{\rm B}\le 0.5$ & 1\,798\,830 & 1.33 & 0.269 & $0.488 \pm 0.023$ & $-0.008 \pm 0.02$ \smallstrut\\
3   & $0.5<z_{\rm B}\le 0.7$ & 3\,638\,808 & 2.35 & 0.290 & $0.667 \pm 0.026$ & $-0.015 \pm 0.02$ \smallstrut\\
4   & $0.7<z_{\rm B}\le 0.9$ & 2\,640\,450 & 1.55 & 0.281 & $0.830 \pm 0.012$ & $+0.010 \pm 0.02$ \smallstrut\\
5   & $0.9<z_{\rm B}\le 1.2$ & 2\,628\,350 & 1.44 & 0.294 & $0.997 \pm 0.011$ & $+0.006 \pm 0.02$ \smallstrut\\
\hline
all & $0.1<z_{\rm B}\le 1.2$ & 11\,733\,942 & 7.38 & 0.283 & $0.714 \pm 0.025$ & \smallstrut\\
\hline
\end{tabular}
\tablefoot{The effective number density in column 4 corresponds to the \citet{heymans/etal:2012} definition. The ellipticity dispersion in column 5 is reported for one component. The $m$-bias (column 7) is defined in Eq.~\ref{eq:mc}, and its estimation with image simulations is described in Sect.~\ref{sec:m_bias}.}
\end{table*}

The fifth high-redshift bin added here contributes an additional 22\% (by \emph{lens}fit weight; see Sect.~\ref{sec:lensfit}) of source galaxies to the lensing measurement. Due to their high redshift these sources carry a large cosmic shear signal and contribute over-proportionally to the signal-to-noise ratio of the measurement presented in Sect.~\ref{sec:results}. Increasing the redshift baseline and adding five more 2-point shear correlation functions (the auto-correlation of the fifth bin as well as the four cross-correlations of the fifth bin with the four lower-redshift bins) hence increases the precision of the cosmological inference. In order to exploit this additional statistical power it is important to ensure that systematic errors are under tight control, for these faint high-redshift sources in particular.

\subsection{Redshift calibration}
\label{sec:DIR}
As in \citetalias{hildebrandt/etal:2017}, we follow redundant approaches to calibrate the KV450 photo-$z$, i.e. to estimate the redshift distributions of the galaxies in the five tomographic photo-$z$ bins. In this section we describe our fiducial technique, dubbed DIR, to estimate the redshift distributions. It relies on a direct estimate of the redshift distributions from deep spectroscopic surveys. It makes few assumptions and is straightforward in its application, which makes it our first choice for this calibration. Some alternatives are discussed in Appendix~\ref{app:photoz_tests}. These are a smoothed version of the DIR approach (sDIR, Appendix~\ref{sec:sDIR}), clustering redshifts using small scales (CC, Appendix~\ref{sec:CC}), and an optimal quadratic estimator of clustering redshifts at large scales (OQE, Appendix~\ref{app:OQE}).

For the DIR method, KiDS- and VIKING-like observations have been obtained in the COSMOS, DEEP2, GAMA-G15Deep, CDFS, and VVDS-2h fields (see Sect.~\ref{sec:spec}). This KiDS+VIKING-like multi-band photometry is used to provide a proper weight for the spectroscopic catalogues and in this way make them more representative of the whole KV450 lensing catalogue. The method, which is based on a $k$th nearest neighbour ($k$NN) approach, is described in detail in \cite{lima/etal:2008} and section 3 of \citetalias{hildebrandt/etal:2017}. In some of these fields, the NIR data is considerably deeper than VIKING. We add noise to those additional deep NIR data to represent the VIKING depth. Running the DIR calibration twice, once with the deeper and once with shallower photometry, yields basically identical results (mean redshifts differ by $\la 0.002$). The $k$NN assignment seems to be very stable under the addition of noise to the photometry of the reference sample. In the end we use the deeper data for the fiducial DIR calibration.

The most important difference with respect to our previous analysis \citepalias{hildebrandt/etal:2017} is that the weights are estimated from density measurements in nine dimensions ($ugriZYJHK_{\rm s}$-magnitude space) instead of four dimensions ($ugri$). This makes the colour-redshift relation that we are trying to calibrate here less degenerate. In the redshift range of interest, which is set by the KiDS $r$-band magnitude limit, colour-redshift degeneracies \citep[for an explanation see][]{benitez:2000} are considerably reduced when using a 9-band filter set spanning the wavelength range $0.3-2.3\mu{\rm m}$ and KiDS/VIKING-like photometric quality. This is also reflected in the comparison of KiDS+VIKING 9-band photo-$z$ and spec-$z$ from the literature as presented in \citetalias{wright/etal:2018}.

The four-dimensional magnitude space of KiDS-450 was quite densely populated with spectroscopic objects given our calibration sample. This density was sufficient to estimate the density of the spectroscopic catalogue in this space by measuring the distance to the $k$th nearest neighbour. Keeping $k$ constant, we also measured the corresponding density in the photometric catalogue. We found that this approach becomes unstable in the more sparsely populated nine-dimensional magnitude space of KV450. Hence we use a `constant volume' approach as suggested by \citet{lima/etal:2008}. For each object in the spectroscopic catalogue we measure the distance to the fourth-nearest spectroscopic neighbour. Then we count the number of objects (weighted by their \emph{lens}fit weight; see Sect.~\ref{sec:shapes}) in the photometric catalogue within the nine-dimensional hyper-sphere of that radius. This density estimate is more stable and can be used to define the spectroscopic weights.

Another difference between KiDS-450 and KV450 is that we include more spectroscopic data. While the \citetalias{hildebrandt/etal:2017} DIR estimate was based on COSMOS, DEEP2, and CDFS data alone, here we add 6480 spec-$z$ from the GAMA-G15Deep and VVDS-2h fields (a 34\% increase in terms of numbers). By increasing the number of independent lines-of-sight we reduce shot noise and sample variance, and make the whole DIR calibration less susceptible to selection effects in the individual surveys.

In KiDS-450 we applied the redshift weighting procedure to the full photometric catalogue and then applied $z_{\rm B}$ photo-$z$ cuts to the weighted spectroscopic catalogue. Here we turn this around and apply the $z_{\rm B}$ photo-$z$ cuts to the photometric catalogue first and perform the re-weighting for each tomographic bin individually. This results in a less noisy DIR estimate as the $z_{\rm B}$ cuts are applied to the larger photometric catalogue.

Shot noise in the DIR redshift distributions, estimated from a bootstrap analysis over the objects in the spectroscopic catalogue, is quite small due to the large number of objects in the calibration sample. However, one of the major unanswered questions about the KiDS-450 DIR calibration was how much the estimate of the redshift distributions was affected by sample variance. This sample variance can be of cosmological origin (large-scale structure) or due to selection effects (e.g. colour pre-selection) and unsuccessful redshift measurements that are different for the different spectroscopic surveys that contribute to our calibration sample. We expect that in our case selection effects and variable redshift success rates are dominant, as we have a large number of spec-$z$ from several different lines-of-sight, which suppresses large-scale structure.

In order to better account for sample variance and selection effects in the KV450 redshift calibration we adopt a spatial bootstrapping approach. For the bootstrap resampling, we split our calibration sample into ten subsamples of equal size (in terms of the number of objects) along the RA direction. Then we draw 1000 bootstrap samples from these subsamples, and estimate the uncertainties of the DIR $n(z)$ from the scatter between the bootstrap samples. This approach yields a more realistic error estimate, and the error on the mean redshift based on this bootstrap resampling is reported in Table~\ref{tab:tomobins}. 

\begin{figure*}
  \includegraphics[width=\textwidth]{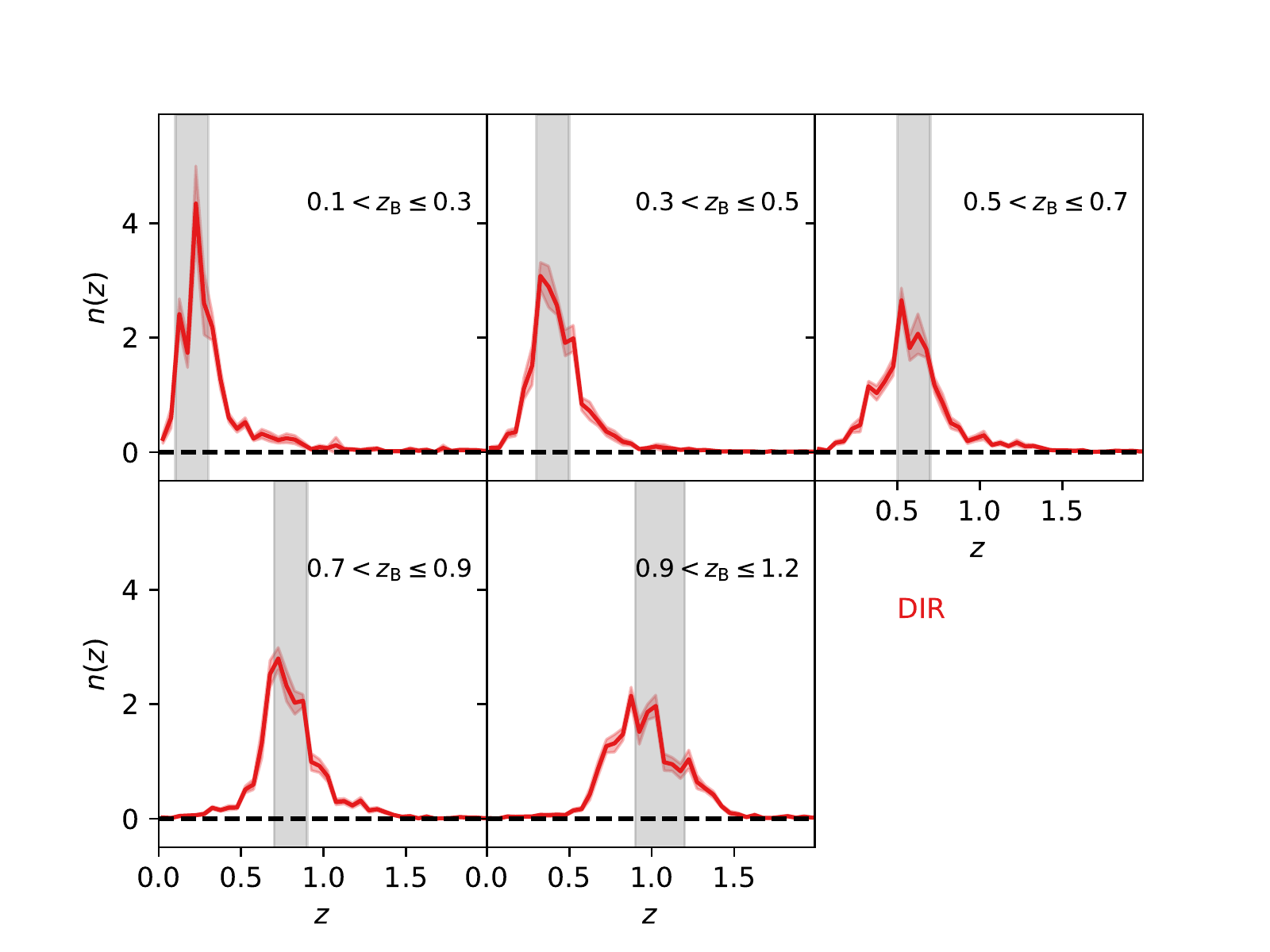}
  \caption{\label{fig:zdist} Redshift distribution estimates for the five tomographic bins used in the KV450 cosmic shear analysis with the DIR technique. The uncertainties shown correspond to the 68\% confidence intervals as estimated from a spatial bootstrap resampling of the spec-$z$ calibration sample.}
\end{figure*}

The resulting DIR redshift distributions are shown in Fig.~\ref{fig:zdist} with their bootstrap uncertainties. We neglect any covariance in the uncertainties of the mean redshifts between the tomographic bins. The small-scale structure that is still visible (and looks somewhat significant) in the $n(z)$ is a sign that the bootstrap resampling method still slightly underestimates the errors. The spurious structures can be attributed to residual large-scale structure, due to the small area on the sky of the spec-$z$ surveys, and especially also selection effects in the different spec-$z$ samples. In order to explore further whether the errors are severely underestimated we report results from an alternative `quasi-jackknife' procedure described below as well as a smoothing method (sDIR, Appendix~\ref{sec:sDIR}) and different clustering-$z$ estimates (Appendix~\ref{sec:CC}~\&~\ref{app:OQE}).

We allow for nuisance parameters $\delta z_i$ in each tomographic bin $i$ that linearly shift the $n_i(z) \rightarrow n_i(z+\delta z_i)$ when modelling the 2-point shear correlation functions. The Gaussian priors for these parameters (see Table~\ref{tab:params}) correspond to the bootstrap errors reported in column 6 of Table~\ref{tab:tomobins}.

A linear shift certainly does not capture the full variance of the $n(z)$. Fluctuations in the high-$z$ tails can have important consequences for the mean redshifts and also the model predictions. Also the errors might be slightly underestimated as discussed above. In order to study the possible extremes of sample variance and selection effects in greater detail, we also estimate redshift distributions for several reduced sets of the calibration sample excluding galaxies from different lines-of-sight. We build these different subsamples by omitting the following data subsamples one at a time:
\begin{enumerate}
\item DEEP2
\item zCOSMOS
\item VVDS
\item zCOSMOS and VVDS
\end{enumerate}
These samples were chosen on the one hand to still give a fair coverage of magnitude space but on the other hand maximise sample variance. We estimate the cosmological parameters (Sect.~\ref{sec:results}) for redshift distributions based on these four reduced calibration samples as well as for the full sample. The differences in the parameter estimates then give an indication of the extremes of the sample variance in the redshift calibration for the cosmological conclusions of this work. It should be noted that this sample variance is not entirely cosmological but also due to different galaxy selections and spectroscopic success rates in the different spec-$z$ surveys.

In a recent paper \citep{wright/etal:2019} we study the coverage of the KV450 source sample in 9D magnitude space by the combined spectroscopic sample with self-organising-maps \citep[SOM;][]{masters/etal:2015}. There we find that DEEP2 is the single most important contribution to the calibration sample, i.e. DEEP2 uniquely calibrates the largest fraction of KV450 sources. We attribute this to the fact that DEEP2 is the highest-redshift survey in our calibration sample. Based on these findings it can be expected that the cosmological conclusions are most affected if DEEP2 is excluded from the calibration. It also means that the whole DIR calibration presented here crucially hinges on the validity of the DEEP2 redshifts.

We further test this by creating mock samples resembling our sources and the different spec-$z$ calibration samples in the MICE simulation (\citealt{fosalba/etal:2015}; \citealt{crocce/etal:2015}; van den Busch et al. in prep.). Running the DIR method on these mock catalogues yields very similar results, i.e. also in the simulations the mock DEEP2 sample is the most important one for the calibration, and excluding it from the mock calibration sample yields biased results.

The same simulation setup allows us to study sample variance. We find that even for a single field like COSMOS, sample variance is negligible after DIR re-weighting. Mean redshifts scatter by only $\sigma_{<z>}\sim 0.005$ for different lines-of-sight for the calibration field. Detailed results are presented in \citet{wright/etal:2019}.

Another somewhat complementary test is carried out with the high-quality photo-$z$ catalogue that is available in the COSMOS field \citep[][called COSMOS-2015 in the following]{laigle/etal:2016} as the calibration sample for DIR. This catalogue is based on an extensive set of photometric measurements over the 2~deg$^2$ COSMOS field. It is complementary to the spec-$z$ calibration sample discussed above because it does not suffer from faint-end incompleteness. There is a redshift estimate for each object down to the magnitude limit of the KiDS data. However, the photo-$z$ are not perfect. While the photo-$z$ scatter is very low, the error distribution is highly non-Gaussian and there is a significant fraction of outliers of $\sim6\%$ at $23<i<24$. This outlier fraction is considerably higher than the spectroscopic failure fraction of $\sim1\%$ in our spec-$z$ calibration sample. Hence, using the COSMOS-2015 photo-$z$ catalogue instead of the combined spec-$z$ sample means trading very low outlier rate and multiple lines-of-sight for higher faint-end completeness. In comparison to the fiducial DIR method with the full spec-$z$ sample we find the mean redshifts for all tomographic bins to be considerably lower when we use the COSMOS-2015 catalogue, with shifts of \smash{$\Delta z=-0.04, -0.07, -0.09, -0.06, -0.04$} for the five bins, respectively.\footnote{This trend is similar to the findings of \citet{amon/etal:2018}, who calibrate the KiDS-$i$-800 redshift distribution with spec-$z$ as well as the COSMOS-2015 catalogue.} The resulting $n(z)$ are shown in Fig.~\ref{fig:zdist_OQE} in comparison to the fiducial $n(z)$.

Using COSMOS-2015 for the DIR method is very similar to the redshift calibration chosen for the DESy1 cosmological analysis \citep{hoyle/etal:2018} and the HSC-DR1 cosmic shear analysis \citep{hikage/etal:2018}. Results will be shown and compared to the DIR with the full spec-$z$ sample in Sect.~\ref{sec:results} and discussed in detail in Sect.~\ref{sec:discussion}. The mean and median redshifts of each redshift distribution discussed here are reported in App.~\ref{sec:meanz}.

\subsection{Blinding}
\label{sec:blinding}

In the KiDS-DR2 analyses \citep[e.g.][]{viola/etal:2015} and the KiDS-450 project \citepalias{hildebrandt/etal:2017} we implemented a blinding scheme to suppress confirmation bias. The ellipticity measurements were coherently perturbed by an external blind-setter, and three catalogues (four in the case of DR2) -- the original catalogue as well as two slightly perturbed catalogues -- were analysed by the team simultaneously without knowledge about the identity of the catalogues. Unblinding happened only shortly before submission of the papers, and -- most importantly -- after the analysis pipelines had been frozen.

We cannot use the same blinding scheme here again as the ellipticity measurements described in Sect.~\ref{sec:shapes} are identical to the ones used in KiDS-450 and have been unblinded for that project. Instead we decided to blind ourselves to the redshift distributions, which -- unlike the ellipticities -- changed from KiDS-450 to KV450. In a very similar way as before, the spectroscopic redshift catalogue used for the DIR method was sent to an external blind-setter, who returned a catalogue with three different spectroscopic redshift columns, two of which are slightly perturbed. The original merged catalogue was deleted before reception of the blinded catalogue to avoid accidental unblinding. 

The amplitude of the perturbation was chosen such that the highest and lowest blinding would differ by roughly $1\sigma$ in terms of $S_8$. Our blinding scheme displaces the mean redshift of each tomographic bin, but does not significantly alter the shape. The mean redshift of the five tomographic bins differs from the truth (as revealed after unblinding) by 0.015 in the lowest redshift bin to 0.04 in the highest redshift bin, for one blind. For the second more extreme blind, the five tomographic bins differ from the truth by 0.03 in the lowest redshift bin to 0.08 in the highest redshift bin. 

It is important that these displacements are internally consistent with one another. For a fiducial cosmology mock data vector created using the true $n_i(z)$, where $i$ indicates the tomographic bin number, a good fit must also be provided when using the displaced blinded $n_i(z)$, but for a different value of $S_8$. This consistency between the blinded redshift bins prevents the likelihood inference automatically unblinding our analysis when nuisance parameters $\delta z_i$ are included to characterise our uncertainty on the mean redshift of each tomographic bin $i$. As the nuisance parameters are treated as uncorrelated parameters and are poorly constrained, they favour the peak of the chosen Gaussian prior with $\delta z_i = 0$ for all bins.  If the cosmological constraints from each of the tomographic bins are consistent with one another when $\delta z_i = 0$, there is no incentive for the chain to explore the extremes of the prior distribution and shift the displaced blinded $n_i(z)$ back to the truth. This demonstrates that marginalising over $\delta z_i$ nuisance parameters will not be able to identify coherent systematic biases across all the tomographic redshift distributions (for example the biases that our blinding introduced). These nuisance parameters are therefore only useful to detect when one or two tomographic redshift bins are outliers and inconsistent with the rest of the data set.

We performed the main cosmological analysis with all three blinded sets of $n(z)$ and all other tests (see Sect.~\ref{sec:MCMC_nz}) with one randomly chosen blinding. We only unblinded at the very end of the project when, again, the analysis pipeline was already frozen. Details on this approach and all steps taken after unblinding are described in Appendix~\ref{app:history}.

\section{Shape measurements}
\label{sec:shapes}

The catalogue of ellipticity measurements of galaxies used here is identical to the one used in \citetalias{hildebrandt/etal:2017}. In Sect.~\ref{sec:lensfit} we summarise its main properties and highlight how the weights that accompany these ellipticities have changed since then. We also motivate why new image simulations are necessary to calibrate the multiplicative shape measurement bias. 

These simulations are described in detail in Sect.~\ref{sec:m_bias}. There we describe how we improve on previous studies by basing our simulations on high-resolution data from the Hubble Space Telescope. This leads to realistic correlations between observables in the simulations and, crucially, allows for photo-$z$ cuts in order to better emulate what has been done to the data. This is again supplemented by a robustness analysis, trying many different setups for the simulations, that lets us arrive at solid estimates for the uncertainty of our multiplicative bias estimates.

Section~\ref{sec:c_term} describes a novel treatment of the additive shape measurement bias term that takes into account new findings about electronic effects in CCD cameras and related insights about weak lensing B-modes. While this treatment does not have any significant effect on the measurements presented here it will become important for future experiments.

\subsection{Shape measurements with \emph{lens}fit}
\label{sec:lensfit}
The shapes of galaxies, described by the two ellipticity components $\epsilon_1$ and $\epsilon_2$, were measured from THELI-reduced individual $r$-band exposures with the self-calibrating version of the \emph{lens}fit algorithm \citep{miller/etal:2007,miller/etal:2013}. The shear biases were determined using image simulations described in \citet[][hereafter FC17]{fenechconti/etal:2017}, where the input galaxy catalogue was constructed using the \emph{lens}fit priors. The shear biases for the different tomographic bins were determined by resampling the simulated catalogues so that the output distributions matched the observed signal-to-noise ratio (SNR) and size distributions. In doing so, \citetalias{fenechconti/etal:2017} assumed that the ellipticities do not correlate with other parameters, and that those galaxy parameters did not explicitly depend on redshift. The resampling corrections were significant for faint, small galaxies, i.e. the highest tomographic bins, resulting in increased systematic uncertainties in the calibration.

To take advantage of the fifth tomographic bin we created a new suite of image simulations that are based on VST and HST observations of the COSMOS field that are discussed in Sect.~\ref{sec:m_bias}. In the process we also corrected the calculation of the \emph{lens}fit weights \citep[see][hereafter K19]{kannawadi/etal:2018}. We find that the shape measurement pipeline yields multiplicative and additive shear biases that are close to zero for all tomographic bins. The updated \emph{lens}fit weights result in negligible B-modes in the data, but do not reduce the overall additive bias that was observed in \citetalias{hildebrandt/etal:2017}; in Sect.~\ref{sec:c_term} we discuss our updated empirical correction.

\subsection{Calibration with image simulations}
\label{sec:m_bias}

Reliable shear estimates are essential for cosmic shear studies, but difficult in practice because the galaxies of interest are faint and small. As a consequence noise and the convolution with the (anisotropic) PSF bias the measurements. Moreover, already during the object detection step biases are introduced (\citetalias{fenechconti/etal:2017}, \citetalias{kannawadi/etal:2018}). To quantify the biases and thus calibrate the shape measurement pipeline it is essential that the algorithm performance is determined using mock data that are sufficiently realistic \citep[e.g.][]{miller/etal:2013,hoekstra/etal:2015}. This is put in a more formal framework in Sect.~2 of 
\citetalias{kannawadi/etal:2018}.

The image simulations presented in \citetalias{fenechconti/etal:2017} used an input catalogue that was based on the \emph{lens}fit priors, which in turn are based on observable properties of galaxies. Although fairly realistic, the image simulations did not reproduce the observed distributions of faint, small galaxies. As the biases are predominantly a function of signal-to-noise (SNR) and size, the shear biases for the different tomographic bins were determined by resampling the simulated catalogues so that the output distributions matched the observations. The resampling procedure used in \citetalias{fenechconti/etal:2017} implicitly assumed that SNR and size (or resolution) are the only parameters to be considered, and that parameters do not explicitly depend on redshift. \citetalias{fenechconti/etal:2017} showed that the calibration was robust for the first four tomographic bins, but it was found to be too uncertain for the calibration of a fifth bin. 

To improve and extend the calibration to the full range of sources we created a new suite of image simulations that are described in detail in \citetalias{kannawadi/etal:2018}. Here we highlight the main changes and present the main results. The simulation pipeline is based on the one described by \citetalias{fenechconti/etal:2017}, but we introduced a number of minor improvements to better reflect the actual data analysis steps. The main difference is our input catalogue, which enables us to  \emph{emulate} VST observations of the COSMOS field under different observing conditions. 

The input catalogue for the image simulations is derived from a combination of VST and VISTA observations of the COSMOS field and a catalogue of S\'{e}rsic parameter fits to the HST observations of the same field by \citet{griffith/etal:2012}. The sizes, shapes, magnitudes, and positions of the galaxies in the simulations are therefore realistic. This captures the impact of blending and clustering of galaxies, as well as correlations between structural parameters. \citetalias{kannawadi/etal:2018} find evidence for correlations between the ellipticity and galaxy properties, whereas \citetalias{fenechconti/etal:2017} assumed these to be uncorrelated. 
Importantly, the KiDS-like multi-band imaging data in 9-bands enables us to assign photometric redshifts to the individual galaxies. The variation of galaxy parameters with redshift is thus also included naturally. Stars are injected as PSF images at random positions, with their magnitude distribution derived from the Besan\c{c}on model \citep{robin/etal:2003}. The realism of the input catalogue marks one of the major improvements over the shear calibration carried out in \citetalias{fenechconti/etal:2017}, and as shown in \citetalias{kannawadi/etal:2018} the simulated data match the observations (of the full KV450 data set) faithfully.  

The overall simulation setup is similar to that used in \citetalias{fenechconti/etal:2017} for KiDS-450, except that we do not generate a random catalogue of sources, but instead simulate KiDS $r$-band observations of the COSMOS field under different seeing conditions (we did not vary the background level). As is the case for the actual survey, we create five OmegaCam exposures, with the exposures dithered with the same pattern as used in KiDS \citep{dejong/etal:2015}. The images are rendered using the publicly available \textsc{GalSim} software \citep{rowe/etal:2015}.  The simulated exposures are split into 32 subfields, corresponding to the 32 CCD chips in the OmegaCam instrument. The individual chips are then co-added using \textsc{SWarp} \citep{bertin:2010}, on which \textsc{SExtractor} is run to obtain a detection catalogue, which is then fed to \emph{lens}fit. As was the case in  \citetalias{fenechconti/etal:2017}, each exposure has a different, spatially constant PSF, but the sequence of PSF parameters is drawn from the survey to match a realistic variation in observing conditions.  A total of thirteen PSF sets\footnote{These are the same as the PSF sets used in \citetalias{fenechconti/etal:2017}.} are simulated, where each PSF set corresponds to a set of thirteen PSF models from successive observations.  In order to be able to measure the shear bias parameters, eight spatially constant reduced shears of magnitude $|g|=0.04$ and different orientations are applied to each set in the simulation that differ in the PSF. Although the volume of the simulated data is much smaller than the observed data, the statistical uncertainty in the bias is reduced by employing a shape-noise cancellation scheme, where each galaxy is rotated by 45, 90 and 135 degrees \citep{massey/etal:2007}. 

To estimate the shear, the ellipticities of the galaxy models are combined with a weight that accounts for the uncertainty in the ellipticity measurement. This leads to a bias in the shear estimate that is sensitive to the ellipticity distribution \citepalias{fenechconti/etal:2017}. To account for this, the KV450 catalogues are divided into $3\times4\times4$ sub-catalogues based on PSF size and the two components of the complex ellipticity, and weight recalibration is performed on each of the sub-catalogues. In the simulations, the individual \emph{lens}fit catalogues, corresponding to the four rotations and eight shears for a given PSF set are combined, and a joint weight recalibration is performed for each PSF set separately.\footnote{This is different from what was done in \citetalias{fenechconti/etal:2017}. The current approach better reflects what is done in the actual analysis and improves the agreement with the observations.}

To quantify the shear bias, we adopt the commonly used linear parametrisation \citep{heymans/etal:2006,massey/etal:2007}, expressed as a multiplicative bias term $m_i$ and an additive bias term $c_i$,
\begin{equation}
\label{eq:mc}
    \hat{g_i} = (1+m_i)g_i^{\text{true}} + c_i,
\end{equation}
where $i=1,2$ refers to the two components of the reduced shear $g$, and $\hat{g}$ is the observed shear.\footnote{In the following we ignore the small difference between shear and reduced shear.} The best-fit straight line to the components of the estimated shear as a function of the input shear gives the multiplicative and additive biases. 

The additive biases are small in the image simulations \citepalias[see section 6.2 of][]{kannawadi/etal:2018} with the average $c_1=(1.1\pm 0.9)\times 10^{-4}$ and $c_2=(7.9\pm0.9)\times 10^{-4}$. Similar amplitudes were found by \citetalias{fenechconti/etal:2017}. Interestingly, the bias in $c_2$ is noticeably larger than $c_1$, similar to what was is observed in the KiDS data \citepalias[][and Sect.~\ref{sec:c_term}]{hildebrandt/etal:2017}. Also the amplitudes  in the simulations and the data compare well ($\langle c_1\rangle=(2.1\pm0.7)\times 10^{-4}$ and $\langle c_2\rangle=(4.8\pm0.5)\times 10^{-4}$ for the KV450 data). However, the image simulations may not include all sources of additive bias, and instead we estimate the residual additive bias from the data themselves (see Sect.~\ref{sec:c_term}).

Although the linear regression is performed to the two components independently, the multiplicative bias is isotropic in practice, i.e., $m_1 \approx m_2$ and we use $m = (m_1+m_2)/2$. Galaxies in the KV450 catalogue are assigned a value for $m$ based on which $z_{\rm B} - \text{SNR}-\mathcal{R}$ bin they belong to, where
\begin{equation}
    \mathcal{R} = \frac{r^2_\text{PSF}}{r^2_{ab}+r^2_\text{PSF}}
\end{equation}
is the resolution parameter and $r_{ab}$ is the circularised size of the galaxy calculated as the geometric mean of \emph{lens}fit measured semi-major and semi-minor axes, and $r_\text{PSF}$ is the size of the PSF. The bias in each tomographic bin is simply the weighted average of the individual $m$ values. 

A notable improvement is that we first split the simulated galaxies into their respective tomographic bins, based on their assigned $z_{\rm B}$ values. Although the size and SNR distributions match the data well (as do the distributions of inferred \emph{lens}fit parameters), we reweight the simulated catalogues so that they match the observed distributions in SNR and $\mathcal{R}$. \citetalias{kannawadi/etal:2018} found that the ellipticity distributions differ slightly between tomographic bins, which highlights the importance of redshift information in the image simulations. Reweighting before dividing the sample in redshift bins shifts the value of $m$ by about $-0.02$ for the first two bins.

The main role of the reweighting is to capture the variation in observing conditions that are present in the KV450 data, which affects the SNR and size distributions. The adjustments are small overall, owing to the overall uniformity of the KiDS data and the realism of the image simulations. The mean multiplicative biases for the five tomographic bins are found to be $m=-0.017,-0.008,-0.015,+0.010,+0.006$. \citetalias{kannawadi/etal:2018} test the robustness of the image simulations and find that the results are not very sensitive to realistic variations in the input catalogues. The two highest redshift bins may be somewhat affected by how galaxies below the detection limit are modelled. The image simulations, however, do not capture variations in the photometric redshift determination that are also expected. \citetalias{kannawadi/etal:2018} show that the impact is expected to be small, but could introduce a bias as large as 0.02 for the lowest and highest redshift bin. They therefore estimate a conservative systematic uncertainty of $\sigma_m=0.02$ per tomographic bin.\footnote{The biases are determined per tomographic bin. Although we expect certain assumptions to lead to correlations between the bins, variations as a function of redshift should be treated as independent.}

Note that in \citetalias{hildebrandt/etal:2017} we estimated $\sigma_m=0.01$. Here we are more conservative with an uncertainty that is twice as large given the new findings of \citetalias{kannawadi/etal:2018}, who include realistic photometric redshifts and correlations between observables in the image simulations for the first time. This extra level of sophistication makes the simulations slightly more sensitive to input parameters, hence the increased uncertainty. The sensitivity of these biases to input parameters is expected to be reduced further by simulating multi-band observations to realistically capture the photometric redshift determination.

\subsection{Additive shear measurement bias}
\label{sec:c_term}

\citetalias{hildebrandt/etal:2017} observed a significant additive shear bias (also called $c$-term). To account for this, \citetalias{hildebrandt/etal:2017} estimated the value for $c_i$ per tomographic bin and per patch by averaging the $\epsilon_{1,2}$ measurements. These mean ellipticity values were used to correct the measurements before 2-point shear correlation functions were estimated. The size and error of this correction also determined the upper limit for the angle $\theta$ used to measure the correlation functions. 

Although the image simulations suggest that a large part of the bias may arise from the shape measurement process, additional sources of bias were identified in \citetalias{hildebrandt/etal:2017} (e.g. asteroid trails, etc.). Without a full model for the c-term \citetalias{hildebrandt/etal:2017} accounted for the additive bias using a purely empirical approach. We use the same approach for KV450 but now also propagate the uncertainty in the $c$-correction into the model. For this we introduce a nuisance parameter $\delta c$ to forward-model this effect (see Table~\ref{tab:params}).

We also introduce a position dependent additive bias pattern that is based on the analysis of detector and readout electronics effects in the OmegaCam instrument (Hoekstra et al. in prep.). For instance, the brighter-fatter effect \citep{antilogus/etal:2014} can affect the PSF sizes and ellipticities (as the effect is typically stronger in the parallel readout direction) for bright stars.  A study of the magnitude dependence of the residuals between the mean (i.e. averaged over magnitude) PSF model and the individual stars' ellipticity measurements revealed a significant trend for faint stars for $\epsilon_1$ (no significant signal was found for $\epsilon_2$). Further study suggests that this is the result of charge trailing during the readout process. Although the effect is small for most chips, one CCD chip stood out (ESO\_CCD\#74). A corresponding increase in $c_1$ at the location of this detector was indeed measured when stacking all the KiDS-450 ellipticities in the detector frame. Hence we expect a pattern in the $c_1$-bias that corresponds to the detector layout. 

\citet{asgari/etal:2018} showed that such a repeating pattern can lead to B-modes in the ellipticity distribution and found a hint of such a pattern in the KiDS-450 data. In combination with the findings about OmegaCam discussed above we decided to correct for such a repeating pattern in the data.

\begin{figure}
\includegraphics[width=\hsize]{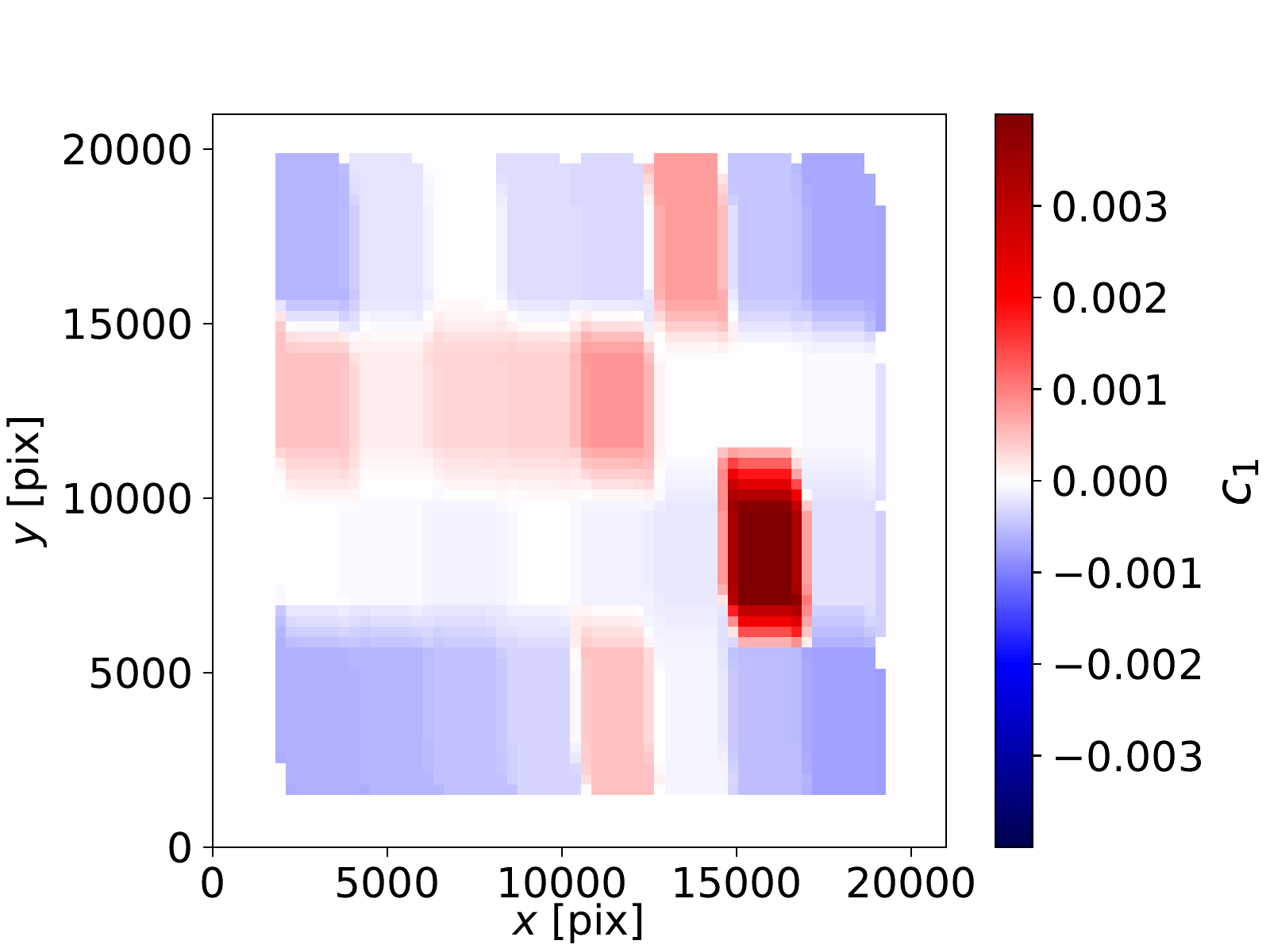}
\caption{\label{fig:c1_map} Map of the predicted $c_1$ term at $r=24$ based on the findings of Hoekstra et al. (in prep.) about electronic effects in the OmegaCam instrument (the $c_2$ pattern is insignificant). We calculate the corresponding 2-point shear correlation function of this pattern and add this to our model via a free nuisance amplitude $A_c$.}
\end{figure}

The trend in the $\epsilon_1$ ellipticity component as a function of magnitude is similar for the different chips but the amplitude of the effect differs. If we assume a linear trend we can extrapolate this behaviour to the faint magnitudes of most of the KiDS galaxies. This is done independently for each chip. Using this prediction directly to correct the galaxies could be problematic as the trends might not be linear as assumed above. A weaker assumption is that the relative amplitudes between the chips are mostly independent of magnitude. So instead we use our star measurements of this effect to predict a map of the $c_1$-term, $c_{1}(x,y)$, shown in Fig.~\ref{fig:c1_map}, and fit these to the observed galaxy ellipticity component $\epsilon_1$ (after subtraction of the global $c_1$-term) averaged in cells in pixel space:
\begin{equation}
\langle \epsilon_{1} \rangle (x,y) = \beta_{1} \, c_{1}(x,y) + \alpha_{1}\,.
\end{equation}

We find $\beta_1=1.01\pm0.13$ and $\alpha_{1}$ consistent with zero for the full shear catalogue. This means that the 2-dimensional structure predicted by the residual stellar ellipticities is also seen in the galaxy $\epsilon_1$ ellipticity component. 

We use this measurement of $\beta_1$ together with the fitting errors to introduce a nuisance parameter $A_c$ and a Gaussian prior that are included in the model to account for this position-dependent additive bias (see Table~\ref{tab:params}). For this we measure the 2-point shear correlation function of the pattern shown in Fig.~\ref{fig:c1_map} by assigning each galaxy the $c_1$ value from this map at its pixel position and run \textsc{treecorr} as described in Sect.~\ref{sec:xipm}. We then scale the contribution of this spurious signal to the overall model by $A_c$ and add it to the cosmological signal (Sect.~\ref{sec:CS_signal}).

\section{Correlation functions and covariance matrix}
\label{sec:cov}

\subsection{2-point shear correlation functions}
\label{sec:xipm}

The 2-point shear correlation function between two tomographic bins $i$ and $j$ is measured with the public \textsc{treecorr} code \citep{jarvis/etal:2004}, which implements the following estimator:
\begin{equation}
\hat{\xi}_{\pm}^{ij}(\theta) = \frac{\sum_{ab} w_a w_b \left[ \epsilon_\rmn{t}^i (\x_a) \epsilon_\rmn{t}^j (\x_b) \, \pm \, \epsilon_\times^i (\x_a) \epsilon_\times ^j(\x_b)
\right]}{
\sum_{ab} w_a w_b } \,
\label{eq:xi_est}
\end{equation}
where $\epsilon_\mathrm{t,\times}$ are the tangential and cross ellipticities of a galaxy measured with respect to the vector $\vec x_a - \vec x_b$ connecting the two galaxies of a pair $(a,b)$, $w$ is the \emph{lens}fit weight, and the sums go over all galaxy pairs with an angular separation $|\vec x_a - \vec x_b|$ in an interval $\Delta\theta$ around $\theta$ (see Sect.~\ref{sec:CS_signal} for a discussion on how to model the signal in such a broad $\theta$ bin). There are five auto-correlations for the five tomographic bins and ten unique cross-correlations between tomographic bins for $\xi_+$ and $\xi_-$ each.

We analyse the same angular scales as in \citetalias{hildebrandt/etal:2017}, i.e. we define nine logarithmically spaced bins in the interval $[0\farcm5, 300']$ and use the first seven bins for $\xi_+$ and the last six bins for $\xi_-$. These limits are chosen such that on small scales the contribution from baryon feedback in the OWLS-AGN \citep{vandaalen/etal:2011} model is less than $\sim$20\% to the overall signal and on large scales the constant $c$-term (see Sect.~\ref{sec:c_term}) if uncorrected for, would still be smaller than the expected cosmic shear signal for a fiducial WMAP9~+~BAO~+~SN cosmology \citep{hinshaw/etal:2013}. These criteria yield the same angular scales for the new fifth tomographic bin as for the lower-redshift bins, which are more similar to the ones used before.

\begin{figure*}
\includegraphics[width=\hsize, clip=true, trim=1cm 0cm 4cm 0cm]{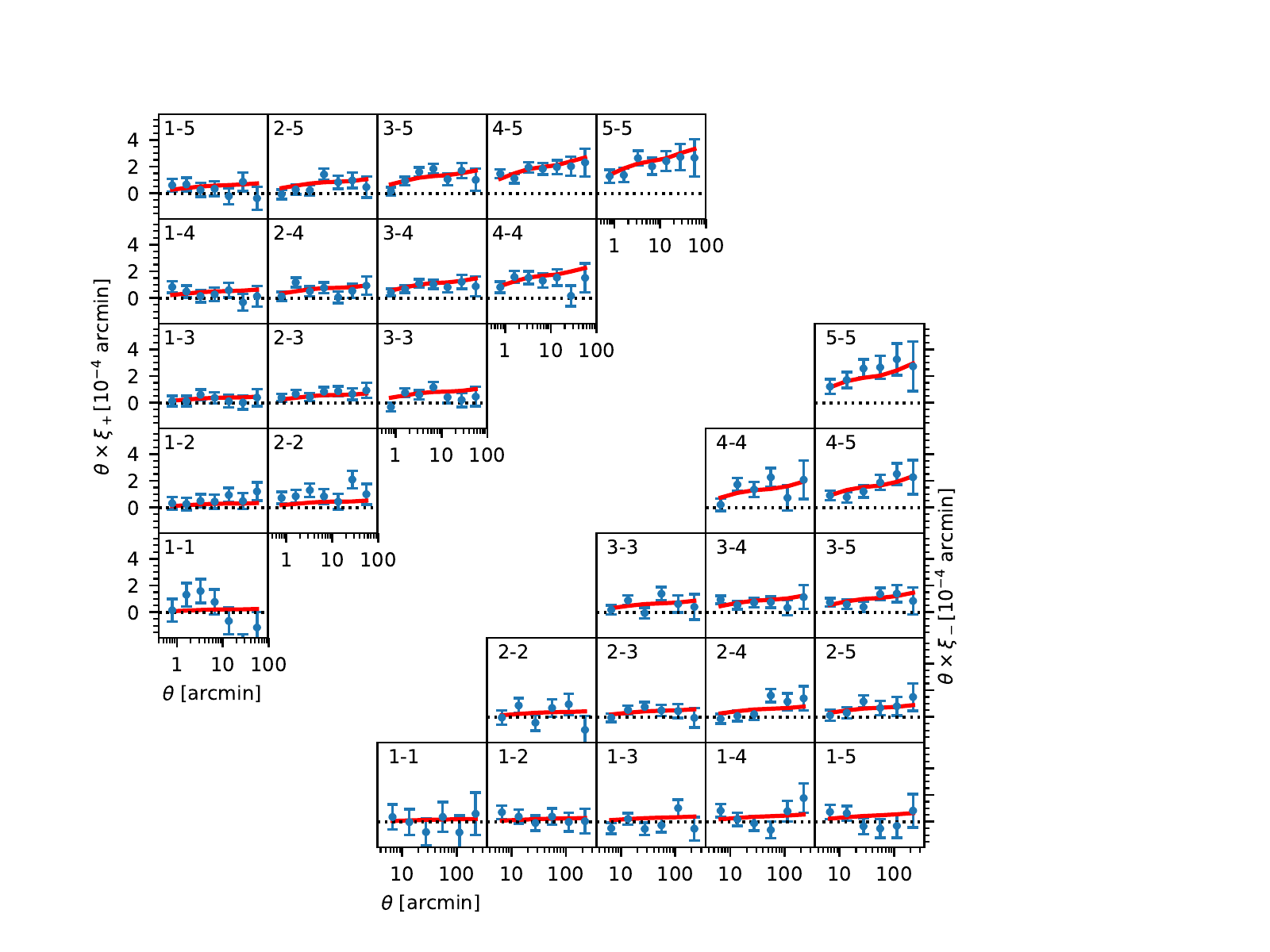}
\caption{\label{fig:xipm} KV450 2-point shear correlation functions $\xi_+$ (\emph{upper-left}) and $\xi_-$ (\emph{lower-right}) plotted as $\theta\times\xi_\pm$. The errors shown represent the square root of the diagonal of the analytical covariance matrix. These errors are significantly correlated between scales and redshift bins. The solid red line corresponds to the best-fit (maximum likelihood) fiducial model from Sect.~\ref{sec:results} including baryon feedback, intrinsic alignments, and all corrections for observational biases.}
\end{figure*}

Given these scale cuts and the 15 distinct correlation functions that we measure, the KV450 cosmic shear data vector consists of $(7+6)\times15=195$ data points, which are shown in Fig.~\ref{fig:xipm}.

\subsection{Covariance matrix}
\label{sec:cov_ana}
We estimate the covariance matrix for our data vector with an analytical recipe. Details of this approach can be found in \citetalias{hildebrandt/etal:2017}. Here we describe a number of changes/improvements that are implemented for this study.

For KV450 we update the footprint according to figure~1 of \citet{wright/etal:2018} and use that information to calculate the coupling of in-survey and super-survey modes. The effective area decreases from 360.3~deg$^2$ to 341.3~deg$^2$ due to incomplete VIKING coverage.

Furthermore, we change the way in which the uncertainty in the multiplicative shear measurement bias \citep[estimated to be $\sigma_m=0.02$][]{kannawadi/etal:2018} is accounted for. We propagate this uncertainty into the covariance matrix \citepalias[see equation 12 of][]{hildebrandt/etal:2017} but we now calculate this contribution using a theoretical data vector instead of the observed data vector. Hence we follow \citet{troxel/etal:2018b} in this aspect. The theoretical data vector is based on the same cosmology as the analytical covariance, namely a WMAP9 + SN + BAO model \citep{hinshaw/etal:2013}. We check for the cosmology dependence of the covariance and find that this choice is neglibible for our results (see setup no.~26 in comparison to the fiducial setup in Sect.~\ref{sec:results}).

The most important change, however, is that we use the actual galaxy pair counts as measured from the data in the calculation of the shape noise contribution to the covariance matrix. A more accurate treatment of the impact of survey geometry effects on the estimate of the shape noise term was shown to have significant impact on the goodness of fit \citep{troxel/etal:2018b}. Shape noise increases on large scales as the number of pairs of galaxies was previously overestimated when survey boundaries and smaller-scale masks were ignored. Using the actual pair counts is simpler and more accurate than the explicit modelling of the mask and clustering performed in \citet{troxel/etal:2018b}, but we find good agreement with their approach. While our method in principle introduces noise into the covariance, this effect is negligible due to the high number density of weak lensing samples. Using the actual number of pairs also naturally accounts for the effect of varying source density on the covariance matrix.

We note that neither \citet{troxel/etal:2018b} nor we account for survey geometry and clustering effects on pair counts in the covariance term that mixes shape noise and sample variance contributions. While inconsistent, we expect the modifications in the mixed term to be subdominant to those in the pure shape noise term. This, and a more accurate treatment of survey geometry effects on the sample variance contribution, will be addressed in future work.

Finally we use the linear mid-point of the $\theta$ bins for the theoretical covariance calculation, which is close to the weighted mean pair separation that was suggested by \citet{joudaki/etal:2017b} and \citet{troxel/etal:2018b}, instead of the logarithmic mid-point that was used in \citetalias{hildebrandt/etal:2017}. Note that for the model (Sect.~\ref{sec:CS_signal}) we go beyond these approximations and integrate over the $\theta$ bins. For the covariance estimate such a level of sophistication is not needed.

\subsection{E-/B-mode decomposition}
\label{sec:EB}

Gravitational lensing only creates curl-free E-modes in the galaxy ellipticity distribution to first order. Cosmological B-modes can be produced from higher-order terms beyond the first-order Born approximation  \citep{schneider/etal:1998,hilbert/etal:2009}, source clustering of galaxies and intrinsic alignments \citep{schneider/etal:2002}, as well as cosmic strings and some alternatives to $\Lambda$CDM \citep[see for example][]{thomas/etal:2017}. All of these produce B-modes that are statistically negligible for the current generation of cosmic shear surveys.

Separating the cosmic shear signal into E- and B-modes is, however, an important check for residual systematics. Much work has been carried out to understand which statistics are most useful for this purpose and what one can learn from a non-zero B-mode signal about systematic errors. Here we follow the work by \citet{asgari/etal:2016,asgari/etal:2018} and use the COSEBIs \citep[Complete Orthogonal Sets of E-/B-mode Integrals;][]{COSEBIS} 2-point statistics to cleanly separate E- from B-modes on a given finite angular range.

B-modes are estimated from the five tomographic bins (i.e. all 15 auto and cross-combinations) using the $\log$-COSEBIs for modes $n\le20$ over an angular baseline of $0\farcm5<\theta< 300'$, spanning the $\theta$ range probed by our correlation function measurement. Consistency with a zero signal is quantified by a $\chi^2$ test using the shape-noise part of the analytical covariance discussed in Sect.~\ref{sec:cov_ana}. This analysis is carried out for all possible intervals $[n_{\rm min}, n_{\rm max}]$ with $1\le n_{\rm min} \le n_{\rm max}\le 20$. The $p$-values from the $\chi^2$ test are almost all well above 1\% indicating no significant B-modes. For only four out of 210 tested intervals $[n_{\rm min}, n_{\rm max}]$ we find $p$-values slightly below 1\%, but all of these are still well above 0.1\%.

We repeat this test for other $\theta$ ranges, $0\farcm5 < \theta < 40'$, $0\farcm5<\theta<72'$, $40'<\theta<100'$, and $8'<\theta<300'$. Results yield even higher $p$-values for these more restricted angular intervals, with only one out of 840 tests showing a $p$-value below 1\%. From these tests we conclude that there are no significant systematic errors that would produce a B-mode signal in a tomographic analysis of the KV450 E-modes over scales $0\farcm5<\theta<300'$. This is a significant improvement over KiDS-450, and we attribute this change to the improved \emph{lens}fit weights. We confirm this finding by an independent Fourier-space analysis of B-modes with band powers \citep[see][for details of this technique]{vanuitert/etal:2018} that also finds no significant signal.

While this is a necessary condition, showing consistency with zero for COSEBIs B-modes is not sufficient to conclude that a correlation function analysis over the same scales is unaffected by B-mode systematics. The correlation functions $\xi_\pm$ also pick up so-called ambiguous modes for which one cannot decide whether they represent E- or B-modes when measurements span only a finite interval in $\theta$ (Schneider in prep.).
These ambiguous modes are not contained in the clean COSEBIs E-/B-mode measurements. Thus for a cosmological analysis with $\xi_\pm$ one implicitly has to assume that the ambiguous modes are pure E-modes. In order to address this concern we also analyse the COSEBIs E-mode signal, which is free from ambiguous modes, with a Gaussian covariance matrix (missing the super-sample covariance but including shape noise, sample variance, and mixed terms) and compare to the results based on correlation function measurements. These results are reported in Sect.~\ref{sec:results}. Note that one possible incarnation of systematic, ambiguous modes is a constant shear. Such a constant pattern would be corrected for by our estimate of the $c$-term and the corresponding nuisance parameter $\delta c$.

\section{Theoretical modelling}
\label{sec:theory}

The theoretical modelling of the cosmic shear signal and the various systematic effects discussed below are carried out with the setup discussed in \citet{kohlinger/etal:2017}. This setup is based on the nested sampling algorithm \textsc{MultiNest} \citep{feroz/etal:2009} as implemented in the \textsc{python} wrapper \textsc{PyMultiNest} \citep{buchner/etal:2014} that is included in the \textsc{MontePython} package \citep{audren/etal:2013,brinckmann/etal:2018}. This is a deviation from the setup used in \citetalias{hildebrandt/etal:2017} but we show in  Appendix~\ref{app:codecomp} that essentially identical cosmological constraints result from using the \textsc{CosmoLSS}\footnote{\url{https://github.com/sjoudaki/cosmolss}} software developed for \citetalias{hildebrandt/etal:2017} and also a third implementation using \textsc{CosmoSIS} \citep{zuntz/etal:2015}. 

\subsection{Cosmic shear signal}
\label{sec:CS_signal}
The estimated quantities $\hat{\xi}_{\pm}$ (Eq.~\ref{eq:xi_est}) are directly related to cosmological theory and can be modelled via
\begin{equation}
\label{eq:xi_pm}
\xi_\pm^{ij}(\theta) = \frac{1}{2\pi}\int_0^\infty \d\ell \,\ell \,P^{ij}_\kappa(\ell) \, J_{0,4}(\ell \theta) \, , 
\end{equation}
where $J_{0,4}$ are Bessel functions of the first kind, $P_\kappa$ is the convergence power spectrum, and $i$ and $j$ are the indices of the tomographic bins that are being cross-correlated. Using the Kaiser-Limber equation 
and the Born approximation one finds
\begin{equation}
P^{ij}_\kappa(\ell) = \int_0^{\chi_{\rm H}} \d \chi \, \frac{q_i(\chi)q_j(\chi)}{[f_K(\chi)]^2} \, P_\delta \left( \frac{\ell+1/2}{f_K(\chi)},\chi \right),
\label{eq:Pkappa} 
\end{equation}
with $P_\delta$ being the non-linear matter power spectrum, $\chi$ being the comoving distance, $\chi_\mathrm{H}$ the comoving horizon distance, and $q$ the lensing efficiency
\begin{equation}
\label{eq:q_of_chi}
q_i(\chi) = \frac{3 H_0^2 \Omega_{\rm m}}{2c^2} \frac{f_K(\chi)}{a(\chi)}\int_\chi^{\chi_{\rm H}}\, \d \chi'\ n_{\chi,i}(\chi') 
\frac{f_K(\chi'-\chi)}{f_K(\chi')} \, , 
\end{equation}
which depends on the redshift distribution of the sources $n_i(z)\mathrm{d}z=n_{\chi,i}(\chi)\mathrm{d}\chi$. The integral over the redshift distribution is carried out by linearly interpolating the mid-points of the histogram (bin width $\Delta z=0.05$) that comes out of the DIR calibration method.

The total matter power spectrum is estimated with the Boltzmann-code \textsc{CLASS} \citep{blas/etal:2011,audren/etal:2011,lesgourgues/etal:2011} with non-linear corrections from \textsc{HMCode} \citep{mead/etal:2015}. The effect of massive neutrinos is included in the \textsc{HMCode} calculation \citep{mead/etal:2016}. We assume two massless neutrinos and one massive neutrino fixing the neutrino mass of this massive neutrino at the minimal mass of $m=0.06~{\rm eV}$. We do not marginalise over any uncertainty in the neutrino mass in our fiducial setup but additionally report results for $m=0~{\rm eV}$ and $m=0.26~{\rm eV}$, the latter corresponding to the 95\% upper limit from Planck-Legacy \citep[TT,TE,EE+lowE; ][]{planck/cosmo:2018}.

Our \textsc{CLASS} and \textsc{MontePython} setup probes a slightly different parameter space than \textsc{CAMB} and \textsc{CosmoMC} \citep{lewis/etal:1999,lewis/bridle:2002} that were used for the \textsc{CosmoLSS} pipeline of \citetalias{hildebrandt/etal:2017}. Here we use as our five primary cosmological parameters the cold-dark-matter density parameter $\Omega_{\rm CDM}$, the scalar power spectrum amplitude $\ln(10^{10}A_{\rm s})$, the baryon density parameter $\Omega_{\rm b}$, the scalar power spectrum index $n_{\rm s}$, and the scaled Hubble parameter $h$. The priors on these parameters are equivalent to the ones in \citetalias{hildebrandt/etal:2017} and reported in Table~\ref{tab:params}. Several of these priors are informative because cosmic shear alone cannot constrain some of the parameters sufficiently well. However, we take care to include all state-of-the-art measurements in the prior ranges, e.g. distance-ladder measurements as well as Planck CMB results in the prior for $h$, CMB as well as big-bang-nucleosynthesis results for $\Omega_{\rm b}h^2$. For a full discussion of the priors, see section~6 of \citetalias{hildebrandt/etal:2017} and \citet{joudaki/etal:2017}.

Values for other cosmological parameters of interest, e.g. $\Omega_{\rm m}$, $\sigma_8$, and $S_8$, and their marginal errors are calculated from the chains after convergence.

\begin{table}
\caption{Model parameters and their priors for the KV450 cosmic shear analysis.}
\label{tab:params}
\begin{tabular}{lll}
\hline
\hline
Parameter & Symbol & Prior\smallstrut\\
\hline
CDM density & $\Omega_{\rm CDM} h^2$ & $[0.01, 0.99]$\smallstrut\\
Scalar spectrum ampl. & $\ln(10^{10}A_{\rm s})$ & $[1.7, 5.0]$\smallstrut\\
Baryon density & $\Omega_{\rm b} h^2$ & $[0.019, 0.026]$ \smallstrut\\
Scalar spectral index & $n_{\rm s}$ & $[0.7, 1.3]$ \smallstrut\\
Hubble parameter & $h$ & $[0.64, 0.82]$ \smallstrut\\
\hline
IA amplitude & $A_{\rm IA}$ & $[-6, 6]$ \smallstrut\\
Baryon feedback ampl. & $B$ & $[2.00, 3.13]$ \smallstrut\\
Constant $c$-term offset & $\delta c$ & $0.0000\pm0.0002$ \smallstrut\\
2D $c$-term amplitude & $A_c$ & $1.01\pm0.13$ \smallstrut\\
Redshift offset bin 1 & $\delta z_1$ & $0.000\pm0.039$ \smallstrut\\
Redshift offset bin 2 & $\delta z_2$ & $0.000\pm0.023$ \smallstrut\\
Redshift offset bin 3 & $\delta z_3$ & $0.000\pm0.026$ \smallstrut\\
Redshift offset bin 4 & $\delta z_4$ & $0.000\pm0.012$ \smallstrut\\
Redshift offset bin 5 & $\delta z_5$ & $0.000\pm0.011$ \smallstrut\\
\hline
\end{tabular}
\tablefoot{The first five lines represent the primary cosmological parameters whereas the following nine lines correspond to the nuisance parameters used in our model. Brackets indicate top-hat priors whereas values with errors indicate Gaussian priors.}
\end{table}

The nine $\theta$ bins in which we estimate $\xi_{\pm}$ are relatively broad so that it is non trivial to relate the model to the data.\footnote{Note that this is less of a problem for the 20 narrower bins used by DES in \citet{troxel/etal:2018a}.} \citet{joudaki/etal:2018} and \citet{troxel/etal:2018b} discuss using the average weighted pair separation instead of the logarithmic mid-point \citepalias[as it was done in ][]{hildebrandt/etal:2017} of the bin to calculate the model. In \citet[][their appendix A]{asgari/etal:2018} it was shown that both approaches are biased. The \citetalias{hildebrandt/etal:2017} approach biases the model for $\xi_\pm$ slightly high and hence $S_8$ is biased low. The \citet{joudaki/etal:2018} and \citet{troxel/etal:2018b} approach tried to correct for this but instead biases $\xi_{\pm}$ low to a similar degree and hence $S_8$ is biased high. Here we integrate $\xi_\pm$ over each $\theta$ bin, which yields results that correspond to the red lines in figure~A.1 of \citet{asgari/etal:2018}. This unbiased approach has the disadvantage of requiring an additional integration in the likelihood. However, since this is a rather fast step in the likelihood evaluation, the computational overhead is minimal.

\subsection{Intrinsic alignments}
\label{sec:IA}
We use the same `non-linear linear' intrinsic alignment model as in \citetalias{hildebrandt/etal:2017}, which modifies the 2-point shear correlation functions by adding two more terms describing the II and GI effects \citep{hirata/seljak:2004}:
\begin{equation}
\hat \xi_\pm = \xi_\pm + \xi_\pm^{\rm II} + \xi_\pm^{\rm GI}\,,
\end{equation}
where $\xi_\pm^{\rm II}$ and $\xi_\pm^{\rm GI}$ are calculated from the II and GI power spectra
\begin{equation}
\begin{split}
\label{eq:P_II_P_GI}
P_{\rm II} (k,z) = F ^2 (z) P_\delta(k,z)\\
P_{\rm GI} (k,z) = F(z) P_\delta(k,z)
\end{split}
\end{equation}
in a similar way as $\xi_\pm$ is calculated from $P_\delta$ (see Eqs.~\ref{eq:xi_pm}-\ref{eq:q_of_chi}) with
\begin{equation}
F(z) = -A_{\rm IA}C_1\rho_{\rm crit}\frac{\Omega_{\rm m}}{D_+(z)}\,\left(\frac{1+z}{1+z_0}\right)^\eta\,,
\end{equation}
where $C_1=5\times10^{-14}h^{-1}M_\odot^{-1}\mathrm{Mpc}^3$, $\rho_{\rm crit}$ is the critical density today, and $D_+(z)$ is the linear growth factor. More details can be found in equations 6--11 of \citetalias{hildebrandt/etal:2017}. Note that we do not include a redshift or luminosity dependence of $F(z)$ in our fiducial cosmological model, i.e. we set $\eta=0$. The mean luminosity of the sources increases with redshift, but the overall redshift dependence of $A_{\rm IA}$ is not well constrained. We run one test where $\eta$ is allowed to vary (and the pivot redshift is set to $z_0=0.3$), which implicitly includes the effect that an increasing luminosity would have on the IA amplitude.

There has been some discussion about whether the linear or non-linear matter power spectrum should be used in Eq.~\ref{eq:P_II_P_GI}. For our fiducial setup we opt to be consistent with previous work and use the non-linear power spectrum and a broad prior $A_{\rm IA}\in[-6, 6]$. We also run a setup where we switch to the linear matter power spectrum, i.e. the standard linear alignment model. As this model has less power on small, non-linear scales, we use it to test our sensitivity to the large uncertainty in the currently poor constraints on small-scale intrinsic alignments, e.g. in the behaviour of satellite galaxy populations. Furthermore, we present results for the default non-linear model with a more informative Gaussian prior for the intrinsic alignment amplitude $A_\mathrm{IA}=1.09 \pm 0.47$, based on the results from \citet{johnston/etal:2018} for the full galaxy sample that they analysed. \citet{johnston/etal:2018} saw a pronounced dichotomy in the alignments of late- and early-type galaxies. Their full galaxy sample is almost equally split between early and late types, while the KV450 tomographic samples have early-type galaxy fractions of $\sim25\%$, and even less in the lowest redshift bin. However, \citet{johnston/etal:2018} found indications that galaxy type alone does not fully describe the observed variability in alignment amplitudes (suggesting a dependence on satellite/central galaxy fractions), and coincidentally the intrinsic alignment amplitude we adopt is close to a prediction from measurements on purely early- and late-type samples \citep[see figure 6 of][]{johnston/etal:2018}.

\subsection{Baryon feedback}
\label{sec:baryons}
The non-linear matter power spectrum is modified by baryonic feedback processes on small scales. This modification has to be taken into account to avoid biases in cosmological parameters estimated from cosmic shear \citep{semboloni/etal:2011}. Currently, however, these feedback processes are not very well understood and different hydrodynamical simulations yield considerably different results \citep{chisari/etal:2018}. There is no clear consensus yet which models are realistic, but it should be noted that many of the hydrodynamical simulations reported in the literature were not run to yield a good match to weak lensing observations but rather to study feedback effects and/or galaxy formation.

The uncertainty in the baryon feedback has to be propagated into the confidence intervals of cosmological parameters that are sensitive to those small scales. In \citetalias{hildebrandt/etal:2017} we took a conservative approach by marginalising over a very wide range of possible small-scale modifications of the matter power spectrum. This prior range clearly included unrealistic models but was chosen to let the data constrain the baryon feedback strength. However, in the end it effectively led to the eradication of small-scale information. As was shown by \citet{troxel/etal:2018b} removing the small $\theta$ bins does not significantly change the constraining power of KiDS-450 because of this effect.

Here we try to improve on this by using a slightly more informative prior for the baryon feedback, enabled by the most recent cosmological hydro simulation results. Instead of marginalising over unrealistic models with \emph{more} small-scale ($k\approx10~{\rm Mpc}^{-1}$) structure than a pure dark matter model or extremely strong AGN feedback recipes that are ruled out by observations of baryonic probes, we adopt a more informative prior that is consistent with such observations. 

Baryon feedback is modelled with \textsc{HMCode} \citep{mead/etal:2015}, which implements a halo model description, using the same dependence of the two parameters $B$ (amplitude of the halo mass-concentration relation) and $\eta_0$ (halo bloating parameter) as in \citet{joudaki/etal:2018}: $\eta_0=0.98-0.12B$. Hence we only include $B$ as a free nuisance parameter in our model adopting a flat prior $B\in[2.0,3.13]$ that spans the range from the most aggressive feedback model in the OWLS simulation suite \citep[][AGN, $B\sim2$]{vandaalen/etal:2011} to the dark matter-only case ($B=3.13$). This range of feedback is consistent with the range allowed by the more modern and well-calibrated BAHAMAS suite \citep[][Mead et al. in prep.]{mccarthy/etal:2017}. Extending the prior further towards models that have enhanced power compared to the dark matter-only case ($B>3.13$) or unrealistically strong feedback \citep[$B<2$, see e.g.][]{yoon/etal:2019} is not justified in our opinion. If the cosmic shear measurements are not able to self-calibrate the feedback then a more informative prior like the one used here should be the preferred choice.

\citet{huang/etal:2018} suggest that \textsc{HMCode} does not model all possible feedback scenarios well and therefore introduces biases. They suggest a different approach that is not based on a halo model but on a principal component analysis of different hydrodynamical simulations. It is important to note though that the biases they see are limited to $\Omega_{\rm m}$ and $\sigma_8$ (and the dark energy equation of state), but $S_8$ is very robust against the choices presented in their paper. This further supports the approach we are taking here as our main result will be the value of $S_8$ from this cosmic shear measurement. In order to alleviate any concerns about this baryon feedback modelling we also present results with very wide, uninformative priors on \emph{both} \textsc{HMCode} parameters in Sect.~\ref{sec:results}.

It should be noted that, unlike other observational biases (multiplicative shear measurement bias, redshift bias) that affect all angular scales equally, the baryon feedback only affects the smallest angular scales. Thus the overall effect of the choice of feedback model and prior is relatively unimportant for the conclusions of this work that tries to constrain standard flat-$\Lambda$CDM cosmological parameters. Constraining possible scale-dependent effects like massive neutrinos or warm dark matter is more degenerate with the baryon feedback and certainly needs more care, which we defer to future work.

\section{Results}
\label{sec:results}

\subsection{Fiducial cosmological results}
\label{sec:fiducial}

\begin{figure*}
\includegraphics[width=0.49\hsize]{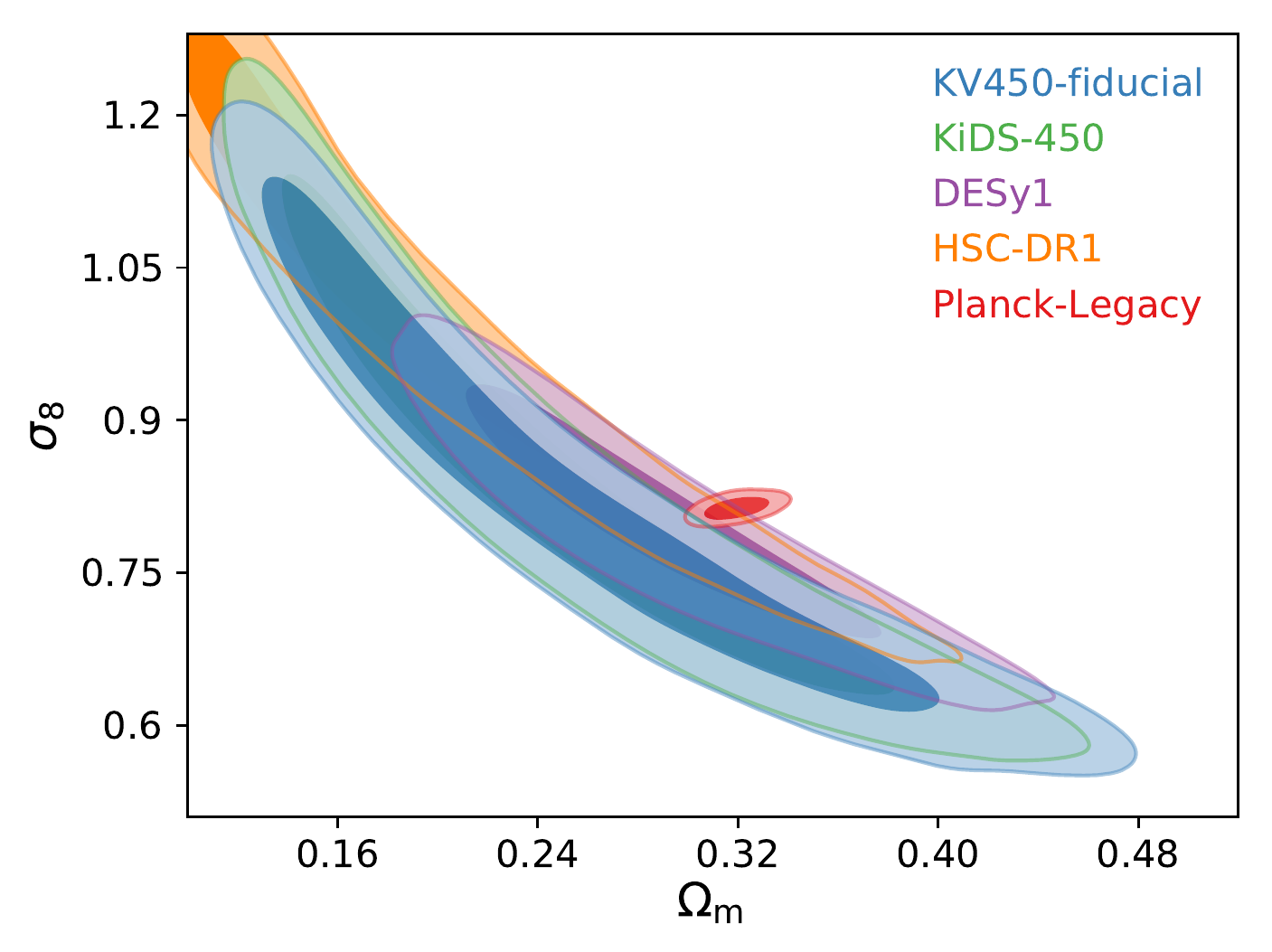}
\includegraphics[width=0.49\hsize]{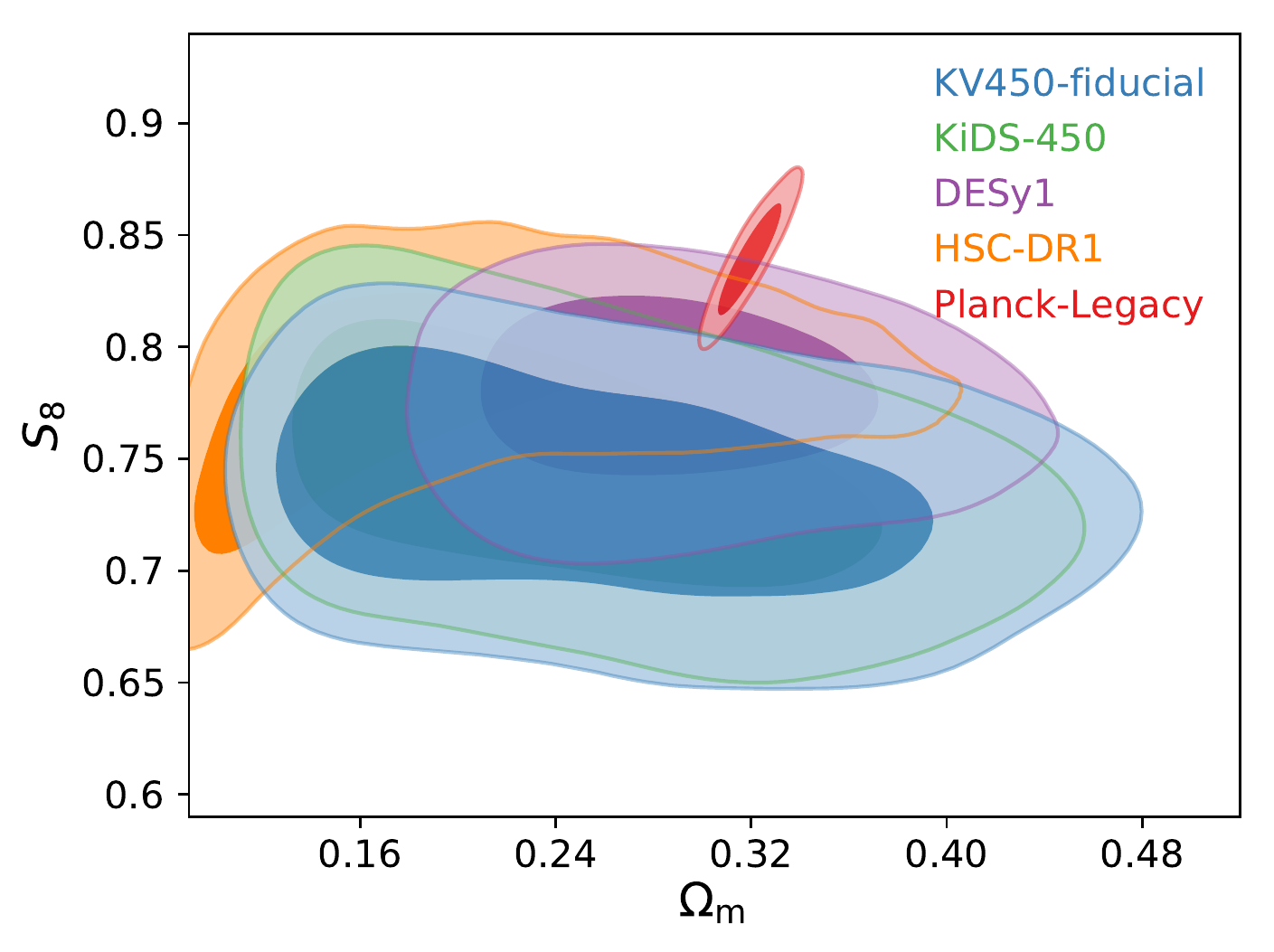}
\caption{\label{fig:Om_s8_fidDz} Marginalised posterior contours (inner 68\% confidence level, outer 95\% confidence level) in the $\Omega_{\rm m}$-$\sigma_8$ plane (\emph{left}) and the $\Omega_{\rm m}$-$S_8$ plane (\emph{right}) for the fiducial KV450 setup (blue), the optical-only KiDS-450 analysis from \citetalias{hildebrandt/etal:2017} (green), DESy1 using cosmic shear only \citep[purple;][]{troxel/etal:2018a}, HSC-DR1 cosmic shear \citep[orange;][]{hikage/etal:2018}, and the Planck-Legacy analysis \citep[red;][TT+TE+EE+lowE]{planck/cosmo:2018}.}
\end{figure*}

Results for the fiducial flat $\Lambda$CDM model (Sect.~\ref{sec:theory}) fitted to the KV450 cosmic 2-point shear correlation functions (Fig.~\ref{fig:xipm}) are presented in Fig.~\ref{fig:Om_s8_fidDz} showing 2D projections into the most relevant cosmological parameters $\Omega_{\rm m}$, $\sigma_8$, and $S_8$ and in Table~\ref{tab:res_fidDz} for all parameters.\footnote{2D projections of all parameters are shown in Fig.~\ref{fig:triangle}.} This fiducial model accounts for and marginalises over uncertainties in the baryon feedback (Sect.~\ref{sec:baryons}), intrinsic alignments (Sect.~\ref{sec:IA}), additive (Sect.~\ref{sec:c_term}) and multiplicative (Sect.~\ref{sec:m_bias}) shear measurement bias, and mean redshifts of the five tomographic bins estimated with the DIR method (Sect.~\ref{sec:DIR}). The KV450 confidence contours are compared to results from cosmic shear with KiDS-450 \citepalias{hildebrandt/etal:2017}, the DESy1 cosmic shear-only analysis \citep{troxel/etal:2018a}, HSC-DR1 cosmic shear \citep{hikage/etal:2018} as well as all primary CMB probes of the Planck-Legacy data set \citep[][TT+TE+EE+lowE]{planck/cosmo:2018}.

\begin{table}
\caption{\label{tab:res_fidDz}Fiducial result for the KV450 cosmic shear measurement.}
\begin{tabular}{llr}
  \hline\hline
  Parameter & Symbol & Value \bigstrut\\
  \hline
  CDM density              & $\Omega_{\rm CDM} h^2$ &  $ 0.118_{-0.066}^{+0.038}$ \bigstrut\\
  Scalar spectrum ampl.    & $\ln 10^{10}A_{\rm s}$ &  $ 3.158_{-1.426}^{+1.154}$ \bigstrut\\
  Baryon density           & $\Omega_{\rm b} h^2$   &  $ 0.022_{-0.004}^{+0.003}$ \bigstrut\\
  Scalar spectral index    & $n_{\rm s}$            &  $ 1.021_{-0.141}^{+0.149}$ \bigstrut\\
  Hubble parameter         & $h$                    &  $ 0.745_{-0.043}^{+0.073}$ \bigstrut\\
  \hline
  IA amplitude             & $A_{\rm IA}$           &  $ 0.981_{-0.678}^{+0.694}$ \bigstrut\\
  Baryon feedback ampl.    & $B$                    &  $ 2.484_{-0.475}^{+0.189}$ \bigstrut\\
  Constant $c$-term offset & $\delta c$             &  $ 0.000_{-0.0002}^{+0.0002}$ \bigstrut\\
  2D $c$-term amplitude    & $A_c$              &  $ 1.022_{-0.125}^{+0.129}$ \bigstrut\\
  Redshift offset bin 1    & $\delta z_1$           &  $-0.007_{-0.034}^{+0.034}$ \bigstrut\\ 
  Redshift offset bin 2    & $\delta z_2$           &  $-0.010_{-0.021}^{+0.019}$ \bigstrut\\ 
  Redshift offset bin 3    & $\delta z_3$           &  $ 0.013_{-0.021}^{+0.020}$ \bigstrut\\
  Redshift offset bin 4    & $\delta z_4$           &  $ 0.001_{-0.011}^{+0.012}$ \bigstrut\\
  Redshift offset bin 5    & $\delta z_5$           &  $-0.001_{-0.011}^{+0.011}$ \bigstrut\\ 
  \hline
  Matter density           & $\Omega_{\rm m}$       &  $ 0.256_{-0.123}^{+0.064}$ \bigstrut\\
  Power spectrum amplitude & $\sigma_8$             &  $ 0.836_{-0.218}^{+0.132}$ \bigstrut\\
  $\sigma_8\sqrt{\Omega_{\rm m}/0.3}$ & $S_{8}$     &  $ 0.737_{-0.036}^{+0.040}$ \bigstrut\\
\hline
\end{tabular}
\tablefoot{Reported are the mean posterior values and 68\% confidence intervals. The first five entries represent the standard cosmological parameters used in our model, which are separated by a horizontal line from the nine nuisance parameters. Another horizontal line separates three derived parameters. Most of these constraints are prior-dominated, the most important exception being $S_8$.}
\end{table}

The KV450 results agree very well with the optical-only KiDS-450 cosmic shear results, which were based on the same ellipticity measurements but used a subset of the KV450 sources (but slightly more area), a different photo-$z$ setup, different tomographic binning, and redshift calibration, different shear calibration, and different \emph{lens}fit weights (apart from more subtle changes described in the preceding sections). This agreement between KV450 and KiDS-450 also means that a tension remains between this most recent cosmic shear measurement and the results from the Planck-Legacy primary CMB measurement. Concentrating on just the $S_8$ parameter, Planck-Legacy (TT+TE+EE+lowE) and KV450 are discrepant at the $2.3\sigma$ level (assuming Gaussian posteriors). Comparing to Planck-2015 (TT+lowP) we find a slightly larger tension in $S_8$ of $2.5\sigma$. The DESy1 cosmic shear result, which exhibits a 25\% smaller error on $S_8$ than KV450, and the HSC-DR1 result lie in the middle with their measured $S_8$ values being $0.9\sigma$ and $0.8\sigma$, respectively, higher than the KV450 value.\footnote{For a further discussion of the KV450 results in comparison to other cosmological probes we refer the reader to \citetalias{hildebrandt/etal:2017} and \citet{mccarthy/etal:2018}. As KV450 and KiDS-450 results are very similar, all comparisons between KiDS-450 and other probes in these two papers also apply to KV450.} When comparing constraints from the different cosmic shear surveys one should keep in mind that the different surveys model some systematic errors with (partly the same) nuisance parameters that are not constrained by the data, blowing up the uncertainties. So the agreement is not as good as it seems if some of the systematic effects bias the different surveys in the same way.

The marginal error on $S_8$, $\sigma_{S_8}=0.038$, is similar to KiDS-450 ($\sigma_{S_8}=0.039$). This does not mean that the high-redshift galaxies in the fifth tomographic bin do not add any statistical power.\footnote{Note that some of these galaxies (roughly half) were already included in the KiDS-450 analysis, but there they constituted the high-$z$ tails of the redshift distributions of the four lower-redshift tomographic bins.} The two analyses differ in quite a few aspects (e.g. number and kind of nuisance parameters, priors) so that a direct comparison of the errors down to the second counting digit is not meaningful and does not reflect the power of the data sets but rather differences in the two analyses (see also Sect.~\ref{sec:MCMC_sys}).

The model fits the data very well yielding a $\chi^2=180.6$ for 181 degrees of freedom\footnote{Note that many of our parameters are constrained by the priors. So this somewhat na\"ive estimate of the degrees of freedom is just an approximation.}. This is a significant improvement compared to \citetalias{hildebrandt/etal:2017}, which we attribute mostly to the more accurate covariance matrix \citep[see also][]{troxel/etal:2018b} but possibly also to better internal consistency (see Sect.~\ref{sec:consistency}).

Looking at the other model parameters we find that the intrinsic alignment amplitude is consistent with unity, in very good agreement with \citetalias{hildebrandt/etal:2017}, \citet{troxel/etal:2018a}, and \citet{hikage/etal:2018}. The value and error of the baryon feedback amplitude $B$ indicate a significant departure from a dark matter only scenario \citep[similar to ][who use a wider prior but also additional data]{joudaki/etal:2018}. The two $c$-term nuisance parameters $\delta c$ and $A_c$ are not constrained by the data and the five $\delta z_i$ parameters are all consistent with zero but also strongly prior-dominated.

In the following we describe results from further tests that divert from the fiducial setup or change the selection of the data vector. This is done to check the robustness of the fiducial results against different choices that were made in the analysis and to relate our results more easily to literature measurements. Table~\ref{tab:MCMC} summarises the different setups that we test via additional MCMC runs. For most of these setups we vary only one aspect at a time to keep things comparable. The resulting $S_8$ values for all setups are shown in Fig.~\ref{fig:S8}, and some additional parameters of interest are reported in Table~\ref{tab:S8}.

\begin{table*}
\centering
\caption{\label{tab:MCMC} Setups for further MCMC test runs.}
\begin{tabular}{rll}
\hline
\hline
no. & Setup & Difference w.r.t. fiducial setup \smallstrut\\
\hline
1  & sDIR                 & sDIR (smoothed version of DIR) $n(z)$ \smallstrut\\
2  & DIR-w/o-COSMOS       & DIR $n(z)$ based on all spec-$z$ except COSMOS \smallstrut\\
3  & DIR-w/o-COSMOS\&VVDS & DIR $n(z)$ based on all spec-$z$ except COSMOS and VVDS \smallstrut\\
4  & DIR-w/o-VVDS         & DIR $n(z)$ based on all spec-$z$ except VVDS \smallstrut\\
5  & DIR-w/o-DEEP2        & DIR $n(z)$ based on all spec-$z$ except DEEP2 \smallstrut\\
6  & DIR-C15              & DIR $n(z)$ based on the COSMOS-2015 photo-$z$ \smallstrut\\
7  & CC-fit               & $n(z)$ from GMM fit to small-scale clustering-$z$ (CC; App.~\ref{sec:CC}) \smallstrut\\
8  & CC-shift             & DIR $n(z)$ shifted to best fit CC measurements \smallstrut\\
9  & OQE-shift            & DIR $n(z)$ shifted to best fit large-scale clustering-$z$ (OQE; App.~\ref{app:OQE}) \smallstrut\\
\hline
10 & no-deltaz            & redshift uncertainty switched off, i.e. $\delta z_i=0$ \smallstrut\\
11 & IA-Gauss             & informative Gaussian prior on $A_{\rm IA}$ \smallstrut\\
12 & IA-linear-PS         & using the linear power spectrum in Eqs.~\ref{eq:P_II_P_GI} \smallstrut\\
13 & IA-z-evolution       & allowing for redshift evolution in the IA model \smallstrut\\
14 & no-baryons           & baryon feedback switched off, i.e. $B=3.13$ \smallstrut\\
15 & wide-baryons         & wide prior on baryon feedback, $B\in[1.4,4.8]$, $\eta_0\in[0.4,0.9]$ \smallstrut\\
16 & no-systematics       & no marginalisation over nuisance parameters, no error on $m$ \smallstrut\\
17 & no-systematics\_merr & same as 16 but including a $\sigma_m=0.02$ uncertainty in the $m$-bias \smallstrut\\
18 & all-xip              & all scales $0\farcm5<\theta<300'$ used for $\xi_+$ \smallstrut\\
19 & nu0                  & massless neutrinos \smallstrut\\
20 & nu0p26               & one massive neutrino with $m=0.26~{\rm eV}$ and two massless neutrinos \smallstrut\\
\hline
21 & no-bin1              & using tomographic bins 2, 3, 4, and 5 only \smallstrut\\
22 & no-bin2              & using tomographic bins 1, 3, 4, and 5 only \smallstrut\\
23 & no-bin3              & using tomographic bins 1, 2, 4, and 5 only \smallstrut\\
24 & no-bin4              & using tomographic bins 1, 2, 3, and 5 only \smallstrut\\
25 & no-bin5              & using tomographic bins 1, 2, 3, and 4 only \smallstrut\\
\hline
26 & iterative-covariance & analytical covariance based on the best-fit fiducial cosmology\smallstrut\\
\hline
\end{tabular}
\end{table*}

\subsection{Tests of the redshift distributions}
\label{sec:MCMC_nz}

The most unique aspect of the cosmic shear measurement presented here is the estimate of the redshift distributions that are needed to interpret the signal. In this section we show how different choices for the redshift distribution affect the results and in particular the main conclusion about the tension between KV450 and Planck. As a first set of tests (setups no.~1-9 from Table~\ref{tab:MCMC}) we substitute the DIR $n(z)$ that is based on the full spectroscopic calibration sample with different alternatives as described in Sect.~\ref{sec:DIR}. For these MCMC runs we do not re-calculate the error on the mean redshifts and assume that these errors, which serve to define the Gaussian priors on the $\delta z_i$ parameters, are identical to the DIR bootstrap analysis with the full spec-$z$ sample. This assumption enhances differences rather than diluting them.

\begin{figure*} \includegraphics[width=0.49\hsize]{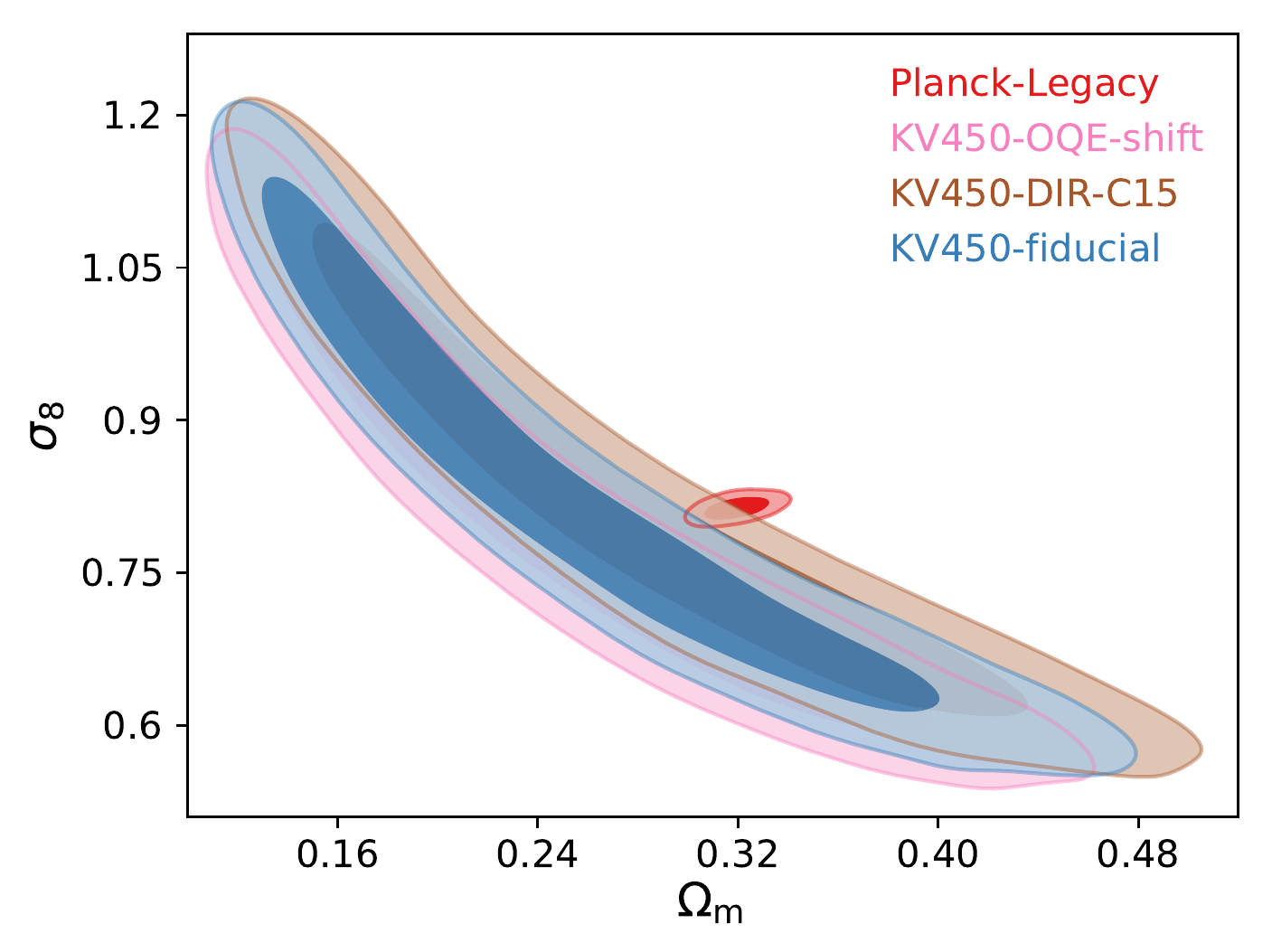} \includegraphics[width=0.49\hsize]{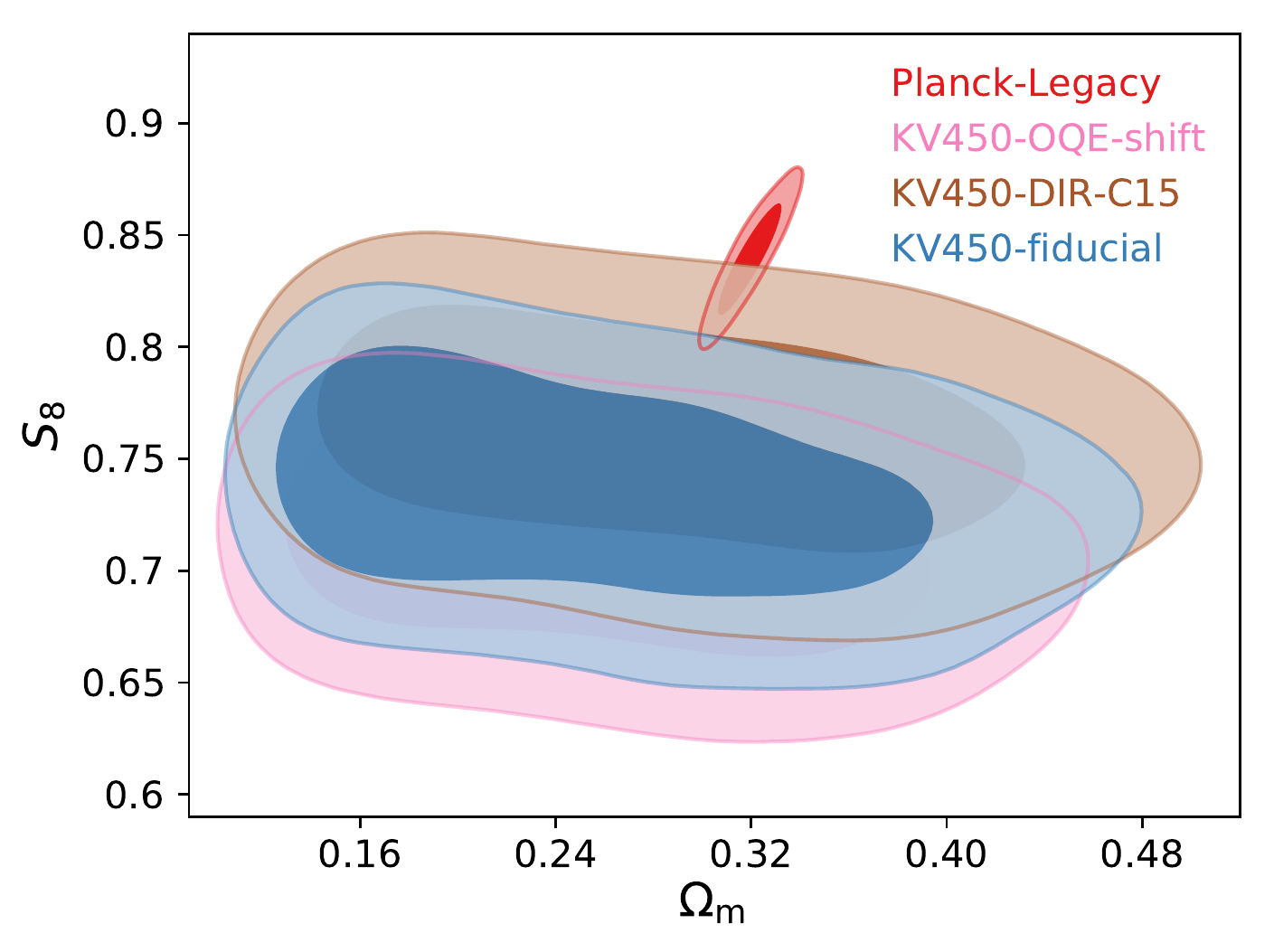} \caption{\label{fig:Om_s8_nz} Same as Fig.~\ref{fig:Om_s8_fidDz} but for the most extreme alternative redshift distributions described in Sect.~\ref{sec:DIR}. Brown contours correspond to the redshift distributions with the highest $S_8$ (DIR-C15) that we argue in Sect.~\ref{sec:DIR} might produce a biased result. The pink contours correspond to the redshift distributions with the lowest $S_8$ (OQE-shift).}
\end{figure*}

Figure~\ref{fig:Om_s8_nz} shows the results in the $\Omega_{\rm m}$-$\sigma_8$ and $\Omega_{\rm m}$-$S_8$ planes for setups no.~6~\&~9, corresponding to the tests with the COSMOS-2015 based DIR $n(z)$ and the OQE-shift $n(z)$ (both redshift distributions are shown in Fig.~\ref{fig:zdist_OQE}), respectively, in comparison to the fiducial setup and Planck. These two setups were chosen because they yield the highest and lowest $S_8$ values, respectively. All other setups no.~1-9 lie in between those extremes. The two extremes with the highest and lowest $S_8$ values are discrepant with Planck at the $1.7\sigma$ and $2.9\sigma$ level, respectively, in terms of their marginal errors on $S_8$. Compared to the fiducial KV450 setup the OQE-shift setup no.~9 yields an $S_8$ that is $0.7\sigma$ lower whereas the DIR-C15 setup no.~6 is $0.6\sigma$ high compared to the fiducial value of $S_8$.

\begin{figure}
\centering
\includegraphics[width=\hsize, clip=true, trim=3.5cm 0cm 3.5cm 1cm]{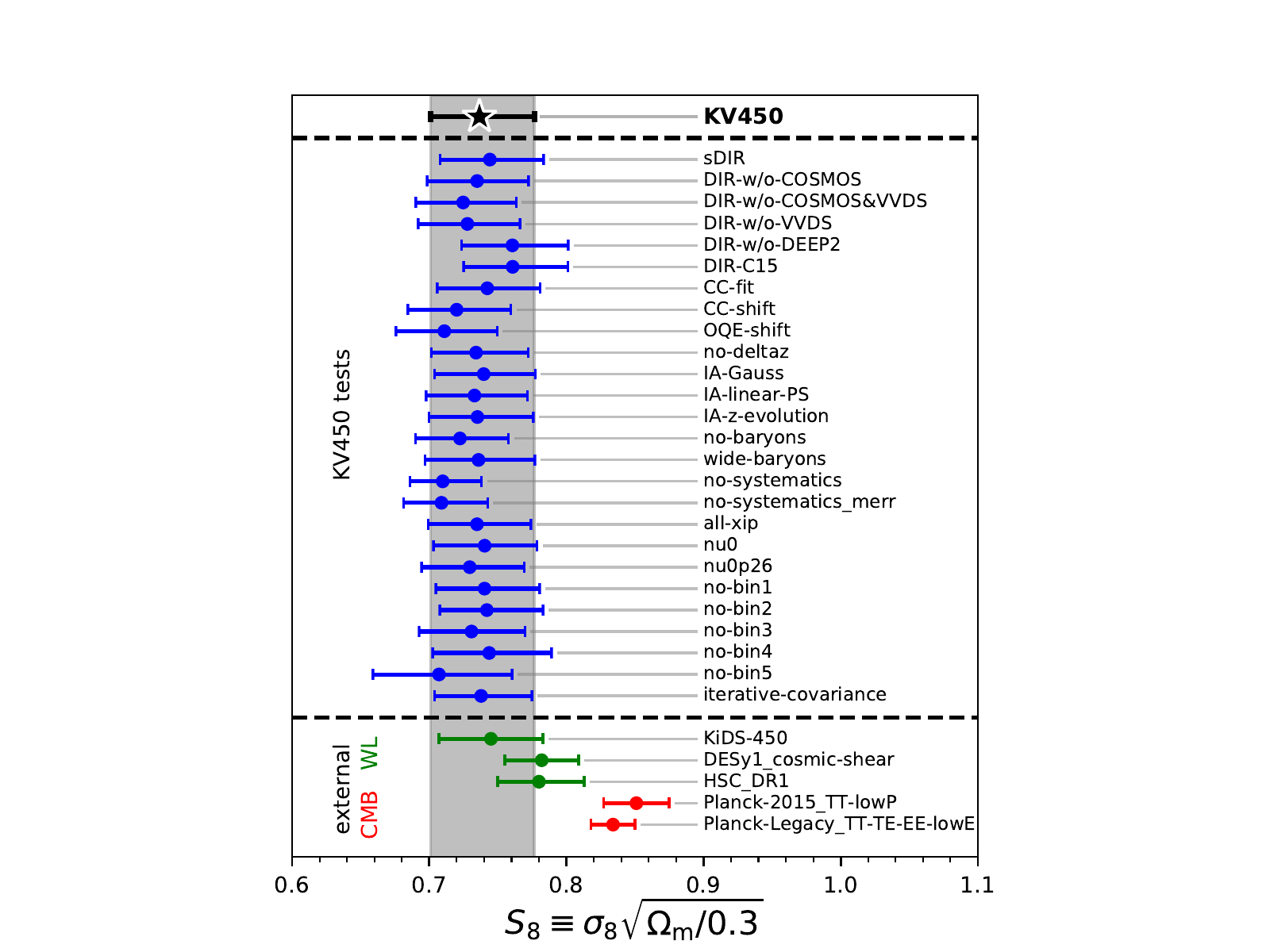}
\caption{\label{fig:S8} Constraints on $S_8$ for the fiducial KV450 setup (black), the different tests described in Sects.~\ref{sec:MCMC_nz}~\&~\ref{sec:MCMC_sys} (blue), other cosmic shear measurements (green) and the Planck CMB results (red). The gray, horizontal lines are only intended to guide the eye.}
\end{figure}

Figure~\ref{fig:S8} and Table~ \ref{tab:S8} show that all redshift distributions tested here yield $S_8$ values that are consistent within $\sim1\sigma$. However, it should be noted that these data points are correlated because a large fraction of the spec-$z$ calibration sample is the same for most setups, the clustering-$z$ setups no.~7--9 and the COSMOS-2015 setup no.~6 being exceptions. The highest $S_8$ values (and correspondingly the lowest mean redshifts) are obtained with the DIR method when using the COSMOS-2015 photo-$z$ catalogue instead of the spec-$z$ catalogue or when excluding DEEP2 (the highest-redshift spec-$z$ catalogue) from the spec-$z$ calibration sample. The lowest $S_8$ values are measured for the DIR $n(z)$ when COSMOS and VVDS are excluded from the spec-$z$ calibration sample and the two setups that are based on shifting the fiducial DIR $n(z)$ to best fit the CC and OQE measurements. The range spanned by these different choices for the $n(z)$ can be regarded as a very conservative estimate of the systematic uncertainty introduced by the redshift distributions.

\begin{table*}
\centering
\caption{\label{tab:S8} Results for further MCMC test runs. }
\begin{tabular}{rlrrrrrrr}
\hline
\hline
no. & Setup & $\Omega_{\rm m}$ & $\sigma_8$ & $S_8$ & $\chi^2$ & $N_{\rm data}$ & $N_{\rm par}$\bigstrut\\
\hline
   & KV450 fiducial       & $0.256_{-0.123}^{+0.064}$ & $0.836_{-0.218}^{+0.132}$ & $0.737_{-0.036}^{+0.040}$ & $180.6$ & 195 & 14 \bigstrut\\\hline
1  & sDIR                 & $0.248_{-0.119}^{+0.058}$ & $0.857_{-0.224}^{+0.168}$ & $0.744_{-0.036}^{+0.039}$ & $178.8$ & 195 & 14 \bigstrut\\
2  & DIR-w/o-COSMOS       & $0.256_{-0.122}^{+0.056}$ & $0.834_{-0.215}^{+0.145}$ & $0.735_{-0.036}^{+0.037}$ & $180.0$ & 195 & 14 \bigstrut\\
3  & DIR-w/o-COSMOS\&VVDS & $0.242_{-0.114}^{+0.049}$ & $0.845_{-0.216}^{+0.135}$ & $0.725_{-0.035}^{+0.039}$ & $181.5$ & 195 & 14 \bigstrut\\
4  & DIR-w/o-VVDS         & $0.245_{-0.114}^{+0.057}$ & $0.842_{-0.215}^{+0.173}$ & $0.728_{-0.036}^{+0.038}$ & $181.2$ & 195 & 14 \bigstrut\\
5  & DIR-w/o-DEEP2        & $0.266_{-0.125}^{+0.056}$ & $0.846_{-0.222}^{+0.180}$ & $0.761_{-0.037}^{+0.041}$ & $179.1$ & 195 & 14 \bigstrut\\
6  & DIR-C15              & $0.286_{-0.118}^{+0.081}$ & $0.814_{-0.212}^{+0.095}$ & $0.761_{-0.036}^{+0.040}$ & $178.7$ & 195 & 14 \bigstrut\\
7  &               CC-fit & $0.288_{-0.107}^{+0.077}$ & $0.786_{-0.189}^{+0.084}$ & $0.742_{-0.036}^{+0.039}$ & $179.5$ & 195 & 14 \bigstrut\\
8  &             CC-shift & $0.243_{-0.114}^{+0.064}$ & $0.838_{-0.202}^{+0.145}$ & $0.720_{-0.036}^{+0.040}$ & $183.0$ & 195 & 14 \bigstrut\\
9  &            OQE-shift & $0.258_{-0.115}^{+0.071}$ & $0.802_{-0.210}^{+0.103}$ & $0.711_{-0.035}^{+0.038}$ & $183.5$ & 195 & 14 \bigstrut\\
10 & no-deltaz            & $0.249_{-0.120}^{+0.056}$ & $0.846_{-0.229}^{+0.156}$ & $0.734_{-0.033}^{+0.038}$ & $179.4$ & 195 &  9 \bigstrut\\
11 & IA-Gauss             & $0.264_{-0.129}^{+0.073}$ & $0.828_{-0.230}^{+0.175}$ & $0.740_{-0.036}^{+0.038}$ & $179.7$ & 195 & 14 \bigstrut\\
12 & IA-linear-PS         & $0.248_{-0.122}^{+0.064}$ & $0.845_{-0.219}^{+0.162}$ & $0.733_{-0.035}^{+0.039}$ & $181.1$ & 195 & 14 \bigstrut\\
13 & IA-z-evolution       & $0.249_{-0.115}^{+0.067}$ & $0.843_{-0.220}^{+0.166}$ & $0.735_{-0.035}^{+0.041}$ & $179.6$ & 195 & 15 \bigstrut\\
14 & no-baryons           & $0.247_{-0.121}^{+0.048}$ & $0.834_{-0.213}^{+0.156}$ & $0.722_{-0.032}^{+0.035}$ & $181.3$ & 195 & 13 \bigstrut\\
15 & wide-baryons         & $0.256_{-0.117}^{+0.058}$ & $0.832_{-0.215}^{+0.119}$ & $0.736_{-0.039}^{+0.041}$ & $180.6$ & 195 & 15 \bigstrut\\
16 & no-systematics       & $0.222_{-0.102}^{+0.035}$ & $0.862_{-0.152}^{+0.180}$ & $0.710_{-0.024}^{+0.028}$ & $180.8$ & 195 &  5 \bigstrut\\
17 & no-systematics\_merr & $0.217_{-0.100}^{+0.030}$ & $0.873_{-0.152}^{+0.205}$ & $0.709_{-0.028}^{+0.034}$ & $180.8$ & 195 &  5 \bigstrut\\
18 & all-xip              & $0.258_{-0.114}^{+0.066}$ & $0.827_{-0.213}^{+0.145}$ & $0.735_{-0.035}^{+0.039}$ & $198.7$ & 225 & 14 \bigstrut\\
19 & nu0                  & $0.246_{-0.117}^{+0.051}$ & $0.856_{-0.217}^{+0.143}$ & $0.740_{-0.037}^{+0.038}$ & $180.1$ & 195 & 14 \bigstrut\\
20 & nu0p26               & $0.275_{-0.125}^{+0.060}$ & $0.795_{-0.205}^{+0.103}$ & $0.730_{-0.035}^{+0.040}$ & $179.9$ & 195 & 14 \bigstrut\\\hline
21 & no-bin1              & $0.249_{-0.118}^{+0.068}$ & $0.850_{-0.216}^{+0.134}$ & $0.740_{-0.035}^{+0.040}$ & $124.0$ & 130 & 13 \bigstrut\\
22 & no-bin2              & $0.236_{-0.109}^{+0.043}$ & $0.874_{-0.199}^{+0.190}$ & $0.742_{-0.034}^{+0.041}$ & $114.6$ & 130 & 13 \bigstrut\\
23 & no-bin3              & $0.289_{-0.099}^{+0.088}$ & $0.775_{-0.189}^{+0.086}$ & $0.731_{-0.038}^{+0.039}$ & $120.5$ & 130 & 13 \bigstrut\\
24 & no-bin4              & $0.284_{-0.132}^{+0.083}$ & $0.801_{-0.213}^{+0.104}$ & $0.744_{-0.041}^{+0.045}$ & $119.0$ & 130 & 13 \bigstrut\\
25 & no-bin5              & $0.261_{-0.111}^{+0.076}$ & $0.791_{-0.213}^{+0.095}$ & $0.707_{-0.048}^{+0.053}$ & $128.7$ & 130 & 13 \bigstrut\\
\hline
26 & iterative covariance & $0.258_{-0.120}^{+0.081}$ & $0.833_{-0.219}^{+0.129}$ & $0.738_{-0.034}^{+0.037}$ & $182.3$ & 195 & 14 \bigstrut\\
\hline
\end{tabular}
\tablefoot{Shown are the matter density (3rd column), the power spectrum amplitude (4th), $S_8$ (5th), $\chi^2$ (6th), the number of data points (7th), and the number of fitting parameters (8th). Note that the parameters values correspond to the mean of the posterior whereas the $\chi^2$ corresponds to the maximum likelihood.}
\end{table*}

As a further test we check the influence of the assumption of uncorrelated $\delta z_i$ uncertainties. The mean redshift estimates of our tomographic bins are indeed significantly correlated as our bootstrap analysis tells us ($\sim90\%$ correlation for neighbouring bins and $\sim40-70\%$ for more widely separated bins). Assuming full correlation the formula presented in \citet{hoyle/etal:2018} suggests an increase of the prior ranges for the $\delta z_i$ - assumed to be uncorrelated - by a factor $\sim 2.4$ to account for the correlation. We conservatively ran a chain where we increased the priors by a factor of three (not shown in Table~\ref{tab:S8} or Fig.~\ref{fig:S8}) and found no change in the central value of $S_8$ and an increase in the $S_8$ uncertainty of $\sim 8\%$.

\subsection{Tests on nuisance parameters, priors, the data vector, and neutrino mass}
\label{sec:MCMC_sys}

As reported in Table~\ref{tab:MCMC} we carry out a number of further tests to check the influence of the systematic effects that we model with nuisance parameters, their priors, the selection of the data vector, and the fixed mass of the neutrinos.

In setup no.~10 we test the influence of the $\delta z_i$ nuisance parameters. When the redshift uncertainties are not marginalised over we find almost identical results to the fiducial setup that includes their marginalisation. The total uncertainty on $S_8$ is reduced by merely $\sim6\%$. This confirms the finding of \citetalias{hildebrandt/etal:2017} that random redshift calibration errors are subdominant to some of the other systematic uncertainties (see below). It should be noted that -- unlike in \citetalias{hildebrandt/etal:2017} -- we explicitly include an estimate of the sample variance (including spectroscopic selection effects) of the $n(z)$ here as our uncertainties are estimated from a spatial bootstrap analysis of the calibration sample. So also this sampling variance is subdominant for KV450. This effect can be compared to the range of results shown in Sect.~\ref{sec:MCMC_nz} suggesting that systematic errors in the redshift calibration dominate over sample variance and shot noise but are hard to quantify.

The choice of the prior for the intrinsic alignment amplitude $A_{\rm IA}$ does not have a large effect on the results either. Using an informative Gaussian prior (setup no.~11) again yields almost identical results to the fiducial setup, with a very similar constraint on the intrinsic alignment amplitude $A_{\rm IA}=1.06_{-0.34}^{+0.37}$ with tighter error compared to $A_{\rm IA}=0.98_{-0.68}^{+0.69}$ for the fiducial setup. Switching from the non-linear to the linear power spectrum to model the GI and II terms in Eq.~\ref{eq:P_II_P_GI} (setup no.~12) does not have an appreciable effect on the results either. Also allowing for redshift evolution in the IA model (setup no.~13) does not change the results in a significant way, meaning that IA modelling and prior choices are currently subdominant in the systematic error budget.

A somewhat larger effect can be seen when baryon feedback is left unaccounted for (setup no.~14). In that case the mean posterior value of $S_8$ is lowered by $\sim0.4\sigma$. This is due to the fact that baryon feedback dilutes structures on small scales ($k\approx 10~h{\rm Mpc}^{-1}$)\footnote{The enhancement of the power spectrum by stellar feedback on very small scales ($k\gg10~h{\rm Mpc}^{-1}$) is unimportant for the $\theta$ range probed by KV450.} and hence lowers the amplitude of the power spectrum. When this is not modelled the power spectrum amplitude increases for a given $S_8$. Thus, a smaller value of $S_8$ is sufficient to describe the observed amplitude of the correlation functions. Allowing for extremely wide priors on the HMCode baryon feedback parameters (setup no.~15) gives consistent results with the fiducial setup. This can be understood in the way that already our slightly informative fiducial prior erases most small-scale information so that even a more conservative prior does not lead to a further loss of statistical power. Alternatively, one could just disregard the smallest scales for $\xi_+$ and not model the baryon feedback at all as it was done by \citet{troxel/etal:2018a} with the DESy1 data. As $\xi_\pm$ mixes all $k$ scales we prefer to model the baryon feedback and properly marginalise over the uncertainty. We would like to stress that these results indicate that the tension seen in $S_8$ between KV450 and Planck cannot be alleviated by a more generous prior on the baryon feedback amplitude. Considering the value of $B\sim1$, favoured by \citet{yoon/etal:2019}, we would note caution as it is unlikely that this level of baryon feedback is physical.

Switching off any marginalisation of systematic errors does not yield any cosmologically meaningful results but can be used to quantify the importance of all systematic effects for the total error budget. Setup no.~16 shows an error $\sigma_{S_8}=0.026$ that is $\sim30\%$ smaller than the fiducial run, $\sigma_{S_8}=0.038$. Na\"ively adding a systematic error of the same size in quadrature to the purely statistical error of setup no.~16 yields a total error that is very close to the fiducial error. This means that in KV450 the marginalisation over systematic uncertainties is approximately equally important as the statistical error for the total error budget. This is similar to the findings of \citetalias{hildebrandt/etal:2017}. Comparing the relative error on $S_8$ for setup no.~16 to the `no-systematics' setup of \citetalias{hildebrandt/etal:2017} reveals a $\sim10\%$ decrease in the uncertainty that we attribute to the statistical power added by the fifth tomographic bin.

The largest single contribution to the systematic error budget as quantified here comes from the uncertainty in the multiplicative shear measurement bias (but see the discussion of the redshift uncertainty in Sect.~\ref{sec:discussion}). This can be seen by comparing setups no.~16~\&~17. The latter includes a propagation of an $m$-bias uncertainty of $\sigma_m=0.02$ into the covariance matrix but is otherwise identical to setup no.~16. Note that this $\sigma_m$ is twice as large as the uncertainty used in \citetalias{hildebrandt/etal:2017} as suggested by the new findings of \citetalias{kannawadi/etal:2018} and hence of increased importance here (see Sect.~\ref{sec:m_bias}).

One extension in the KV450 analysis compared to KiDS-450 is the inclusion of two nuisance parameters that describe the uncertainty in the additive shear measurement bias and the amplitude of the 2D ellipticity pattern imprinted on the data (Sect.~\ref{sec:c_term}). While their influence on the fiducial result is completely negligible these parameters help to make the analysis less susceptible to systematics on large scales. The main reason for restricting the $\xi_+$ measurements to fiducial scales of $\theta<72'$ in the past was the uncertainty in the $c$-term. With these new nuisance parameters properly accounting for this uncertainty there is no longer any strong reason to restrict the analysis to these small scales. In setup no.~18 we explore what happens when also $\xi_+$ is analysed all the way out to $\theta<300'$. Not surprisingly the difference to the fiducial setup is minuscule. The $\delta c$ and $A_c$ nuisance parameters essentially eradicate all information from these scales rendering the choice of an upper $\theta$ cut-off for the $\xi_+$ analysis rather unimportant.

There has been much discussion on the importance of assumptions about the neutrino mass on the results from cosmic shear experiments \citep[e.g.][]{mccarthy/etal:2018}. Here we show two extreme setups (no.~19~\&~20) that support the notion that this is not a major concern for current cosmic shear measurements. Neither assuming massless neutrinos (setup no.~19) nor setting the neutrino mass to the maximum value allowed by Planck ($\Sigma m_\nu<0.26~{\rm eV}$ at 95\% confidence; setup no.~20) changes the KV450 cosmic shear constraints on $S_8$ in a significant way, in agreement with the findings of \citet{joudaki/etal:2017}.

We find no evidence for B-modes in KV450 as described in Sect.~\ref{sec:EB}. In order to further support this we compare our `no-systematics' setup no.~16 to an analysis of COSEBI E-modes finding almost identical results for the central value of $S_8$ ($S_8=0.710$ for $\xi_\pm$ and $S_8=0.700$ for the COSEBIs).\footnote{Note that the errors are not directly comparable yet as we are using slightly different scales and a Gaussian covariance matrix for the COSEBIs analysis (see Sect.~\ref{sec:EB}).} This finding suggests that our correlation function analysis is also not significantly affected by B-mode systematics in the ambiguous modes (Schneider in prep.). 
A more comprehensive analysis of the COSEBIs E-/B-mode signals and a full cosmological analysis will be presented in a forthcoming paper (Asgari et al. in prep.).

\subsection{Consistency tests}
\label{sec:consistency}

\citet{efstathiou/etal:2018} pointed out that the different tomographic bins in KiDS-450 were in slight tension with each other, citing significances of $\sim3\sigma$. In particular the third and fourth bin showed a lower/higher amplitude than expected from the cosmological model fitted to the other three bins. This behaviour is already visible in figure~5 of \citetalias{hildebrandt/etal:2017} where most of the data points including the third bin lie below the best-fit global model. There has been some debate about quantifying the significance of this discrepancy though, also in light of the revisions for the covariance matrix introduced in \citet{troxel/etal:2018b}. In particular, \citet{koehlinger/etal:2018} show that a Bayesian evidence analysis yields significances of $<3\sigma$ and no strong evidence for internal tension in KiDS-450 in contrast to the analysis presented in \citet{efstathiou/etal:2018}. However, \citet{efstathiou/etal:2018} did highlight the importance of performing internal consistency checks as a standard part of any analysis.

We therefore run five more MCMC setups (no.~21-25) where one tomographic bin is rejected at a time. These quick tests show consistency in their $S_8$ values with the fiducial results, with all values lying well within $1\sigma$. Following the methodology of \citet{koehlinger/etal:2018} we also split the data vector and run additional chains with a separate set of parameters for both parts and then report the Bayes factor:
\begin{equation}
R_{01} = \frac{\mathcal{Z}({\rm H_0})}{\mathcal{Z}({\rm H_1})}\,,
\end{equation}
where $\mathcal{Z}$ is the Bayesian evidence, $\rm H_0$ is the null hypothesis that `there exists one common set of parameters that describes the full data vector' and $\rm H_1$ is the alternative hypothesis that `there exist two sets of parameters that each describe one part of the data vector'. We find values of $\ln R_{01} = 3.0; 2.5; 4.5; 4.9; 5.9$ for the five tomographic bins. This can be translated via the `Jeffreys' scale' into `strong evidence' for the null hypotheses for the 1st and 2nd tomographic bins and `decisive evidence' for the 3rd, 4th, and 5th tomographic bins, meaning there is no significant tension between the bins.

As shown by \citet{koehlinger/etal:2018} the evidence ratio test described above is only a necessary condition for internal consistency \citep[see also][]{raveri/etal:2018}. Another possible test is to check the posterior probability distribution of parameter differences between the two parts of the split data vector. Following \citet[][see their table~3]{koehlinger/etal:2018} we look at all combinations of the parameters $S_8$, $\Omega_{\rm m}$, and $A_{\rm IA}$. For all of the combinations we find agreement between the splits with differences at a level of $<1.3\sigma$ (most much smaller) indicating again no significant internal tension for any of the tomographic bins.

We also split the data vector into its $\xi_+$ and $\xi_-$ parts and apply the same formalism as described above. Another test consists in splitting the small-scale ($\theta<5'$) part of $\xi_+$ from all large-scale ($\theta>5'$, $\xi_+$ and $\xi_-$) measurements. Again we find `decisive evidence' for the null hypothesis of no internal tension in both cases, with  $\ln R_{01} = 6.7; 6.2$ for the $\xi_\pm$ and large-scale/small-scale split, respectively. The maximum parameter differences for these two splits correspond to $0.8\sigma$, which means that also these parts of the data vector are fully consistent with each other.

Another test that directly checks for consistency of the redshift distributions is the shear-ratio test, which we present in Appendix~\ref{app:SRT}. The fiducial redshift distributions pass this test easily. However, it should be mentioned that all other redshift distributions used above (scenarios no.~1--9) pass the test equally well so that it must be concluded that given our lens and source samples this shear-ratio test is not discriminative enough yet.

\subsection{Iterative covariance}
As in \citet{vanuitert/etal:2018}, we update the covariance model with the best-fit cosmology of our fiducial run and repeat the parameter inference (setup no.~26 in Tables~\ref{tab:MCMC}~\&~\ref{tab:S8}). The central value for $S_8$ changes by only $\sim0.001$, while the errors on $S_8$ shrink by $\sim5\%$. The latter trend is expected because the parameter values of the original cosmology calculation lie a little above the region of large posterior values in the $\Omega_{\rm m}$-$\sigma_8$ plane, which leads to slightly larger sample variance \citep{reischke/etal:2017}. We retain the more conservative errors of the original run as our main result since these are expected to largely cover the variability of sample variance across the range of cosmologies with high posterior probability.

\section{Discussion}
\label{sec:discussion}

The KV450 analysis presented here is the first wide-field cosmic chear experiment in which photo-$z$ are estimated from well-matched optical+near-infrared photometry spanning the wavelength range $320~{\rm nm}\la\lambda\la2350~{\rm nm}$ \citep{wright/etal:2018}. The fiducial redshift distributions are estimated from a re-weighting technique (DIR) that can take full advantage of the degeneracy-breaking power of 9-dimensional magnitude space and a large and diverse spectroscopic calibration sample. Hence KV450 is arguably the cosmic shear experiment with the best-calibrated redshift distributions to date. The results agree very well with previously published results from KiDS-450 optical-only data \citepalias{hildebrandt/etal:2017} and are consistent with DESy1 \citep{troxel/etal:2018a} and HSC-DR1 \citep{hikage/etal:2018}, with the latter two yielding slightly higher values of $S_8$.

While most of the galaxies used in the KV450 analysis were also used for KiDS-450 and the ellipticity measurements for these galaxies have not changed, this result is non-trivial nevertheless. The value of $S_8$, which essentially determines the position of the confidence contours perpendicular to their degeneracy direction, is itself fully degenerate with the mean redshifts of the tomographic bins. Since the photo-$z$ estimates, the tomographic binning, and the redshift calibration changed from KiDS-450 to KV450 as described in Sect.~\ref{sec:photo} this agreement was not a foregone conclusion. 

We have demonstrated that the tension with the Planck CMB results is robust against all reasonable choices of parameters, priors, and modelling details as summarised in Fig.~\ref{fig:S8}. The only changes to the fiducial setup that show some potential to alleviate the tension with Planck are related to the redshift distribution of the sources. If the highest-redshift spectroscopic survey in our calibration sample (DEEP2) is excluded, the estimated mean redshifts go down and the value for $S_8$ goes up by about $0.6\sigma$. A similar shift can be observed when the weighted direct calibration is used but the spectroscopic calibration sample is substituted by the high-quality COSMOS-2015 photo-$z$ catalogue from \citet{laigle/etal:2016}. 

In this latter case our results agree better with the measurements from DESy1 \citep{troxel/etal:2018a} and HSC-DR1 \citep{hikage/etal:2018}. It is intriguing that these two surveys calibrated their redshift distributions also with the help of the \citet{laigle/etal:2016} COSMOS-2015 photo-$z$ catalogue. This raises the question whether something is special about this catalogue -- compared to our spec-$z$ calibration sample -- that would shift down the mean redshift estimates (and shift up $S_8$) in a systematic way. In \citet{laigle/etal:2016} it is reported that, while the photo-$z$ are truly excellent in comparison to other public photo-$z$ catalogues, there is still a non-negligible fraction of catastrophic outliers present. In the magnitude range of $23<i<24$ this is quantified to be $\eta\sim6\%$, where $\eta$ is the fraction of objects with $|z_{\rm phot}-z_{\rm spec}|/(1+z_{\rm spec})>0.15$.\footnote{Note that the outlier rate reported in \citet{laigle/etal:2016} is a steep function of magnitude with $\sim1\%$ outliers at $22<i<23$ and $>10\%$ outliers at $24<i<25$.} This estimate should be regarded as a lower limit as the spec-$z$ surveys that the COSMOS-2015 photo-$z$ are being compared to are incomplete at these magnitudes. At low COSMOS-2015 photo-$z$ these outliers are almost exclusively objects that are in reality at very high redshift.\footnote{The impact of such catastrophic redshifts outliers on weak lensing studies has already been investigated in \citet{schrabback/etal:2010} and \citet{schrabback/etal:2018} for earlier photo-zs from COSMOS \citep{ilbert/etal:2009} and 3D-HST \citep{skelton/etal:2014}.
} Furthermore, we found a small photo-$z$-dependent bias when comparing the COSMOS-2015 photo-$z$ to spec-$z$, with the COSMOS-2015 photo-$z$ underestimating the true redshifts by $\sim 0.01$. Assuming that this is the case we deliberately added a peak at $z=2$ to the DIR-C15 $n(z)$ coresponding to 5\% of all sources, shifted the $n(z)$ by 0.01, and re-ran the cosmological analysis. In this toy model case, we find the presence of low-level bias and catastrophic outliers reduce the inferred value of $S_8$, constraining $S_8=0.737_{-0.036}^{+0.039}$. It is clear that this example is somewhat simplistic but it illustrates the effect that even a small fraction of outliers and a small bias can have on the inferred cosmological parameters. Just introducing the 5\% outlier peak without shifting the $n(z)$ by 0.01 yields $S_8=0.749_{-0.036}^{+0.040}$.

Hence assuming that the \citet{laigle/etal:2016} photo-$z$ are an unbiased estimate of the true redshift, without correcting for this effect, would lead to an under-estimate of the mean redshifts of all the tomographic bins. As discussed in Sect.~\ref{sec:blinding}, the current approach to marginalising over our uncertainty in the redshift distributions is insensitive to a coherent systematic bias in all tomographic bins.  Furthermore the differences we find between the mean redshifts of the COSMOS-2015 calibrated redshift distributions and our spectroscopic calibration, are significantly larger than the redshift uncertainty allowed for in both the DESy1 \citep{troxel/etal:2018a} and HSC-DR1 \citep{hikage/etal:2018} analyses. We leave a quantification of this effect to a future analysis. It will be very interesting to see whether correcting for these outliers will significantly raise the mean redshifts of DESy1 and HSC-DR1, decrease their $S_8$ estimates, and increase the tension with Planck, possibly solidifying the challenge to the standard flat $\Lambda$CDM cosmological model.

All of that being said it is equally important and prudent to ask whether our DIR calibration with the combined spec-$z$ sample could possibly systematically bias the mean redshifts high (and hence $S_8$ low). One important aspect here is the incompleteness of the spec-$z$ calibration sample at faint magnitudes due to failure of successfully measuring a redshift. This is well documented for zCOSMOS-bright \citep{lilly/etal:2009}, DEEP2 \citep{newman/etal:2013}, and VVDS \citep{lefevre/etal:2013} showing an obvious trend with magnitude. At the faint end of our source sample ($r\sim24$) the spectroscopic success rate is $\sim60-80\%$, meaning that some objects are left unaccounted for. This problem is alleviated somewhat when different surveys are combined as in our spec-$z$ calibration sample. Also, the re-weighting technique can partly overcome this issue if the sample is not missing parts of our 9-dimensional magnitude space entirely. A 9D re-weighting scheme is obviously more robust than a lower-dimensional re-weighting approach. In 9D it is less likely that colour-redshift degeneracies would go undetected due to incompleteness in the calibration sample. Certainly, also in the 9D case the matching in magnitude space is not perfect, mostly due to significant photometric noise. Deeper data in the calibration fields will help to reduce this source of systematic error and help in quantifying the success (or failure) of the DIR method. This problem will be analysed in more detail in a forthcoming paper. However, the most important point about the spectroscopic incompleteness due to a variable success rate is that systematically missing galaxies would most likely bias the redshifts low and $S_8$ high because in general it is harder to secure redshifts of high-$z$ than low-$z$ galaxies (hence the ``redshift desert''). It is hard to imagine that the spectroscopic incompleteness would artificially enhance the tension with Planck. The fully redundant analyses with redshift distributions based on clustering-$z$ (setups no.~7--9) further increase our confidence in the robustness of the DIR method with the full spec-$z$ calibration sample.

The finding that statistical and systematic errors are roughly equal in size sets the agenda for future cosmic shear analyses. The systematic error budget for $S_8$ reported here is dominated by the uncertainty in the estimate of the multiplicative shear measurement bias ($\sigma_m=0.02$). However, there is additional uncertainty in the redshift distribution (discussed at length above), the intrinsic alignment model, and the baryon feedback, all of which are hard to quantify at the moment. We believe it is currently not meaningful to put a concrete number on these effects but these uncertainties might well be comparable to other effects that have been quantified more precisely. 

Taking the redshift distribution uncertainty as an example, it depends on which of the scenarios no.~1--9 are considered realistic. Obviously the scatter in the $S_8$ values between all of these nine alternative setups rivals the $S_8$ statistical error in size. Eliminating some of the setups like DIR-C15 (because of photo-$z$ instead of spec-$z$), DIR-w/o-DEEP2 (because of insufficient high-$z$ coverage), and the CC-shift and OQE-shift setups (because of less mature methodology) greatly decreases the scatter. However, there is no clear consensus yet on which of these setups should be considered realistic. This situation needs to change if future cosmic shear experiments are to reach their full potential.

We take the good fit of the model to the data as an indication that systematic errors are not severely underestimated. The internal consistency between different parts of the data vector is shown to be high. However, it should be noted that any consistency analysis is subject to a-posteriori statistics. While it might be reasonable to test the different tomographic bins against each other as well as $\xi_+$ against $\xi_-$ and large against small scales \citep[as suggested by][]{efstathiou/etal:2018} there is a look-elsewhere effect that is almost impossible to quantify. In the past, some of these consistency analyses were conducted a posteriori, in the case when things did not look quite right. In the future it will be important to define data splits before even looking at the data, as was done here (see Appendix~\ref{app:history}) and by other teams in recent years, to minimise the look-elsewhere effect.

\section{Summary}
\label{sec:summary}

Here we present new cosmic shear results based on a combination of optical data from the Kilo-Degree Survey and near-infrared data from the VISTA Kilo-Degree Infrared Galaxy Survey. For the first time such a combined data set is used in a wide-field cosmic shear experiment addressing one of the largest systematic uncertainties in such measurements, the calibration of the redshifts of the sources. Compared to a previous optical-only study over the same area of 450~deg$^2$ \citep{hildebrandt/etal:2017} we significantly improve this crucial part of analysis and report cosmic shear measurements that are arguably the most robust in terms of the redshift calibration to date.

These unprecedentedly accurate redshift distributions for $\sim 12\times10^6$ galaxies in five tomographic bins out to a redshift of $z\sim1.2$ are combined with state-of-the-art shear measurements of all these sources based on high-resolution images from a telescope and camera that were purpose-built for weak gravitational lensing measurements. An updated suite of image simulations and improved empirical techniques are used to estimate and correct for any residual multiplicative and additive biases in these shear measurements. No significant B-modes are detected in the KV450 data. Together with a careful treatment of all known astrophysical systematic effects (intrinsic alignments, baryon feedback, neutrino mass) as well as observational uncertainties we update the findings of \citet{hildebrandt/etal:2017}. A standard flat $\Lambda$CDM model complemented by several nuisance parameters, which allow us to propagate the uncertainties of all the systematic corrections mentioned above into the model, yields an almost perfect fit to the data. In a blind analysis we find as our main result a value of $S_8\equiv\sigma_8\sqrt{\Omega_{\rm m}/0.3}=0.737_{-0.036}^{+0.040}$, very similar to the optical-only analysis of KiDS by \citet{hildebrandt/etal:2017}. Data  products from the analysis presented here are available at \url{http://kids.strw.leidenuniv.nl}.

This estimate of $S_8$ is discrepant with measurements from the Planck-Legacy analysis at the $2.3\sigma$ level. We test several possible alternative treatments of systematic errors such as the baryon feedback priors, the intrinsic alignment modelling, and the neutrino mass, but the only alternatives that alleviate the tension are related to the redshift distribution. In particular, if the COSMOS-2015 photo-$z$ catalogue \citep{laigle/etal:2016} is used instead of a combined spec-$z$ calibration sample, our $S_8$ estimate increases by $\sim0.6\sigma$ to $S_8=0.761_{-0.036}^{+0.040}$. This value is very close to the cosmic shear results from the Dark Energy Survey's first-year analysis (DESy1; $S_8=0.782$) and the first data release of Hyper SuprimeCam (HSC-DR1; $S_8=0.780$), both of which also use the COSMOS-2015 catalogue for their redshift calibration. Here we speculate that a significant fraction of high-$z$ outliers that are assigned a low photo-$z$ in the COSMOS-2015 catalogue, as reported by \citet{laigle/etal:2016}, could artificially increase the $S_8$ values of DESy1, HSC-DR1, and our alternative analysis where we use COSMOS-2015. If this is found to be correct and accounted for, the DESy1 and HSC-DR1 results for $S_8$ would come down, also increasing their tension with Planck. In that case a combined analysis of all three surveys, based on a consistent and robust approach to calibrate the redshifts, could potentially show a very significant discrepancy with the best current CMB measurements. Alternatively, a bias in our fiducial analysis would be caused if a significant fraction of the redshifts in our spec-$z$ calibration sample were incorrect and/or the incompleteness of the spec-$z$ at the faint end was entirely due to missing low-$z$ galaxies. At the current stage we do not have any indication for such a bias but cannot exclude this possibility without more high-quality spec-$z$.

It remains to be seen if the tension with Planck measurements reported here from KiDS and VIKING and the hints for possible biases in the DESy1 and HSC-DR1 redshift calibration will stand the test of time or if some yet unknown systematic errors are responsible for these puzzling results and will be corrected in the future. Analysing the public DESy1 and HSC-DR1 data with the KV450 pipeline and -- most importantly -- calibrating their redshift distributions with the techniques and calibration samples described here and vice versa would shed some further light on this and either resolve the tension or deepen the mystery.

\section*{Acknowledgments}

We thank Matthias Bartelmann for being our external blinder, revealing which of the three redshift distributions analysed was the true unblinded redshift distribution at the end of this study. We would like to thank George Efstathiou for valuable comments on the first version of the manuscript, some of which were incorporated in the revised version.

We thank Julien Lesgourges, Thejs Brinkmann, Thomas Tram, and Benjamin Audren for developing the \textsc{CLASS} and \textsc{MontePython} codes and Anthony Lewis for developing \textsc{getdist}. We are grateful to the HSC team for providing their likelihood chain to us and to the zCOSMOS team to give us early access to additional deep spec-$z$ that were not availale in the public domain.

We are indebted to the staff at ESO-Garching and ESO-Paranal for managing the observations at VST and VISTA that yielded the data presented here. Based on observations made with ESO Telescopes at the La Silla Paranal Observatory under programme IDs 177.A-3016, 177.A-3017, 177.A-3018, 179.A-2004, 298.A-5015, and on data products produced by the KiDS consortium. 

We acknowledge support from European Research Council grants 647112 (CH, MA, BG, AA), 693024 (SJ), 670193 (JP), and 770935 (HHi, JLvdB, AW), the Deutsche Forschungsgemeinschaft (HHi, CM, MT, grants Hi 1495/2-1 and Hi 1495/5-1 as well as the TR33 `The Dark Universe' program), the Alexander von Humboldt Foundation (KK), the STFC (LM, grant ST/N000919/1), NWO (KK, JdJ, MB, HHo, research grants 621.016.402, 614.001.451 and 639.043.512), the World Premier International Research Center Initiative, MEXT, Japan (FK), NASA (AC, grant 15-WFIRST15-0008), the EU's Horizon 2020 programme (TT, AM, Marie Curie grants 797794, 702971), NSF/AURA (CM, grant 1258333), DoE/SLAC (CM, grant DE-AC02-76SF00515), the DIRAC Institute (CM), the Beecroft Trust (SJ), and the DiRAC Data Intensive service at Leicester (STFC grants ST/K000373/1, ST/R002363/1, ST/R001014/1; SJ). CH acknowledges support from the Max Planck Society and the Alexander von Humboldt Foundation in the framework of the Max Planck-Humboldt Research Award endowed by the Federal Ministry of Education and Research.

We are very grateful to the Lorentz Centre and ESO-Garching for hosting several team meetings. This work was performed in part at Aspen Center for Physics, which is supported by NSF grant PHY-1607611.

{\small \textit{Author Contributions:} All authors contributed to the development and writing of this paper. The authorship list is given in three groups: the lead authors (HHi, FK, JLvdB, BJ, CH, AK, AW), followed by two alphabetical groups. The first alphabetical group includes those who are key contributors to both the scientific analysis and the data products. The second group covers those who have either made a significant contribution to the data products or to the scientific analysis.}

\bibliographystyle{aa}

\bibliography{KiDS-VIKING-450_cosmic_shear_paper_rev3}

\appendix

\section{Changes with respect to KiDS-450}
\label{app:changes}

For the expert readers who are familiar with our previous analysis in \citetalias{hildebrandt/etal:2017}, we provide a concise bullet-point summary of the updates included in this KV450 analysis, as detailed in Sects.~\ref{sec:data}--\ref{sec:cov}.

\subsection{Data}
\begin{itemize}
\item Addition of five VIKING NIR bands to the four KiDS optical bands over $\sim 450$~deg$^2$.
\item \textsc{GAaP} photometry on NIR data at the level of individual VISTA chips.
\item \textsc{BPZ} photo-$z$ based on a newer version of the code (v1.99.3) and an improved prior.
\item Additional spectroscopic data to improve the redshift calibration.
\end{itemize}

\subsection{Tomographic bins \& redshift calibration}
\begin{figure}
  \includegraphics[width=\hsize]{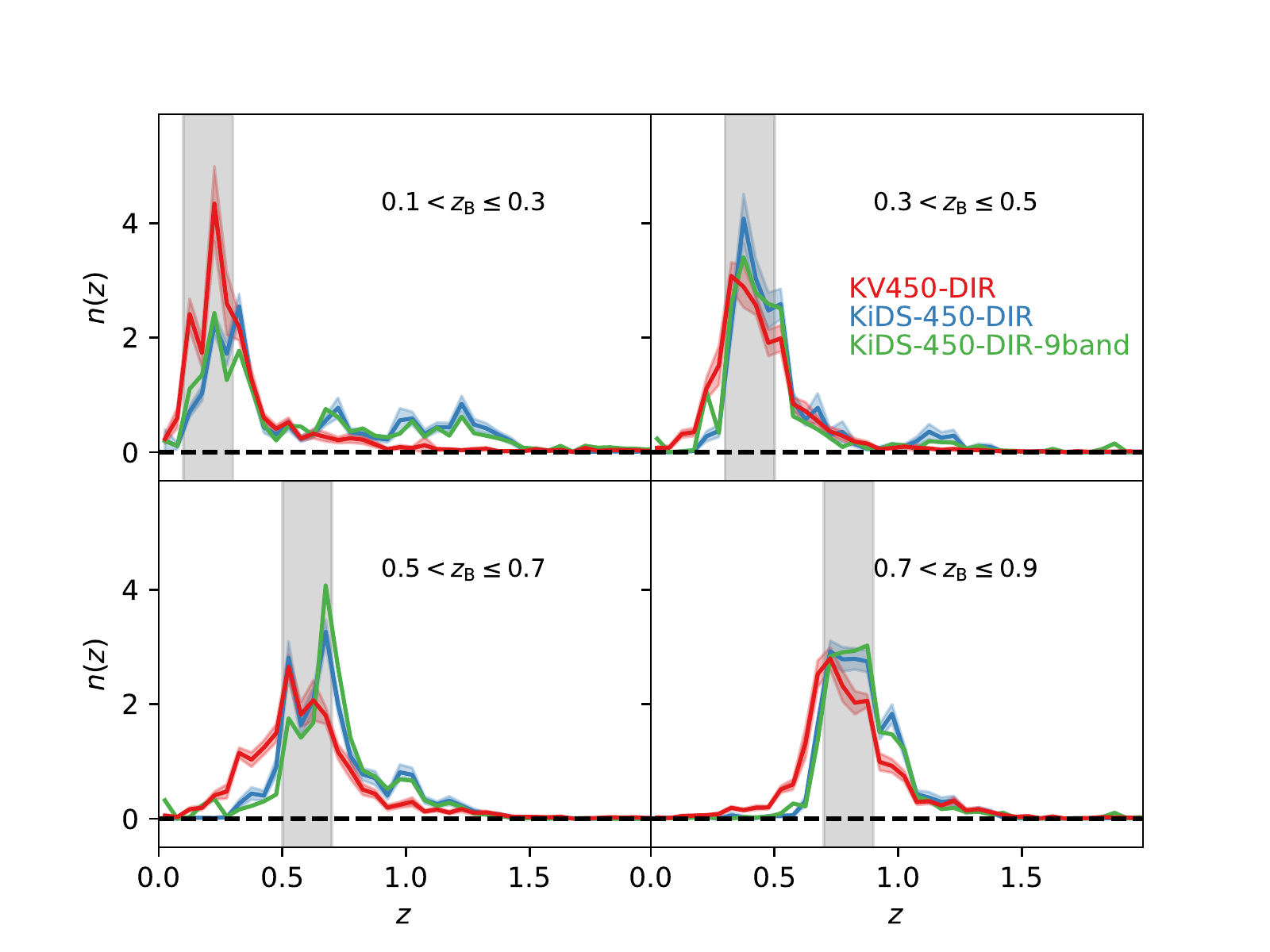}
  \caption{\label{fig:KVvsK}Same as Fig.~\ref{fig:zdist} but showing the $n(z)$ in the first four tomographic bins for KV450 (this study, red), KiDS-450 (\citetalias{hildebrandt/etal:2017}, blue), and KiDS-450 calibrated with 9-band photometry (green). This illustrates the better high-$z$ behaviour of the optical+infrared photo-$z$. Note that the galaxies in the bins are different in the two studies.}
\end{figure}

\begin{itemize}
\item Tomographic binning by new 9-band photo-$z$ resulting in smaller high-$z$ tails (see Fig.~\ref{fig:KVvsK}).
\item New fifth tomographic bin with $0.9<z_\mathrm{B}\le1.2$.
\item DIR calibration: 
  \begin{itemize}
  \item Inclusion of the additional calibration fields VVDS-2h and GAMA-G15Deep ($\sim 6500$ additional galaxies with spec-$z$).
  \item Constant volume approach instead of more unstable constant number of neighbours when performing $k$th nearest neighbour matching in magnitude space.
  \item Photo-$z$ ($z_{\rm B}$) filtering on the photometric catalogue before re-weighting instead of filtering on the weighted spec-$z$ catalogue after re-weighting.
  \item Estimate of sample variance (including selection effects) by \emph{spatial} bootstrap resampling of the calibration sample.
  \item Introduction of five nuisance parameters $\delta z_i$ to account for the uncertainties in the mean redshifts of the tomographic bins.
  \item Quasi-jackknife approach rejecting the different spec-$z$ calibration surveys one at a time to estimate the extremes of the sample variance and selection effects.
  \end{itemize}
\item Blinding at the level of the redshift distributions instead of ellipticities.
\end{itemize}

\subsection{Shape measurements}
\begin{itemize}
\item Improved estimates of the multiplicative bias for all five tomographic bins thanks to new image simulations that allow us to emulate VST observations of the COSMOS field. The improvements include
  \begin{itemize}
  \item The input catalogue is based on structural parameters measured from HST images; this includes realistic clustering and blending.
  \item Realistic correlations between observables (size, ellipticity, SNR, photo-$z$).
  \item Photometric redshifts are assigned to each simulated galaxy, which enables a consistent split into tomographic bins.
  \item Accurate re-calibrated weights for optimal SNR measurements.
  \item Multiplicative shear measurement bias of $m=-0.017,-0.008,-0.015,+0.010,+0.006$ with an estimated uncertainty of $\sigma_m=0.02$ (conservative estimate that is twice as large as for KiDS-450).
  \end{itemize}
\item Propagation of the uncertainty in the $c$-correction into the model via a nuisance parameter $\delta c$.
\item Modelling of a pointing-wide 2D pattern in the galaxy ellipticities discovered in the stellar ellipticities via a nuisance parameter $A_c$ to account for the uncertainty of this correction.
\end{itemize}

\subsection{Correlation functions and covariance matrix }
\begin{itemize}
\item We switch to using \textsc{treecorr} \citep{jarvis/etal:2004} instead of \textsc{athena}\footnote{\url{http://www.cosmostat.org/software/athena/}} to estimate correlation functions.
\item By including a fifth tomographic bin the size of the data vector increases from 130 to 195 elements.
\item In the analytical calculation of the covariance we account for the slightly smaller effective area of KV450 compared to KiDS-450 by using the actual KV450 footprint when calculating the coupling of in-survey and super-survey modes.
\item The uncertainty in the multiplicative shear measurement bias ($\sigma_m=0.02$) is propagated into the covariance by using a theoretical data vector instead of a noisy measurement.
\item The shape noise estimate in the Gaussian part of the covariance is based on the actual measured number of pairs instead of a na\"ive area scaling.
\item The covariance is estimated at the linear mid-point of the $\theta$ bins instead of the logarithmic mid-point. 
\item More comprehensive E-/B-mode decomposition with COSEBIs revealing no significant B-modes in the KV450 tomographic analysis.
\end{itemize}

\subsection{Theoretical modelling}
\begin{itemize}
\item We switch to a new cosmology pipeline based on \textsc{CLASS}, \textsc{HMCode}, and \textsc{MontePython}.
\item We include one massive neutrino with the minimal required mass of $m_\nu=0.06~{\rm eV}$.
\item We follow the arguments by \citet{asgari/etal:2018} and integrate our $\xi_\pm$ models over each broad $\theta$ bin.
\item We use a slightly more informative prior for the baryon feedback amplitude $B$ consistent with the most recent BAHAMAS hydro simulations.
\item As described in Sect.~\ref{sec:c_term} and Sect.~\ref{sec:DIR} we include additional nuisance parameters to marginalise over the uncertainties in the additive shear measurement bias and the mean redshifts.
\end{itemize}

\section{2D projections of cosmological parameter constraints}
\label{app:triangle}
\begin{figure*}
\includegraphics[width=\hsize]{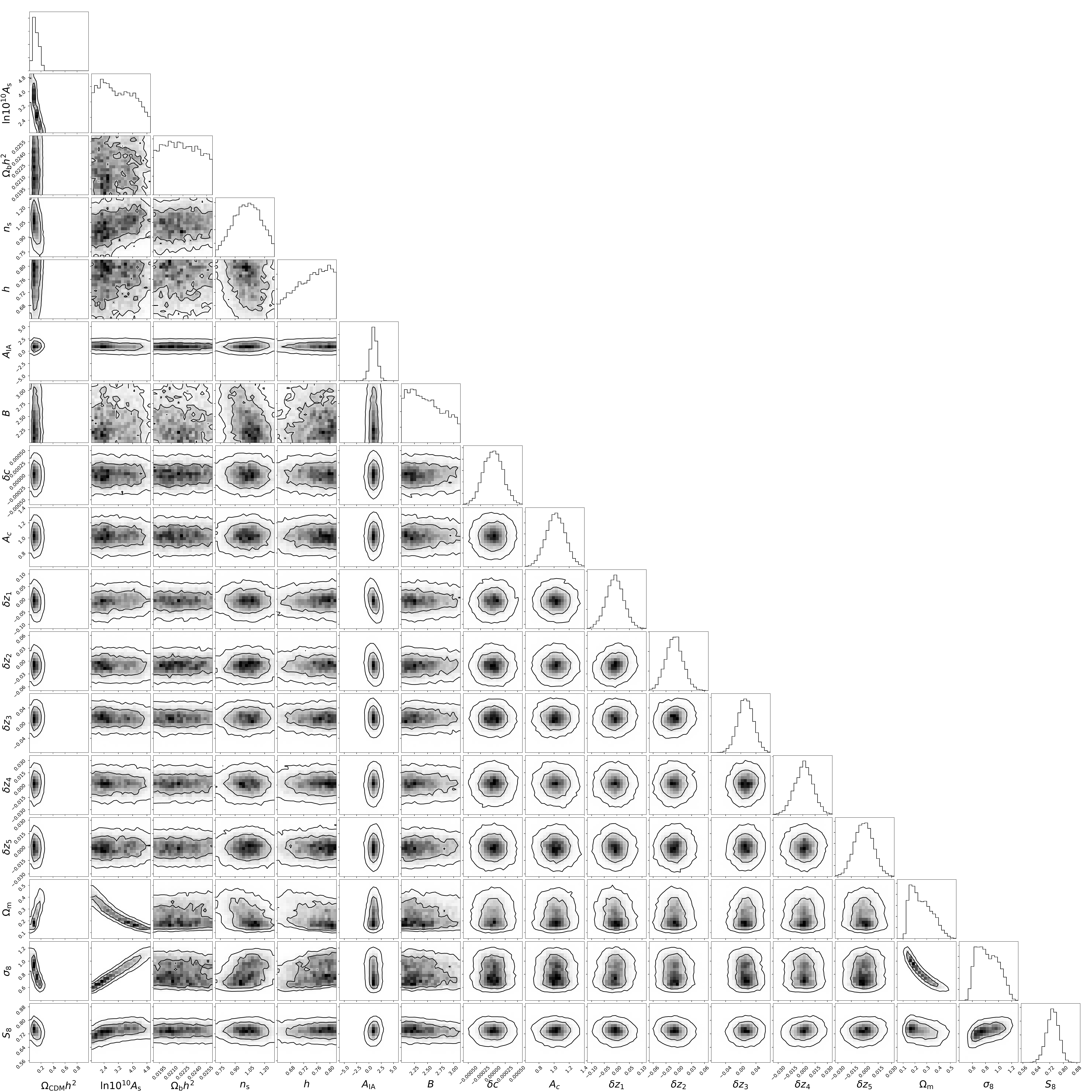}
\caption{\label{fig:triangle}2D projections of the 68\% and 95\% (inner and outer contours) credibility intervals for all parameters of the KV450-fiducial run. Note that the plotting ranges correspond to the prior ranges (except for the three derived parameters $\Omega_{\rm m}$, $\sigma_8$, and $S_8$).}
\end{figure*}

In Fig.~\ref{fig:triangle} we show 2D projections of the confidence regions of all primary and derived parameters used in our model.

\section{Photo-$z$ tests}
\label{app:photoz_tests}

\begin{figure*}
\includegraphics[width=\textwidth,clip=true,trim=0.5cm 0.cm 0.5cm 4cm]{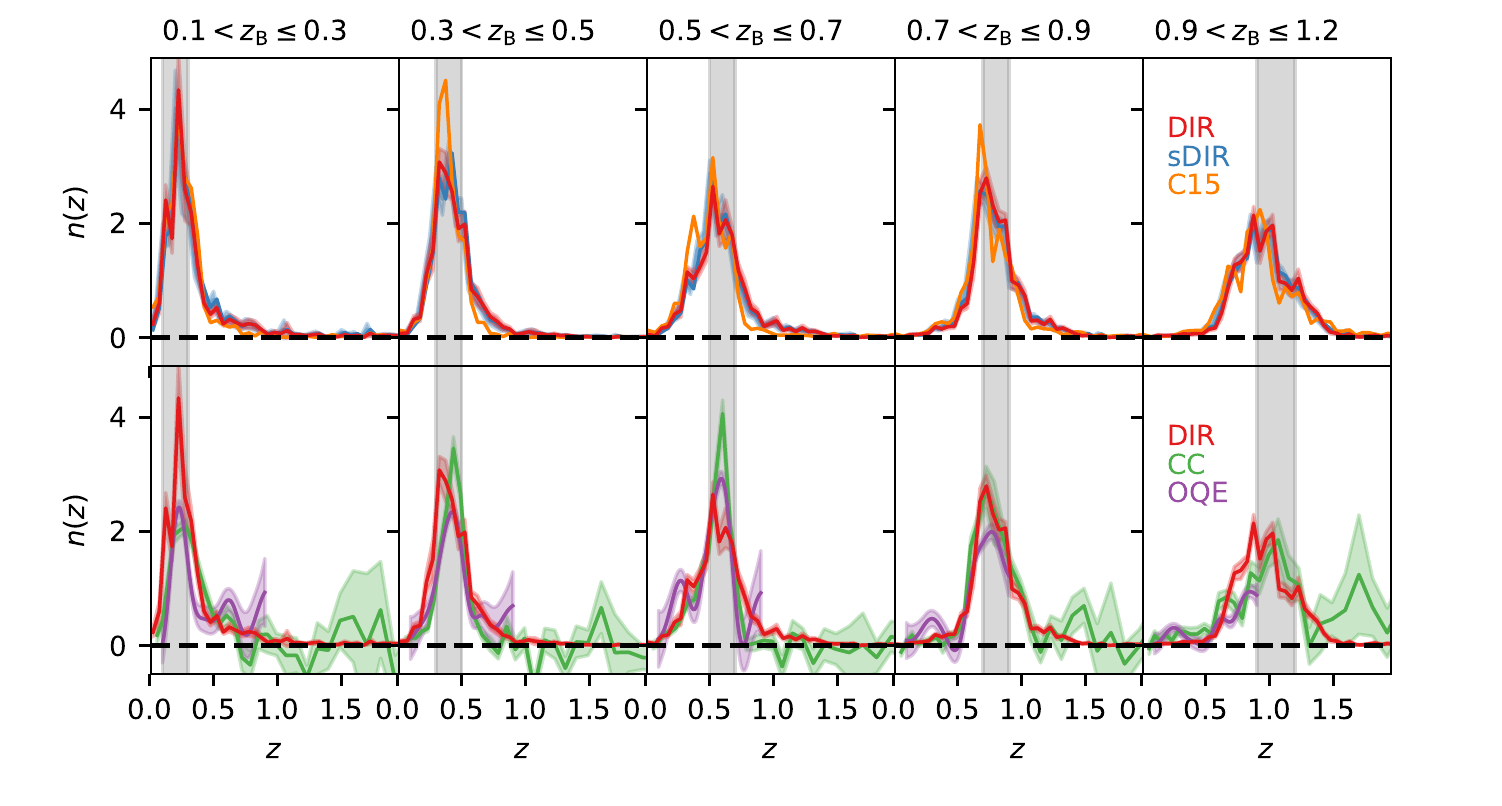}
\caption{\label{fig:zdist_OQE} Similar to Fig.~\ref{fig:zdist} showing the fiducial redshift distributions and also some alternative $n(z)$ estimates. The red lines and their confidence regions correspond to the weighted direct calibration technique (DIR, Sect.~\ref{sec:DIR}), blue corresponds to a smoothed version of the DIR method (sDIR, Appendix~\ref{sec:sDIR}), and green shows the small-scale clustering-$z$ measurements (CC, Appendix~\ref{sec:CC}) after correction for the spectroscopic bias but before fitting with a Gaussian mixture model and correction for the photometric bias (the latter being negligible). The clustering-$z$ $n(z)$ as estimated with the optimal-quadratic-estimator (OQE, Appendix~\ref{app:OQE}) out to $z<0.9$ are shown in purple. Note that the normalisation of the green CC estimate is somewhat ambiguous due to noise and the resulting negative amplitudes. The purple OQE estimates have been normalised to the same area as the CC estimates for the redshift range $z<0.9$. We also include the DIR $n(z)$ that result when the combined spec-$z$ calibration sample is replaced by the COSMOS-2015 photo-$z$ catalogue (orange; shown without uncertainties).}
\end{figure*}

In this appendix we investigate three alternatives to the fiducial redshift distributions described in Sect.~\ref{sec:DIR}. These alternative $n(z)$ are shown in Fig.~\ref{fig:zdist_OQE} alongside the fiducial DIR $n(z)$. There is a great level of redundancy in the CC (Appendix~\ref{sec:CC}) and OQE (Appendix~\ref{app:OQE}) approaches when compared to DIR (Sect.~\ref{sec:DIR}) and sDIR (Appendix~\ref{sec:sDIR}). These clustering-$z$ methods make very different assumptions and rely on different data. The very good agreement of these different methods as reported in Fig.~\ref{fig:S8} and Table~\ref{tab:S8} is hence a strong argument for the robustness of the fiducial results presented in this paper. 

\subsection{Smoothing the DIR redshift distribution (sDIR)}
\label{sec:sDIR}
\begin{figure}
\includegraphics[width=\hsize]{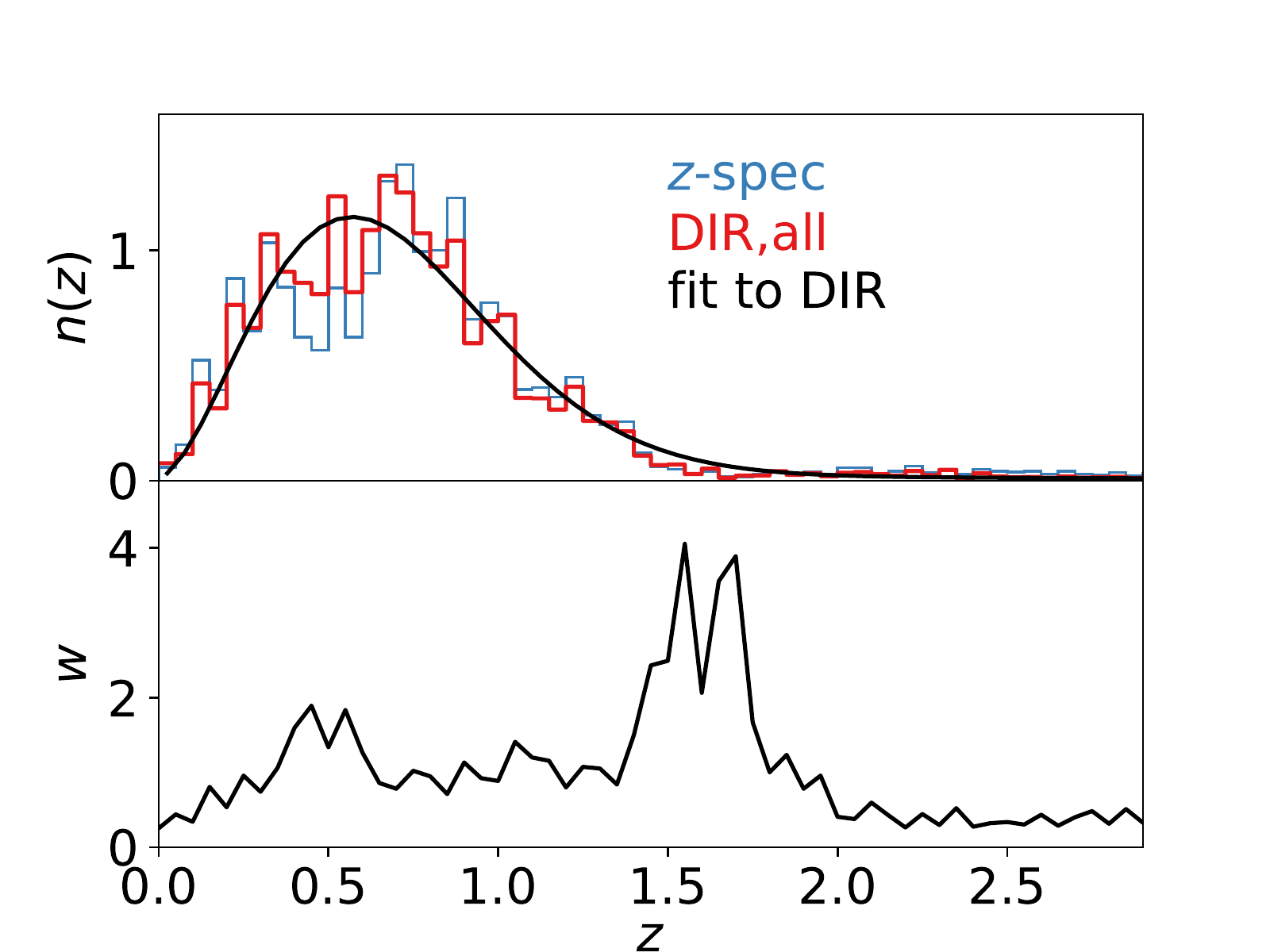}
\caption{\label{fig:sDIR} Illustration of the sDIR method. The blue line represents the unweighted spectroscopic redshift distribution of the calibration sample. The red line is the DIR estimate of the redshift distribution of the full lensing catalogue, $n_\mathrm{DIR,all}(z)$. The black line in the upper panel shows a parametric fit to the red line and the lower panel shows the ratio of this fitted function to the blue line, which is a first guess of the smoothing weight.}
\end{figure}

\begin{figure}
\includegraphics[width=\hsize]{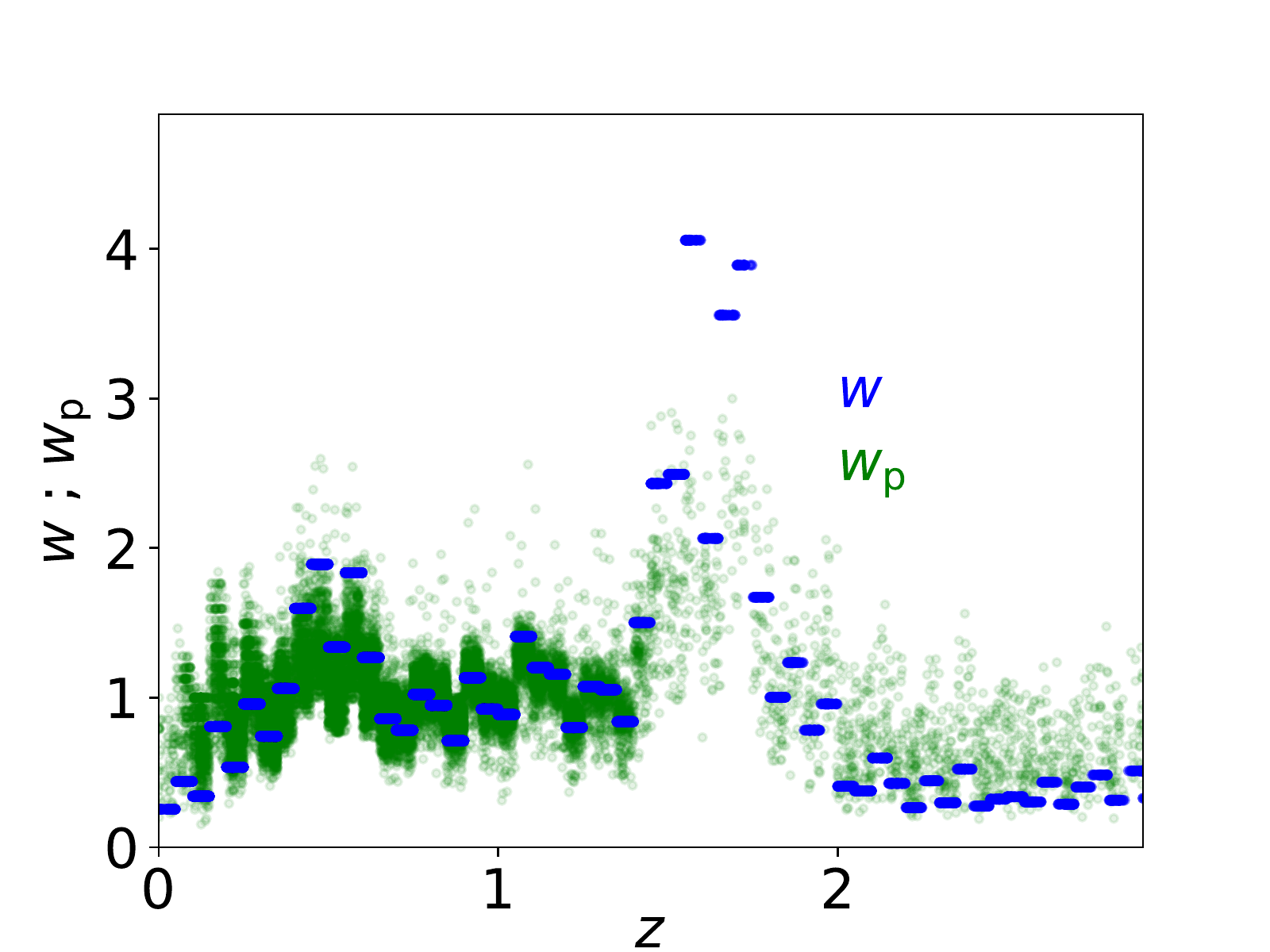}
\caption{\label{fig:sDIR2} Refinement of the smoothing weight for each calibration object. The blue data points represent the initial guess of the smoothing weights, $w_i(z)$, that just depend on redshift (equivalent to the lower panel of Fig.~\ref{fig:sDIR}) while the green data points represent $w_{\mathrm{p},i}$, which is the ratio of $w_i$ and the average $\langle w\rangle_i$ of the $w_j$ of the $k$ nearest neighbours around an object $i$.}
\end{figure}

In the following we describe how we try to suppress the potentially spurious structures in the $n(z)$ estimates, originating from large-scale structure and selection effects in the spectroscopic calibration sample, by extending the DIR methodology and introducing a smoothing scheme. The basic assumption behind this method, which we call sDIR (for smoothed DIR) in the following, is that the true redshift distribution of the tomographic photo-$z$ bins should be smooth. The peaks that can be seen in the red lines in Fig.~\ref{fig:zdist} are structures that are also visible in the unweighted spectroscopic redshift distribution of the calibration sample. The DIR weighting scheme cannot fully remove these structures due to finite sampling of magnitude space and photometric noise. These structures represent either large-scale structure in the spec-$z$ calibration fields or selection effects of the spectroscopic surveys. Such features can be suppressed in an almost unbiased way by applying the following recipe, which is further illustrated in Figs.~\ref{fig:sDIR}~\&~\ref{fig:sDIR2}.

\begin{enumerate}
\item Run DIR for the full lensing source catalogue, i.e. without any photo-$z$ cuts. This yields $n_\mathrm{DIR, all}(z)$ (red line in top panel of Fig.~\ref{fig:sDIR}).
\item Fit a smooth function to $n_\mathrm{DIR, all}(z)$. This yields $n_\mathrm{smooth,all}(z)$ (black line in top panel of Fig.~\ref{fig:sDIR}).
\item Define a weight function as \\$w(z)=n_\mathrm{smooth,all}(z)/n_\mathrm{DIR, all}(z)$ (solid line in bottom panel of Fig.~\ref{fig:sDIR} and blue data points in Fig.~\ref{fig:sDIR2}).
\item For the $i$th spec-$z$ calibration object (with redshift $z_{\mathrm{spec},i}$), find its $k$ nearest spec-$z$ neighbours in magnitude space and calculate the mean $w(z)$ of those $k$ nearest neighbours, $\langle w\rangle_i$ .
\item Calculate $w_{\mathrm{p},i} = w(z_{\mathrm{spec},i}) / \langle w\rangle_i$ , which will scatter around 1 (green data points in Fig.~\ref{fig:sDIR2}).
\item Run an updated DIR for each tomographic bin where the original DIR weights for each $i$th spec-$z$ calibration object are multiplied with $w_{\mathrm{p},i}$ (blue lines in Fig.~\ref{fig:zdist_OQE}).
\end{enumerate}

The second step is crucial here, where a smooth, parametric function has to be chosen. We test the sensitivity of the sDIR technique on simulations and found that it is surprisingly stable against plausible choices of this smoothing function. We try fitting multiple Gaussians to the $n_\mathrm{DIR, all}(z)$ as well as theoretically motivated redshift distributions as used in the literature \citep{brainderd/etal:1996,benitez:2000,schrabback/etal:2010}. The latter have the benefit of a more realistic low-redshift behaviour that accurately reflects the redshift dependence of the cosmological volume element. 

In our analysis we choose to use the sum of the redshift distribution suggested by \citet{benitez:2000} and a single Gaussian with a mean redshift forced at $z>1$. This choice yields a visually good fit to $n_\mathrm{DIR, all}(z)$ accounting for the low- as well as the high-redshift shape of the distribution. We note that this step is somewhat arbitrary but the resulting sDIR $n(z)$ for the tomographic bins is extremely stable against this choice. Mean/median redshifts of the bins scatter by only $\sim0.01$ for plausible choices of this function, which is similar to the $1\sigma$ error of the mean redshifts reported in Table~\ref{tab:tomobins}. The reason for this stable behaviour is that -- unlike in older works in the weak lensing literature -- this parametric function is not used directly as the redshift distribution but only indirectly as a correction to the DIR method. Thus, the bulk of the information for the $n(z)$ comes from the galaxy colours and only a mild smoothing of the strong features (not an overall smoothing that would lead to a broadening of the $n(z)$) is introduced by this parametric fit.

\subsection{Calibration with cross-correlation (CC)}
\label{sec:CC}
As a powerful alternative to the weighted, direct calibration we use angular cross-correlation measurements between spectroscopic calibration samples and the KV450 galaxies to infer the redshift distributions in the tomographic bins. We call this technique CC in the following. This method, originally proposed by \citet{schneider/etal:2006} and \citet{newman:2008} and refined in many follow-up papers \citep{matthews/etal:2010,menard/etal:2013,schmidt/etal:2013,mcquinn/etal:2013,morrison/etal:2017}, makes very different assumptions to the DIR method. Most importantly it does not require the calibration sample to cover the same region in multi-dimensional magnitude space as the lensing catalogue because it uses positional information instead of magnitude/colour information. It is sufficient to cover the full redshift range over which lensing sources are expected, as it is assumed that all galaxies at the same redshift cluster with each other.

We extend the CC analysis in \citetalias{hildebrandt/etal:2017} by using a number of additional wide-area spectroscopic surveys for the KV450 CC implementation that are not used for DIR:
\begin{itemize}

\item GAMA \citep[Galaxy and Mass Assembly,][]{driver/etal:2011}: KV450 overlaps with 91~deg$^2$ (after conservative masking) of GAMA. This dense and highly complete (down to a limiting magnitude of $r<19.8$) spectroscopic survey is ideally suited for CC calibration out to $z\sim0.4$ \citep[see such an application in][]{morrison/etal:2017}.

\item SDSS \citep[Sloan Digital Sky Survey,][]{alam/etal:2015}: The equatorial KV450 fields almost fully overlap with SDSS spectroscopy (198~deg$^2$ after masking), in particular the BOSS \citep[Baryon Oscillation Spectroscopic Survey,][]{dawson/etal:2013}, but we also make use of the SDSS Main Galaxy Sample \citep{strauss/etal:2002} and the QSO sample \citep{schneider/etal:2010}. This data set can be used in a similar way as GAMA for CC, but out to a higher redshift of $z\sim0.7$. There is some information from the QSO sample at even higher redshifts which we also exploit in CC.

\item 2dFLenS \citep[2-degree Field Lensing Survey,][]{blake/etal:2016}: Designed to closely resemble BOSS but situated in the Southern hemisphere, 2dFLenS adds more area (91~deg$^2$ after masking) to the CC calibration out to $z\sim0.8$ and hence helps to reduce shot noise and sample variance out to that redshift.

\item WiggleZ Dark Energy Survey \citep{drinkwater/etal:2010}: This survey overlaps KV450 by $\sim87$~deg$^2$ and contains emission-line galaxies that allow us to calibrate the photo-$z$ with the CC technique with improved signal-to-noise ratio out to even higher redshifts of $z\la 1.1$.

\end{itemize}
Note that none of these wide-area spec-$z$ surveys were used in the KiDS-450 analysis \citepalias{hildebrandt/etal:2017} but were only integrated in our data flow later \citep{morrison/etal:2017}. Hence the CC calibration described here yields considerably more precise results now in comparison to the previous CC analysis presented in \citetalias{hildebrandt/etal:2017}.

The different assumptions as well as the different calibration data make the CC calibration highly complementary to the DIR calibration so that potential systematic errors in the redshift distributions should in principle be reliably identified. Here we apply a method for estimating the CC distribution based on small-scale measurements that optimises the signal-to-noise ratio of the redshift recovery \citep{schmidt/etal:2013}. A comparison with a different method based on large-scale measurements with an optimal quadratic estimator (OQE) as proposed by \citet{mcquinn/etal:2013} and used on KiDS-450 by \citet{johnson/etal:2017} is presented in Appendix~\ref{app:OQE}.

We use the public clustering redshift code \textsc{the-wizz}\footnote{https://github.com/morriscb/the-wizz} \citep{morrison/etal:2017} to estimate the angular cross-correlation function $w_\mathrm{s,p_i}(z)$ of the finely-binned spec-$z$ calibration sample at redshift $z$ and the $i$th tomographic bin of the photometric lensing catalogue. We measure the cross-correlation in a single bin of comoving separation $100~{\rm kpc}<r<1~{\rm Mpc}$. The spec-$z$ sample is split up into areas of homogeneous coverage, i.e. an area that is covered by 2dFLenS or SDSS only, an area that is covered by GAMA and SDSS, an area that is covered by GAMA, SDSS, and WiggleZ, etc. Furthermore, each of these subsamples is split into spatial bootstrap regions defined by the boundaries of the KiDS pointings. Then, \textsc{the-wizz} is run separately for all these 397 regions (due to segmentation by the GAMA/SDSS/WiggleZ geometry) to mitigate the effects of observational density variations.
The average redshift distribution (before bias correction; see below) in each $i$th tomographic bin, $\hat n_i(z)$, and their errors are then estimated from 1000 bootstrap realisations:
\begin{equation}
\hat n_i(z) = A\,w_\mathrm{s,p_i}(z)\,,
\end{equation}
where $A$ is a constant that normalises $\hat n_i(z)$.

In order to solidify the estimate at high redshift we also run CC on the small, deep spec-$z$ fields of COSMOS, DEEP2, and VVDS using a few bootstrap regions of $\sim0.1~{\rm deg}^2$ (17 in total). This independent estimate of the redshift distribution from the deep fields is then combined with the estimate from the wide fields, described above, by integrating them in the bootstrap resampling (i.e. drawing randomly from a total of $397+17=414$ regions). In this way the wide fields contribute the bulk of the information at low redshift whereas the deep fields constrain the high-$z$ tails.

The redshift distributions $\hat n_i(z)$ constructed directly from the cross-correlation amplitudes still suffer from degeneracies with the unknown galaxy bias of the spectroscopic and photometric samples, $b_{\rm s}(z)$ and $b_{\rm p}(z)$,\footnote{Note that the functions $b_{\rm s}(z)$ and $b_{\rm p}(z)$ should not be confused with the linear bias parameters of these galaxy samples. These functions implicitly include the non-linear structure growth with redshift.} respectively, as well as redshift-dependent selection effects $\beta(z)$:
\begin{equation}
\hat n_i(z) = n_i(z)\, b_{\rm s}(z)\, b_{\rm p}(z)\, \beta(z) \equiv n_i(z) \, b_{\rm s}(z)\, \mathcal{B}(z)\, . 
\end{equation}
We summarise the redshift-dependent selection effects and the galaxy bias of the photometric sample in the function $\mathcal{B}(z)$. 

\begin{figure*}
\includegraphics[width=\textwidth]{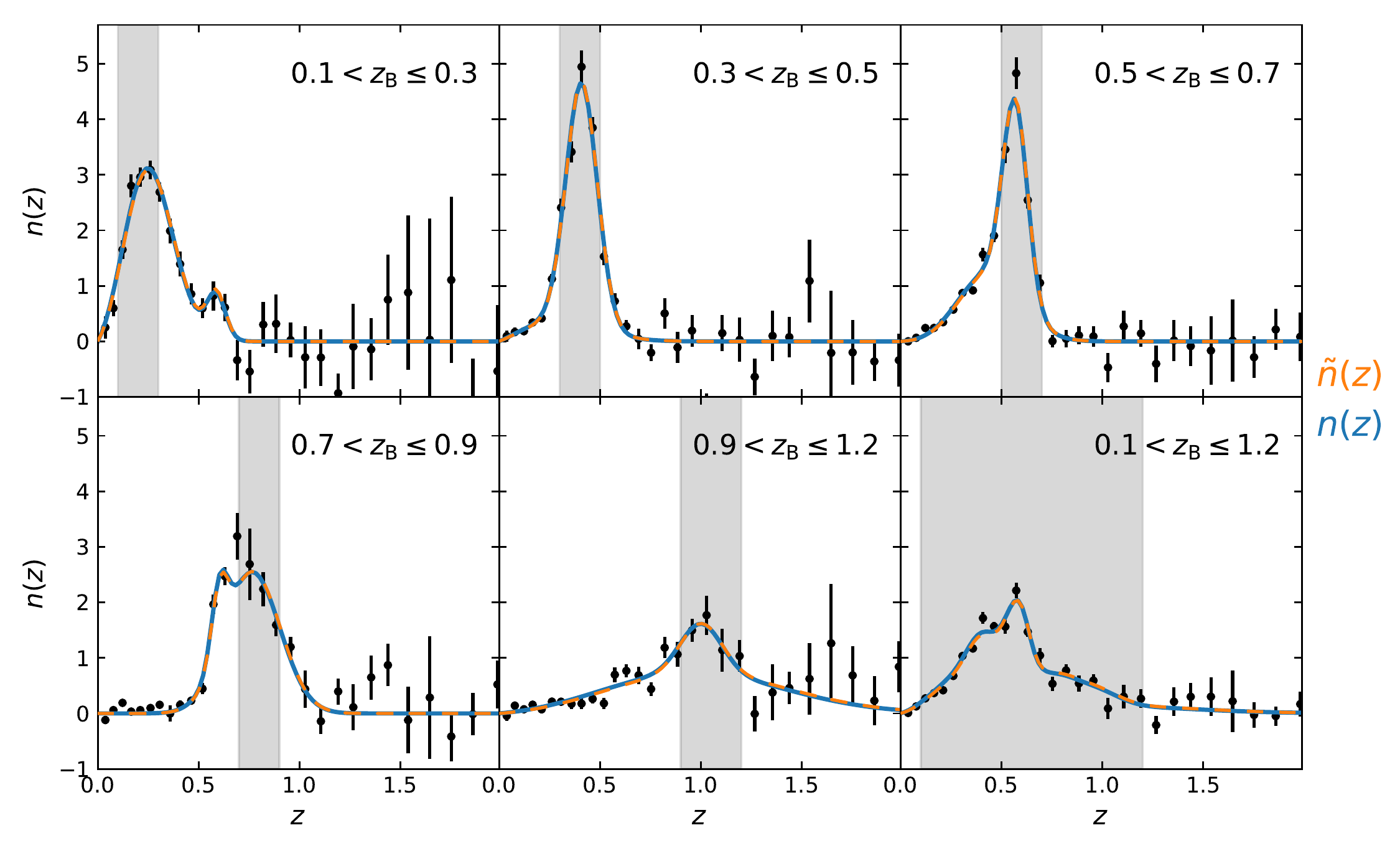}
\caption{\label{fig:CC_bestfit} Clustering-redshift (CC) measurements for the five tomographic bins and a broad bin with $0.1<z_{\rm B}\le 1.2$. The data points represent $\tilde n(z)$, i.e. the redshift distribution after correction for the spectroscopic bias $b_{\rm s}(z)$, and the orange lines show the best-fit Gaussian-mixture-model (GMM) to these data. The blue lines show the same GMM but after correcting for the bias function $\mathcal{B}(z)=(1+z)^\alpha$ with the best-fit $\alpha=0.232\pm0.441$, showing a gentle skewing to lower redshift.}
\end{figure*}

First we estimate the functional form of the spectroscopic bias $b_{\rm s}(z)$ by measuring the projected auto-correlation function of the spectroscopic sample as a function of redshift, again in a single bin of projected comoving separation \citep[this approach is similar to the one described by][]{rahman/etal:2015,scottez/etal:2017}. We use the same scales ($100~{\rm kpc}<r<1~{\rm Mpc}$) and the same redshift binning as for the cross-correlation measurements. Note that we are not interested in the absolute value of the galaxy bias but just in its relative evolution with redshift, as the $n(z)$ are normalised after bias correction anyway. Hence choosing the same scales for the auto- and cross-correlations ensures that no inconsistency is introduced. We correct the $\hat n$ with the spectroscopic bias estimate to yield
\begin{equation}
\tilde n_i(z) \equiv \frac{\hat n_i(z)}{b_{\rm s}(z)}=n_i(z)\,\mathcal{B}(z)\,.
\end{equation}
This mitigation of the spectroscopic bias evolution is fully integrated in the spatial bootstrap resampling and implemented for each individual spectroscopic reference survey and is propagated to the combined redshift distributions $\tilde n_i(z)$ of the tomographic bins. The resulting $\tilde n_i(z)$ data points are shown in Fig.~\ref{fig:CC_bestfit}.

The photometric bias and selection effects are corrected with a different method that is described in \citet{schmidt/etal:2013} and \citet{morrison/etal:2017}. If $\mathcal{B}(z)$ is not constant with redshift this will skew the inferred redshift distributions. In order to correct for these redshift-dependent biases we exploit the fact that the redshift evolution of such biases is less pronounced for narrower redshift bins. If one can pre-select narrow redshift bins (e.g. via photo-$z$) the $\tilde n_i(z)$ are only affected by the relative bias evolution over the redshift-width of the bin. The overall normalisation is fixed by the fact that one estimates a probability density so that the absolute value of the bias function $\mathcal{B}$ becomes unimportant.

This behaviour can be exploited by comparing the clustering redshift results of a broad bin  $\tilde n_\mathrm{all}(z)$  that is strongly affected by evolving bias and/or selection effects with the combined results of narrow bins that make up the broad bin and are much less strongly affected, $\sum_i w_i \tilde n_i(z)$. The relative normalisation of the biases between the narrow bins is given by the weighted counts of galaxies in these bins, $w_i$. Here we choose the five tomographic bins defined above (see Table~\ref{tab:tomobins}) as the narrow bins and one broad bin with $0.1<z_\mathrm{B}\le1.2$.

This means that in practice we use five relatively independent data points to constrain the function $\mathcal{B}(z)$ over the redshift range of interest. We choose a smooth function $\mathcal{B}_\alpha(z)=(1+z)^\alpha$ and fit the parameter $\alpha$ to effectively minimise the following quantity
\begin{equation}
\label{eq:CC_resid}
\Delta = \sum_{j=1}^{j_\mathrm{max}} \left[ \mathrm{norm}\left(\frac{\tilde n_\mathrm{all}(z_j)}{\mathcal{B}_\alpha(z_j)}\right) - \sum_{i=1}^5 w_i \,\mathrm{norm}\left(\frac{\tilde n_i(z_j)}{\mathcal{B}_\alpha(z_j)}\right) \right]^2 \,,
\end{equation}
where $\mathrm{norm}\left(n(z)\right) = \frac{n(z)}{\int_0^\infty \mathrm{d}z\,n(z)}$ expresses the necessity to re-normalize the bias corrected redshift distributions, $w_i$ is the relative weight of the $i$th tomographic bin (column 4 of Table~\ref{tab:tomobins}), $z_j$ is the mean redshift of the $j$th redshift bin of the cross-correlation measurement, and $j_\mathrm{max}$ is the index of the highest redshift bin that is used for the measurement. 

One problem here is that the $\tilde n(z)$ need to be normalised, which is ambiguous for noisy data points. Another complication is that due to noise (and possibly imperfect masking or other systematic effects) the cross-correlation functions $w_\mathrm{s,p_i}(z)$ can attain negative values at some redshifts.

One method to solve both problems simultaneously involves fitting a Gaussian-mixture model (GMM), i.e. a sum of an arbitrary number of Gaussians with free positive amplitudes, to the noisy $\tilde n(z)$ data points. This model is positive-valued by construction, and we modify the model by multiplying with $z$ to allow for correct low-$z$ behaviour. The fitting is done at the same time as the fit for the bias correction described above. The covariance matrix used in this fit is assumed to be diagonal, with the diagonal elements corresponding to the bootstrap errors. The optimal number of Gaussians is chosen independently for each tomographic bin with the help of the Bayesian Information Criterion \citep[BIC;][]{schwarz1978} resulting in two components for the first, second, and fifth bin, three components for the third, and four components for the fourth bin. Minimising $\Delta$ from Eq.~\ref{eq:CC_resid} we find a value of $\alpha=0.232 \pm 0.441$ for the free parameter of the bias model, which indicates no significant redshift evolution of the bias of the photometric sample. The best-fit models for the five bins and the broad bin are shown in Fig.~\ref{fig:CC_bestfit}.

In general the GMM fit to the cross-correlation data points yields significantly lower mean redshifts than the DIR and sDIR methods as it suppresses the noisy high-redshift data points, which do not contain enough information to justify further Gaussian components at high-$z$ according to the BIC. However, the core of the distribution is constrained very well and closely resembles the results from the DIR and sDIR methods (see Fig.~\ref{fig:zdist_OQE}).

An error analysis with bootstrapping is complicated by the fact that for each bootstrap sample the number of Gaussians changes, which makes the whole procedure unwieldy and slow. Here we do not use these errors and leave a more thorough error estimate to future work. Instead we use the Gaussian priors from the DIR method for the $\delta z_i$ nuisance parameters in the cosmological analysis with the GMM set of $n(z)$ (setup no.~7 called `CC-fit' in Table~\ref{tab:MCMC}).

Once the bias function $\mathcal{B}(z)$ is established by the fit one can also go back a step, correct the noisy $\tilde n(z)$ data points with $\mathcal{B}(z)$, and directly compare this noisy CC $n(z)$ estimate to the DIR result. We do this by fitting a linear shift in redshift that yields the best agreement between the two estimates. This method is very similar to what was done in the DESy1 analysis presented in \citet{davis/etal:2017,davis/etal:2018}. We find shifts of $\Delta z=0.043, 0.049, 0.000, -0.008, -0.005$ for the five tomographic bins, meaning that the red DIR lines in Fig.~\ref{fig:zdist_OQE} are to be shifted up in the first two bins and slightly shifted down in bins 4 and 5. Note that all these shifts are insignificant within the combined errors of DIR and CC. Applying these shifts to the DIR $n(z)$ we run another setup (no.~8 called `CC-shift') to check the influence on the cosmological results. Both setups, `CC-fit' and `CC-shift', yield consistent cosmological results.

For the CC distributions we used a slightly different blinding approach, as there the redshift is used for the calibration method itself (unlike for DIR, which only uses colour information at runtime). Instead of starting with a blinded spec-$z$ catalogue we blinded the resulting $n(z)$ with the same perturbation factors as for DIR. This was done by co-authors J.~L.~van~den~Busch and M.~Tewes, who were hence unblind but also not responsible for the cosmological analysis. During the whole blinding process we made sure that the people who worked on the implementation of the CC blinding were different from the ones that carried out the cosmological fits.

\subsection{CC with an optimal quadratic estimator}
\label{app:OQE}

In order to test the scale dependence and the overall implementation of the clustering redshifts we also apply the method from \citet{mcquinn/etal:2013} and \citet{johnson/etal:2017} that implements an optimal quadratic estimator (OQE) to measure the clustering-$z$ distributions. The setup is equivalent to the one presented in \citet{johnson/etal:2017} and uses the wide-area spectroscopic reference surveys only (2dFLenS, SDSS, GAMA, WiggleZ). This is necessary to limit the method to the linear-bias regime, which cannot be measured efficiently from the deep, small-area spec-$z$ surveys. However, this means that we have little information at $z>0.9$ and thus we only measure the OQE $n(z)$ out to this redshift. Amplitudes are fitted to the cross-correlation functions between 0.01 and 1 degree.  The smoothing is implemented by a Gaussian process.

We further assume that the photometric bias function $\mathcal{B}(z)$ is the same as for the CC method described above and correct the OQE data points, which are already corrected for the spectroscopic bias $b_{\rm s}(z)$, with $\mathcal{B}(z)=(1+z)^{0.232}$. Results are included in Fig.~\ref{fig:zdist_OQE}. In a next step we again determine a linear shift for each tomographic bin that gives the best fit between the OQE and DIR $n(z)$ finding $\Delta z=-0.020, 0.044, -0.015, -0.012, -0.034$ for the five tomographic bins. The shifted DIR $n(z)$ are then used for setup no.~9 (`OQE-shift') from Table~\ref{tab:MCMC} yielding consistent results with the other CC-based setups (no.~7~\&~8) and the fiducial setup.

\subsection{Shear-ratio test}
\label{app:SRT}
One method to test the accuracy of redshift distributions is the so called `shear-ratio' test \citep[][]{jain/etal:2003,heymans/etal:2012,kitching/etal:2015,schneider:2016}. Assuming the shear measurements are accurate, one can use a lens galaxy sample and compare the galaxy-galaxy lensing signal of two or more different source samples behind these lenses to test the $n(z)$. The $n(z)$ of the sources and lenses predict the shear ratios, i.e. the ratios of the $\gamma_\mathrm{t}(\theta)$ signals, for the different source samples with very weak dependence on cosmology. It is then tested whether the measured shear ratios are compatible with the predictions within errors.

Here we follow the procedure already used in \citetalias{hildebrandt/etal:2017} using spectroscopic lenses from GAMA and SDSS:
\begin{enumerate}
\item The lens samples are divided into thin redshift-subsamples with a width of $\Delta z=0.1$ out to $z\le0.5$ and $z\le0.7$ for GAMA and SDSS, respectively.
\item We measure $\gamma_\mathrm{t}(\theta)$ for all five tomographic bins around all lens subsamples in four logarithmically-spaced $\theta$ bins in the range $2'<\theta<30'$. Hence we measure 25 lens-source combinations for GAMA (5 lens subsamples times 5 tomographic bins) and 35 for SDSS (7 lens subsamples times 5 tomographic bins).
\item We estimate a boost correction \citep[see e.g.][]{hoekstra/etal:2015} from the angular cross-correlation function of lenses and sources to account for $\theta$-dependent changes of the $n(z)$ in the vicinity of the lenses.
\item We subtract the signal around random points with the same footprint on the sky as the lens samples \citep{mandelbaum/etal:2005,singh/etal:2017}.
\item A maximally flexible model with free parameters for the amplitude at each angular scale for each lens subsample is fitted to the data vector. This is done separately for GAMA and SDSS. The relative amplitude for the different source samples is predicted based on the $n(z)$. Hence for GAMA we fit 100 data points with 20 parameters and for SDSS we fit 140 data points with 28 parameters.
\item We calculate the $p$-value for each fit based on its $\chi^2$ and 80 (112) degrees-of-freedom for GAMA (SDSS).
\end{enumerate}

\begin{figure*}
\includegraphics[width=\hsize]{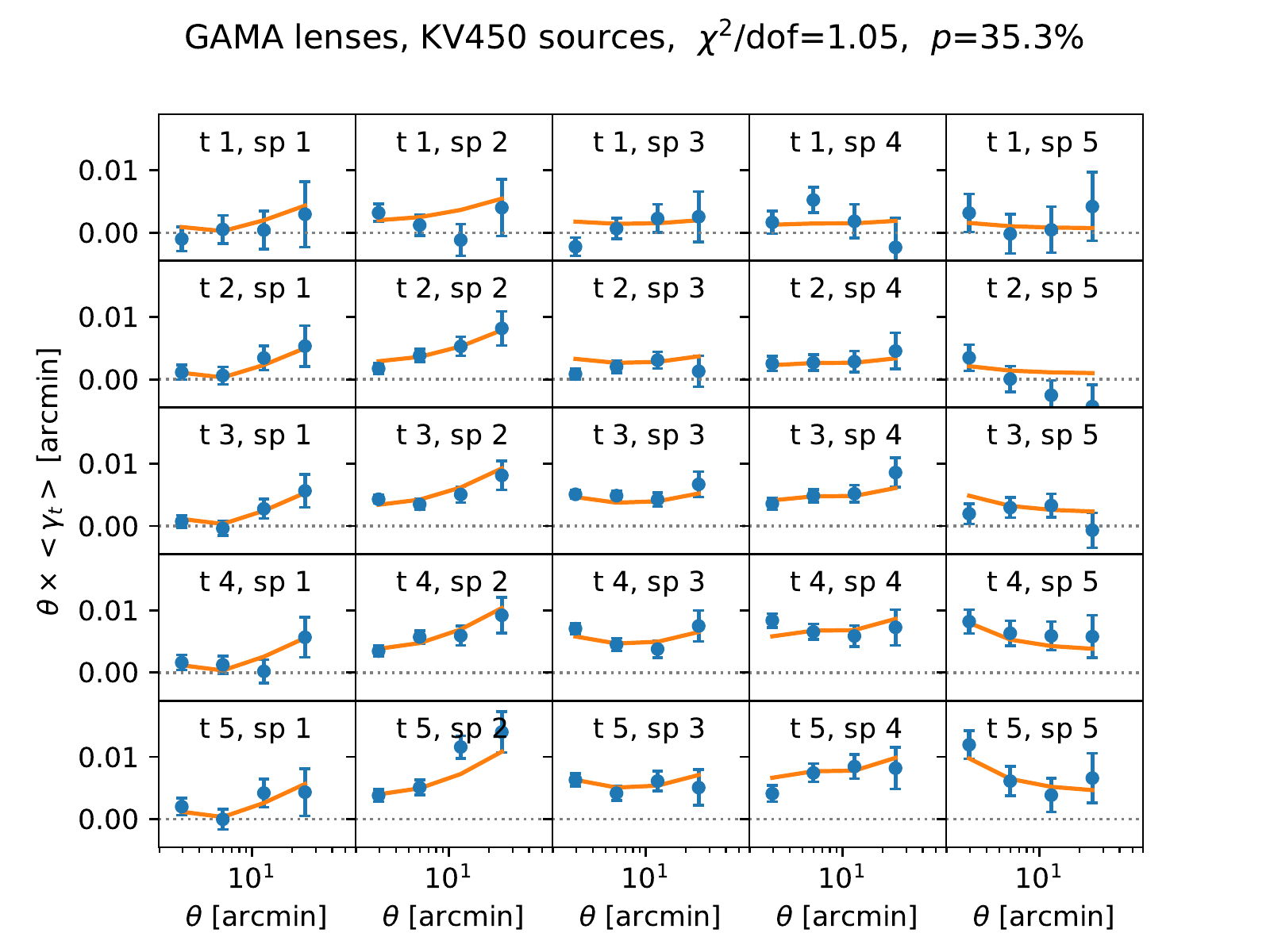}
\caption{\label{fig:SRT} Shear-ratio test showing the galaxy-galaxy-lensing signal of GAMA lenses in five different redshift bins of width $\Delta z=0.1$ starting at redshift $z=0$ (columns labelled sp 1--5) and the KV450 sources in the five tomographic bins (rows labelled t 1--5). Errors are estimated from bootstrapping. The best-fit model (shown by the green line) fitted to the data has 20 free parameters, one free amplitude for each angular scale of each lens sample. The \emph{relative} amplitude for the different tomographic bins is calculated from the DIR redshift distributions of the sources and the spectroscopic redshift distribution of the lenses.}
\end{figure*}

We find acceptable $p$-values of $p>10\%$ for all redshift distributions that we tested. Hence within the precision of this test we cannot distinguish whether some set of $n(z)$ is more accurate than another. Figure~\ref{fig:SRT} shows one set of measurements for the fiducial DIR $n(z)$ and the GAMA lens sample.

\subsection{Statistics for the different redshift calibration setups}
\label{sec:meanz}
In this section we present some statistics for the different redshift calibration setups. In Table~\ref{tab:meanz} we report the mean and median redshifts for the five tomographic bins. It is currently still hard to check the consistency of the different methods on the level of the $n(z)$. The main reason for this is that the error estimates, especially for the clustering-$z$ methods, are not fully reliable. A better quantification of these errors will be presented in a forthcoming publication (van den Busch et al., in prep.).

\begin{table*}
  \caption{\label{tab:meanz}Mean and median redshifts of the five tomographic bins estimated from the different redshift calibration setups.}
  \begin{tabular}{rl|rr|rr|rr|rr|rr}
    \hline
    \hline
    no. & Setup & \multicolumn{2}{c|}{bin 1} & \multicolumn{2}{c|}{bin 2} & \multicolumn{2}{c|}{bin 3} & \multicolumn{2}{c|}{bin 4} & \multicolumn{2}{c}{bin 5} \\
    & & $<z>$ & $\mathrm{med}(z)$ & $<z>$ & $\mathrm{med}(z)$ & $<z>$ & $\mathrm{med}(z)$ & $<z>$ & $\mathrm{med}(z)$ & $<z>$ & $\mathrm{med}(z)$ \\
    \hline
0 &                KV450  & 0.394 & 0.257  & 0.488 & 0.414  & 0.667 & 0.577  & 0.830 & 0.777  & 0.997 & 0.957  \bigstrut\\
1 &                 sDIR  & 0.437 & 0.293  & 0.518 & 0.445  & 0.690 & 0.597  & 0.862 & 0.809  & 1.035 & 0.999  \bigstrut\\
2 &       DIR-w/o-COSMOS  & 0.337 & 0.254  & 0.455 & 0.406  & 0.627 & 0.575  & 0.813 & 0.775  & 0.985 & 0.957  \bigstrut\\
3 & DIR-w/o-COSMOS\&VVDS  & 0.349 & 0.255  & 0.472 & 0.404  & 0.661 & 0.582  & 0.836 & 0.787  & 0.997 & 0.967  \bigstrut\\
4 &         DIR-w/o-VVDS  & 0.410 & 0.257  & 0.519 & 0.414  & 0.708 & 0.591  & 0.850 & 0.789  & 1.005 & 0.961  \bigstrut\\
5 &        DIR-w/o-DEEP2  & 0.390 & 0.257  & 0.481 & 0.405  & 0.654 & 0.549  & 0.816 & 0.733  & 1.006 & 0.916  \bigstrut\\
6 &              DIR-C15  & 0.354 & 0.250  & 0.420 & 0.372  & 0.576 & 0.520  & 0.774 & 0.722  & 0.962 & 0.928  \bigstrut\\
7 &               CC-fit  & 0.296 & 0.274  & 0.398 & 0.402  & 0.511 & 0.534  & 0.746 & 0.739  & 1.012 & 0.997  \bigstrut\\
8 &             CC-shift  & 0.437 & 0.300  & 0.537 & 0.463  & 0.667 & 0.577  & 0.822 & 0.769  & 0.992 & 0.952  \bigstrut\\
9 &            OQE-shift  & 0.374 & 0.237  & 0.532 & 0.458  & 0.652 & 0.562  & 0.818 & 0.765  & 0.963 & 0.923  \bigstrut\\
  \end{tabular}
\end{table*}

\section{Code comparison}
\label{app:codecomp}

\begin{figure}
  \includegraphics[width=\hsize]{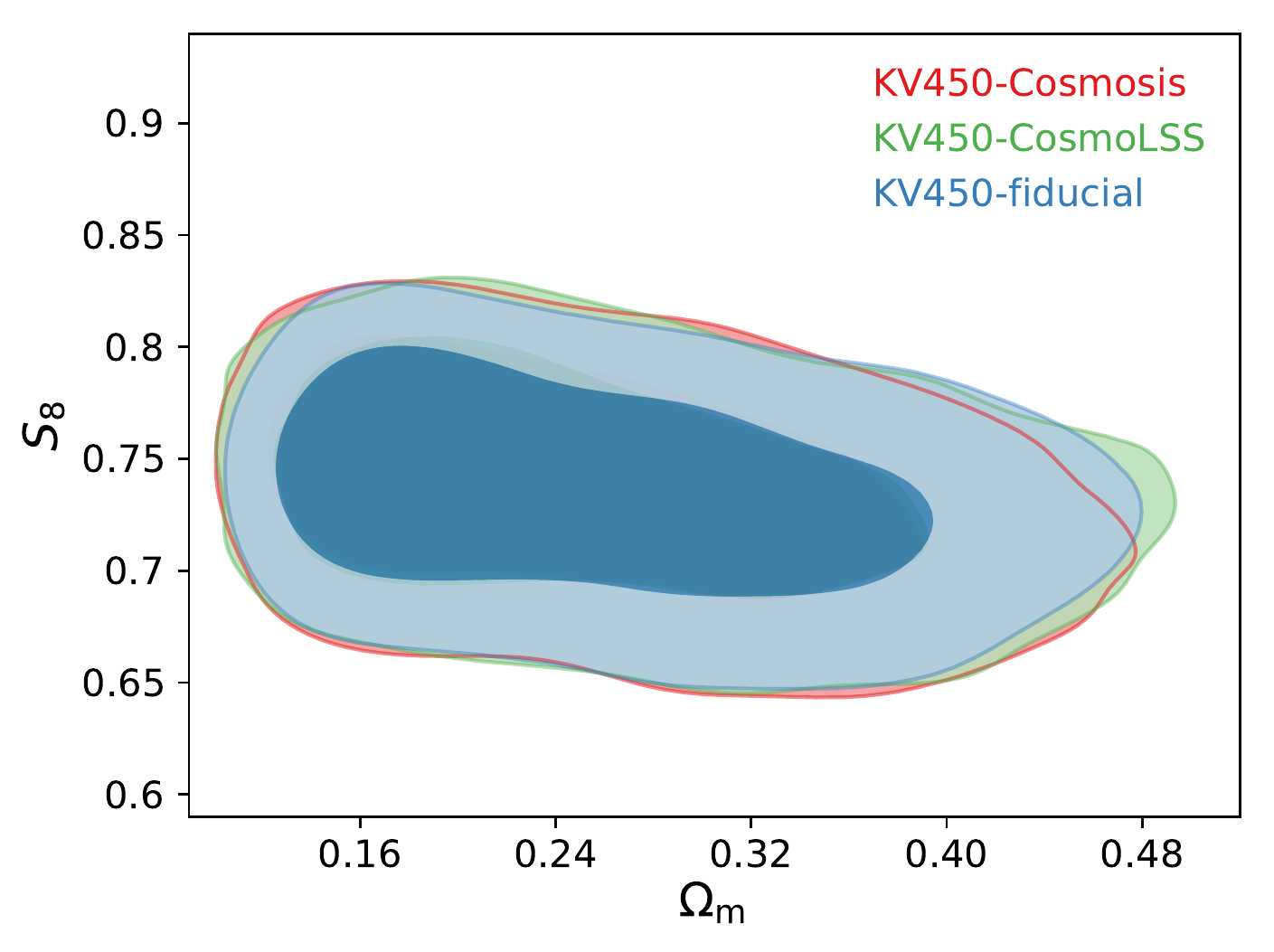}
  \caption{\label{fig:Om_s8_nosys_SJTT} Same as the right panel of Fig.~\ref{fig:Om_s8_fidDz} but showing the fiducial KV450 setup (blue) in comparison to the same setup run with the \citetalias{hildebrandt/etal:2017} \textsc{CosmoLSS} pipeline (based on \textsc{CosmoMC} and \textsc{CAMB}) in green and a \textsc{CosmoSIS}-based pipeline using the \textsc{emcee} sampler in red.}
\end{figure}

We run the fiducial setup with two different analysis pipelines, the cosmological inference code \textsc{CosmoLSS}\footnote{\textsc{CosmoLSS} moreover forms a benchmark of the \textsc{Core Cosmology Library} (CCL; \url{https://github.com/LSSTDESC/CCL}), which was designed to meet the accuracy requirements of LSST DESC.} that was used for \citetalias{hildebrandt/etal:2017} \citep[based on \textsc{CosmoMC} and \textsc{CAMB};][]{lewis/etal:1999,lewis/bridle:2002} as well as a pipeline based on \textsc{CosmoSIS} \citep{zuntz/etal:2015} using the \textsc{emcee} sampler \citep{foreman-mackey/etal:2013}. Results for $\Omega_{\rm m}$ and $S_8$ are shown in Fig.~\ref{fig:Om_s8_nosys_SJTT} in comparison to the \textsc{CLASS}/\textsc{MontePython} setup used in this paper showing excellent consistency. We find $S_8=0.738^{+0.040}_{-0.036}$ for the \textsc{CosmoLSS} setup and $S_8=0.738^{+0.042}_{-0.036}$ for \textsc{CosmoSIS} in comparison to $S_8=0.737_{-0.036}^{+0.040}$ for the fiducial setup.

\section{Project history}
\label{app:history}

In this section we make transparent the history of the KV450 project in relation to the blinding and any changes that happened after unblinding.

As described in Sect.~\ref{sec:blinding} the redshift distributions were blinded by an external person (M. Bartelmann, University of Heidelberg). The analysis was carried out with three different redshift distributions, two perturbed versions and the original one. The amplitude of the blinding was set such that the two extremes differed by $\sim1\sigma$ in terms of $S_8$.

Code development was carried out with one randomly chosen blinding. Over the course of the project the team looked at several blinded versions of plots like the one shown in Fig.~\ref{fig:Om_s8_fidDz}. This led to the discovery of two errors in our analysis pipeline when, regardless which blinding we looked at, the confidence contours were found to be centred on an unnaturally low matter density ($\Omega_{\rm m}\sim0.15$) and a very high power spectrum amplitude ($\sigma_8\sim1.1$), and did not show the hyperbolic degeneracy so typical of cosmic shear constraints. This episode illustrates that our analysis was never blind against the values of $\Omega_{\rm m}$ and $\sigma_8$, whose values are also affected by our prior choices \citep[for a detailed discussion of this effect see][]{joudaki/etal:2017}. Our blinding strategy was only designed to blind us against the value of $S_8$. This goal was reached as even during the discovery of these bugs the corresponding values of $S_8$ were unsuspicious and showed the $\sim1\sigma$ scaling for the different blindings as described above.

After correcting our analysis errors the fiducial analysis was carried out for all three blindings whereas the other setups (see Table~\ref{tab:MCMC}) were only run for a randomly chosen blinding. At the time of unblinding it was clear that we wanted to apply some further minor changes to the fiducial setup after unblinding (in order to save time). This was communicated to the external blind-setter and all these changes have absolutely no effect on the results presented here. In detail these changes were:
\begin{itemize}
\item For the final fiducial results we ran a longer chain for better convergence.
\item We slightly changed the prior for the $A_c$ nuisance parameter from  $A_c = 0.91 \pm 0.08$ to $A_c = 1.01 \pm 0.13$ after re-fitting the 2D $c$-term. This is completely negligible for the end result.
\item We switched to the non-linear power spectrum with a wide prior on $A_{\rm IA}=[-6, 6]$. Before unblinding we used the linear power-spectrum and an informative Gaussian prior. As can be seen from setups no.~11~\&~12 in comparison to the fiducial setup this does not have any significant effect either.
\item We ran more image simulations (before unblinding we had 5 PSF sets, whereas in the final analysis we use 13), which very slightly modified the $m$-bias values for the five tomographic bins. This was done to reduce the statistical error on these estimates, but certainly also changed the central value ever so slightly. The bias changed from \smash{$m = -0.0174, -0.0079, -0.0147, +0.0098, +0.0057$} to \smash{$m = -0.0128, -0.0104, -0.0114, +0.0072, +0.0061$} for the five tomographic bins, which again does not have any effect on the conclusions of this paper.
\item As the randomly chosen blinding did not correspond to the correct redshift distribution we had to re-run all tests from Table~\ref{tab:MCMC} again with the correct redshift distribution and all the updates discussed above.
\end{itemize}

Apart from the blinding strategy it should be noted that the data splits for the consistency checks in Sect.~\ref{sec:consistency} were defined before the data were inspected. Such an approach minimises the look-elsewhere-effect because it prevents the preferential, a-posteriori treatment of peculiar findings.

  After submission of the first version we noticed two shortcomings that we corrected for in the revised version of the paper:
\begin{itemize}
\item We updated the numbers in columns 3, 4, 5, and 6 of Table~2. By mistake, these were not changed after the last iteration of the source catalogue in the first version of the manuscript.
\item We found a small bug in the treatment of the spectroscopic galaxy bias in the CC method, which required us to re-run this redshift calibration method and the CCfit, CCshift, and OQEshift chains. This results in minor changes to the parameters quoted in Table 6, the power-law index $\alpha$ of the bias function $\mathcal{B}(z)$, the shifts quoted in Appendices C.2 and C.3, the CC and OQE $n(z)$ shown in Figs.~C.1 and C.4, the OQE contours in Fig.~5, and the relevant data points in Fig.~6. All of these changes are very minor and barely visible. Note that the OQEshift $n(z)$ also changes because $\mathcal{B}(z)$, determined from CC, is used to correct the photometric galaxy bias in the OQE method.
\end{itemize}

\end{document}